\documentclass[12pt]{article}
\usepackage{jheppub}
\usepackage{graphicx}
\usepackage{array}
\usepackage{amsfonts}
\usepackage{amssymb}
\usepackage[inline]{enumitem}
\usepackage{amsthm}
\usepackage{subfig}
\usepackage{multirow}
\usepackage{hyperref}
\usepackage{mathrsfs}
\usepackage{accents}
\usepackage{relsize}
\usepackage{slashed}
\usepackage{shuffle}
\usepackage{tikz}
\usepackage[T1]{fontenc} 

\usetikzlibrary{patterns,decorations.markings,decorations.pathmorphing}
\tikzset{box/.pic={\filldraw[fill=black]  (0,0) circle (2.5pt);
				   \filldraw [fill=black] (0.5,0) circle (2.5pt);
			       \draw [line width=5pt] (0,0) -- (0.5,0);}}

\makeatletter
\newcommand \UPlus {\mathop {\operator@font \uplus }\limits }
\makeatother
\makeatletter
\newcommand \Bigcup {\mathop {\operator@font \bigcup }\limits }
\makeatother
\usepackage{kbordermatrix}
\renewcommand{\kbldelim}{(}
\renewcommand{\kbrdelim}{)}
\def\LabelNote#1{}
\def\Label#1{\label{#1}%
\smash{\hbox to0pt{\raise1ex\hbox{\tiny[#1]}\hss}}}
  \def\Cdot{{\cdot}}
  
\def\nn{\nonumber}

\def\eqn#1{eq.~(\ref{#1})}

 \def\Je#1.#2.#3{J_{#1 \otimes #2 \otimes #3}}
 \def\Te#1.#2.#3.#4{\varepsilon_{#1}(#2, #3, #4)}
 \def\Tev#1.#2.#3{\varepsilon_{#1}(#2, #3)}
\def\spa#1.#2{\left\langle#1\,#2\right\rangle}
\def\spb#1.#2{\left[#1\,#2\right]}

\def\be{\begin{equation}}
\def\ee{\end{equation}}
\def\bea{\begin{eqnarray}}
\def\eea{\end{eqnarray}}  

\definecolor{green1}{RGB}{76, 153, 0}

\newcommand{\npre}{\mathcal{N}}         
\newcommand{\num}{N}                         
\newcommand{\degree}{d}         
\newcommand{\reyop}{\mathbb{R}}         
\newcommand{\neco}{\mathbb{L}}          
\newcommand{\mec}{\mathsf{C}}
\newcommand{\Dc}{\text{Dim}(\mec)}
\newcommand{\rmom}{X}
\newcommand{\commut}{\Gamma}

\title{Next-to-MHV Yang-Mills kinematic algebra}
\abstract{Kinematic numerators of Yang-Mills scattering amplitudes possess a rich Lie algebraic structure that suggest the existence of a hidden infinite-dimensional kinematic algebra. Explicitly realizing such a kinematic algebra is a longstanding open problem that only has had partial success for simple helicity sectors. In past work, we introduced a framework using tensor currents and fusion rules to generate BCJ numerators of a special subsector of NMHV amplitudes in Yang-Mills theory. Here we enlarge the scope and explicitly realize a kinematic algebra for all NMHV amplitudes. Master numerators are obtained directly from the algebraic rules and through commutators and kinematic Jacobi identities other numerators can be generated. Inspecting the output of the algebra, we conjecture a closed-form expression for the master BCJ numerator up to any multiplicity. We also introduce a new method, based on group algebra of the permutation group, to solve for the generalized gauge freedom of BCJ numerators. It uses the recently introduced {\it binary BCJ relations} to provide a complete set of NMHV kinematic numerators that consist of {\it pure gauge}.}

\author[a]{Gang Chen,}
\author[b,c]{Henrik Johansson,}
\author[b,d]{Fei Teng}
\author[e]{and Tianheng Wang}
\affiliation[a]{Centre for Research in String Theory, School of Physics and Astronomy, Queen Mary University of London, Mile End Road, London E1 4NS, U.K.}
\affiliation[b]{Department of Physics and Astronomy, Uppsala University, Box 516, 75120 Uppsala, Sweden}
\affiliation[c]{Nordita, Stockholm University and KTH Royal Institute of Technology, Hannes Alfv\'{e}ns v\"{a}g 12, 10691 Stockholm, Sweden}
\affiliation[d]{Institute for Gravitation and the Cosmos, Pennsylvania State University, University Park, PA 16802, U.S.A}
\affiliation[e]{Institut f\"ur Physik und IRIS Adlershof, Humboldt-Universit\"at zu Berlin, Zum Gro{\ss}en Windkanal 6, 12489 Berlin, Germany}

\emailAdd{g.chen@qmul.ac.uk}
\emailAdd{henrik.johansson@physics.uu.se}
\emailAdd{fei.teng@psu.edu}
\emailAdd{tianheng.wang@physik.hu-berlin.de}

\preprint{UUITP-21/21 \\
\phantom{~} \hfill NORDITA 2021-031 \\
\phantom{~} \hfill HU-EP-21/10\\
\phantom{~} \hfill QMUL-PH-21-22}

\begin{document} 
\maketitle
\flushbottom
 
\section{Introduction}

Scattering amplitudes provide a rich source for better understanding of quantum field theory and gravity by exposing hidden structures that are not visible in a conventional Lagrangian approach. A structure that underlies many new insights is the Bern-Carrasco-Johansson (BCJ) duality between color and kinematics~\cite{Bern:2008qj,Bern:2010ue,Bern:2019prr}. 
The duality states that scattering amplitudes in many gauge theories may be organized through cubic diagrams, where each diagram consists of a kinematic numerator and a corresponding color factor that satisfy isomorphic Lie-algebraic relations. Since the color factors inherit their properties from the Lie algebra of the gauge group, the duality implies the existence of a hidden kinematic Lie algebra that builds up the kinematic numerators in a similar fashion. 

For gauge theories that have only massless adjoint fields, such as pure Yang-Mills (YM) theory, color-kinematics duality at tree level can be equivalently phrased as the existence of BCJ amplitude relations~\cite{Bern:2008qj,Stieberger:2009hq,BjerrumBohr:2009rd,Feng:2010my,BjerrumBohr:2010hn}. Color-kinematics duality and BCJ amplitude relations were first identified for pure YM theory~\cite{Bern:2008qj} and related pure supersymmetric theories~\cite{Bern:2010ue,Bern:2010yg,Stieberger:2009hq,BjerrumBohr:2009rd}. Many incarnations of the duality were later discovered in a variety of gauge theories~\cite{Bargheer:2012gv,Huang:2012wr,Broedel:2012rc,Chiodaroli:2013upa,Johansson:2014zca,Chiodaroli:2014xia,Johansson:2015oia,Chiodaroli:2015rdg,Johansson:2017srf,Chiodaroli:2018dbu,Johansson:2018ues,Johansson:2019dnu,Bautista:2019evw,Plefka:2019wyg} as well as scalar effective field theories~\cite{Chen:2013fya,Cheung:2016prv,Carrasco:2016ldy,Mafra:2016mcc,Carrasco:2016ygv,Low:2019wuv,Cheung:2020qxc,Rodina:2021isd}. The duality has been generalized to loop-level amplitudes~\cite{Bern:2010ue, Carrasco:2011mn, Bern:2012uf, Boels:2013bi, Bjerrum-Bohr:2013iza, Bern:2013yya, Nohle:2013bfa, Mogull:2015adi, Mafra:2015mja, He:2015wgf,Johansson:2017bfl, Hohenegger:2017kqy, Mafra:2017ioj, Faller:2018vdz, Kalin:2018thp, Ben-Shahar:2018uie, Duhr:2019ywc, Geyer:2019hnn, Edison:2020uzf, Casali:2020knc, DHoker:2020prr, Carrasco:2020ywq, Bridges:2021ebs} and form factors~\cite{Boels:2012ew,Yang:2016ear,Boels:2017ftb,Lin:2020dyj}, and it appears to transpire to certain curved-space observables~\cite{Adamo:2017nia,Farrow:2018yni,Adamo:2018mpq,Lipstein:2019mpu, Prabhu:2020avf, Armstrong:2020woi,Albayrak:2020fyp,Adamo:2020qru,Alday:2021odx}. The existence of color-kinematics duality and BCJ relations in massless gauge theories has been understood from a variety of different perspectives, including string theory, scattering equations, and positive geometry~\cite{BjerrumBohr:2009rd,Stieberger:2009hq,Cachazo:2012uq,Arkani-Hamed:2017mur,Mizera:2019blq}. 

A major insight from the color-kinematics duality is that gravitational amplitudes have the same diagrammatic structure as gauge theory amplitudes, except that the color factors are replaced by a second copy of kinematic numerators~\cite{Bern:2008qj,Bern:2010ue}. This construction of gravity amplitudes is known as the double copy~\cite{Bern:2010ue}. It clarifies the existence of a general connection between gauge and gravity theories, which first came to light through the Kawai-Lewellen-Tye (KLT) relations~\cite{Kawai:1985xq} between open and closed strings. The double copy provides an efficient tool for loop calculations in a large class of gravity theories~\cite{Bern:2011rj, BoucherVeronneau:2011qv, Bern:2013uka,Bern:2014sna,Chiodaroli:2015wal,Johansson:2017bfl,Chiodaroli:2017ehv,Bern:2018jmv,Bern:2021ppb}. Notable recent results using the double copy includes applications to classical solutions~\cite{Monteiro:2014cda,Luna:2015paa,Luna:2016hge,Bahjat-Abbas:2017htu,Carrillo-Gonzalez:2017iyj,Berman:2018hwd,CarrilloGonzalez:2019gof,Goldberger:2019xef,Huang:2019cja,Bahjat-Abbas:2020cyb,Easson:2020esh,Emond:2020lwi, Godazgar:2020zbv,Chacon:2021wbr,Chacon:2020fmr,Alfonsi:2020lub, Monteiro:2020plf, White:2020sfn, Elor:2020nqe,Pasarin:2020qoa}, black-hole scattering and gravitational wave physics~\cite{Luna:2016due,Goldberger:2016iau,Luna:2017dtq,Shen:2018ebu,Plefka:2018dpa,Bern:2019nnu,Plefka:2019hmz,Bern:2019crd,Bern:2020buy,Almeida:2020mrg,Bern:2021dqo}.

In this paper we are interested in the mathematical consequences of color-kinematics duality for pure YM theory. The duality-satisfying numerators, also known as BCJ numerators, can by now be computed for pure YM theory  in principle up to any multiplicity at tree level~\cite{BjerrumBohr:2010hn,Mafra:2011kj,Fu:2012uy,Mafra:2015vca,Bjerrum-Bohr:2016axv,Du:2017kpo,Chen:2017bug,Fu:2018hpu,Edison:2020ehu,Hou:2021mvg}. However, the details of the underlying kinematic Lie algebra are still understood only at a rudimentary level. General properties of it can be inferred from our knowledge of the BCJ numerators: the algebra should be infinite dimensional, since the structure constants must be parametrized by continuous momenta. It should not be invariant under gauge transformations since the BCJ numerators transform non-trivially. Indeed, gauge transformations may play a central role in the details of the algebra. The structure constants of the algebra should be isomorphic to interaction terms in a cubic Lagrangian that give BCJ numerators through its Feynman rules.  However, there are many unknowns regarding the precise formulation of the algebra. Some details have been exposed by explicit constructions in simple helicity sectors.

The first explicit kinematic-algebra construction was formulated in the self-dual sector of YM theory, by Monteiro and O'Connell~\cite{Monteiro:2011pc}. 
They found an explicit generator representation of the kinematic algebra in this sector, and recognized it as corresponding to area-preserving diffeomorphisms.  A plane wave factor and a derivative operator built up a generator, and the commutator of two generators closed in the Lie algebra with structure constants identified as interaction terms of self-dual YM. Thus the tree-level Feynman diagrams mapped to a Lie bracket structure, explicitly manifesting color-kinematics duality. While these diagrams provide a self-dual solution to the YM field equations, taken on shell they give tree-level scattering amplitudes that vanish. Indeed, only one-loop amplitudes are non-vanishing in self-dual YM theory~\cite{Cangemi:1996rx} (see ref.~\cite{Boels:2013bi} for color-kinematics duality for these amplitudes). The maximally-helicity-violating (MHV) sector of YM gives the simplest non-zero amplitudes in 4D, and while ref.~\cite{Monteiro:2011pc} considered it through a non-local gauge choice, a local kinematic algebra of the MHV sector was unknown for quite some time.

A second example of an explicit construction of a kinematic algebra sub-sector is by Cheung and Shen~\cite{Cheung:2016prv}. They realized a cubic Lagrangian for the $SU(N)$ non-linear sigma model, which directly gave tree-level Feynman rules that obey color-kinematics duality. A curious feature of their construction was that they relied on vector fields for describing a scalar field. Indeed, in later work it was made clear that this Lagrangian is a truncation of YM via dimensional reduction operations~\cite{Cheung:2017yef,Cheung:2017ems}. And in ref.~\cite{Chen:2019ywi} it was observed that the Cheung-Shen Lagrangian indirectly computes local BCJ numerators for the MHV sector of YM theory. 

It is convenient to promote the 4D concept of (Next-to)${}^k$-MHV sectors to general spacetime dimension by recognizing that this grading of YM translates to independent structures in a BCJ numerator~\cite{Chen:2019ywi}. By the MHV sector we mean the numerator terms with the fewest $\varepsilon_i {\cdot} \varepsilon_j$ factors, which in YM implies the schematic form $\sim \varepsilon_i {\cdot} \varepsilon_j \prod \varepsilon_k {\cdot} p_l $.  Such numerator terms are sufficient for computing MHV amplitudes in 4D~\cite{Chen:2019ywi}. It turns out that the BCJ numerator built from these terms is unique at every multiplicity~\cite{Chen:2019ywi}. The Cheung-Shen Lagrangian computes all these terms in the unique BCJ numerator, thus providing a local realization of the kinematic algebra for the MHV sector of YM. Curiously, a closely related kinematic algebra has recently been formulated  as a non-abelian generalization of the Navier-Stokes equation~\cite{Cheung:2020djz} (see also ref.~\cite{Keeler:2020rcv}).

In earlier work, duality-satisfying effective Lagrangians were constructed using non-local terms or auxiliary fields, which gave Feynman rules that computed BCJ numerators up to five~\cite{Bern:2010yg} and six points~\cite{Tolotti:2013caa}. These constructions encountered a proliferation of ambiguities related to the non-uniqueness of the BCJ numerators. We now understand that these ambiguities start in the NMHV sector of the YM kinematic algebra~\cite{Chen:2019ywi}, and they are a reflection of the generalized gauge freedom of the BCJ numerators~\cite{Bern:2008qj,Bern:2010ue}, which include the standard gauge freedom. Because of these difficulties, refined frameworks are needed for getting the kinematic algebra under control. 

A first attempt at formulating a NMHV-sector kinematic algebra, for $D$-dimensional local numerators, was made in our earlier work~\cite{Chen:2019ywi}. The problem was distilled down to a sub-sector of the NMHV numerators, made out of terms proportional to a fixed polarization product $\varepsilon_1 {\cdot} \varepsilon_n$, which effectively meant studying two scalars interacting with the YM field. This bi-scalar sector retains much of the NMHV sector details, and a new framework based of tensor currents and fusion products realized the algebra in this sector. Interestingly, the tensor currents both provided the algebra with generators of enlarged Lorentz tensor structure, and encoded part of the generalized gauge freedom in an enlightening way.

In the first part of this paper, we construct a realization of the kinematic algebra for the complete NMHV sector of $D$-dimensional YM. We expand on the framework of tensor currents and fusion products previously introduced~\cite{Chen:2019ywi}. It is convenient to divide up the computational task into an algebraic construction of a pre-numerator, and subsequently a BCJ numerator is obtained through a nested set of commutators applied to the pre-numerator. This mimics the construction of color factors by expressing them as nested commutators of traces of the gauge group generators. In the bi-scalar subsector, the nested commutator collapsed~\cite{Chen:2019ywi} into an ordered product of generators which meant that the distinction between pre-numerator and numerator was not necessary in that subsector. Compared to the previous work~\cite{Chen:2019ywi}, we have a larger set of tensor currents and somewhat more involved fusion products that defines an algebraic structure in the NMHV sector. After constructing the algebra, we note that the resulting pre-numerator admits an all-multiplicity closed formula. The master numerators (the half-ladder diagram numerators) as well as all the other numerators, are obtained through $(2n-5)!!$ different nested commutators of the pre-numerator, and enjoys a manifest $S_{n-1}$ crossing symmetry.

In the second part of the paper, we observe that the binary BCJ relations introduced in ref.~\cite{Chen:2019ywi} endows the local BCJ numerators with a complete parametrization of the generalized gauge freedom for the NMHV sector. This is an important observation that we expect to have natural generalizations for higher N${}^{k}$MHV sectors, and thus may enable a full control over the ambiguities in a local description of the kinematic algebra or equivalent Lagrangian descriptions. In the present context, we analyse the generalized gauge freedom in great detail by prescribing how to fully classify the polynomials that can appear. They are given by invariant functions under a specific element of the group algebra over the permutation group. This element acts as a simple projector on monomials, which reduces the problem of constructing pure-gauge BCJ numerators to a computationally straightforward task. We explicitly count all the degrees of freedom that can appear in pure-gauge BCJ numerators up to seven points, including the N${}^2$MHV sector.

The paper is organized as follows: In section~\ref{sec:kinAlg}, we introduce the general setup that is based on the analogous structures of matrix products between Lie algebra generators and fusion products of vector and tensor currents. In section~\ref{sec:FusionProd}, we construct the fusion products necessary for describing the NMHV sector pre-numerator. In section~\ref{sec:numerator}, we present a closed formula for the pre-numerator as obtained from the algebraic construction. After completing the particular kinematic algebra, we switch gears and study the more general problem of pure gauge freedom in the NMHV sector BCJ numerators. We introduce essential mathematical background on group algebra in section~\ref{eq:gaaction}. We then show in section~\ref{sec:PureGaugeNMHV} that the NMHV sector pure gauge terms with $S_{n-2}$ crossing symmetry can be classified by the kinematic polynomials that are invariant under certain group algebra action. In section~\ref{sec:Sn-1puregauge}, we discuss the generalization of this construction to $S_{n-1}$ and $S_n$ crossing symmetric pure gauge terms. Conclusion and outlook are in section~\ref{sec:conclusion}.

\section{Kinematic algebra from tensor currents}\label{sec:kinAlg}

Here we review and extend some formal notations that were introduced in ref.~\cite{Chen:2019ywi}.  We denote a generator of a putative algebra as $J_U$, and consider products of two such generators, $J_U\star J_V$, which we refer to as a fusion product. The fusion product takes the form
\begin{align}
J_U \star J_V = \sum_{W} f_{UV}^{W} J_{W},
\end{align}
where the $f$'s are coefficients that encode the algebraic structure. The fusion product is required to be linear in the first argument,
\begin{align}
(c_1 J_{U}+c_2 J_{V})\star J_W&=c_1 (J_U\star J_W)+c_2 (J_V\star J_W)\,.
\end{align}
In principe, it should also be linear in the second argument, but it will not be necessary for the purpose of this paper.

Let us refine the details of the generators, so that the algebraic framework can describe objects of a kinematic algebra for YM numerators. We upgrade the generators to tensor (vector) currents $ J_U \rightarrow J^{(w)}_U(p)$ that are specified by three types of arguments: 
\begin{enumerate*}[label=(\arabic*)]
	\item a Lorentz tensor structure $U$; 
	\item momentum argument $p$;
	\item additional degeneracy index $w$ to distinguish otherwise identical generators.
\end{enumerate*}
The tensor structure is constructed as a tensor product $U=a_1\otimes a_2\otimes\cdots\otimes a_r$, where $a_i$'s are polarization vectors and momenta. We will refer to the number of vectors in $U$ as the \emph{tensor rank} $r$, and a rank-one tensor current is thus a vector current. The coefficients in the fusion product $J^{(w)}_U(p) \star J^{(w')}_V(p') $ should be polynomials of the Lorentz invariant products between the kinematic objects that appear in the arguments of the generators. However for some tensor types we will need a slight generalization, which we will discuss later. 

The language of currents is borrowed from the old topic of current algebras, but unlike that context the fusion products we consider will be strictly \emph{local}, since we are interested in describing the kinematic algebra of local BCJ numerators, which have no poles.  In principle, it would be ideal to label the generators with only the kinematic data of the external states of an amplitude, however, there is no guarantee that these are one-to-one. As already observed in ref.~\cite{Chen:2019ywi}, currents corresponding to internal states of identical tensor and momentum structure may require additional labels for the algebra to close. Thus an additional degeneracy index $w$ is necessary to allow for a richer set of potential fusion products. 
These considerations imply that we have to be careful to distinguish the vector current associated with a (physical, on-shell) external gluon from an internal vector current of a similar type. We denote the vector current associated with a on-shell gluon as $J_{\varepsilon_i}(p_i)$, where $\varepsilon_i$ and $p_i$ are the polarization vector and on-shell momentum of the gluon, and the $w$ index is omitted. 

By analogy with the workings of a Lie algebra, we imagine that the generators of the  kinematic algebra transform in some particular representation that is appropriate for spin-1/2 particles. This implies that we can try to realize an on-shell gluon as a bilinear in the on-shell fermion wave function, as should be very familiar from the spinor-helicity framework. However, we will only require that one gluon, corresponding to leg $n$ of an $n$-point amplitude, is obtained this way, whereas the other gluons $i\in \{1,\ldots, n-1\}$ are obtained by writing down the corresponding generators $J_{\varepsilon_i}(p_i)$. That is, we formally split the polarization vector $\varepsilon_n$ into a spinor $u_n(p_n)=| n \rangle$ and a conjugate spinor $\bar{v}_q(q)= \langle q|$, where the latter carries soft (reference) momentum $q$. Then we use these spinors to sandwich the remaining $n{-}1$ generators that are fused according to an ordering that corresponds to a half-ladder (or multiperiferal) diagram: 
\begin{align}\label{eq:OrderProduct}
\renewcommand{\arraystretch}{1.2}
\begin{array}{>{\centering $}p{4cm}<{$}>{\centering $}p{4cm}<{$}>{\centering\arraybackslash $}p{5cm}<{$}}
\text{color factor} & \multirow{2}{*}{\begin{tikzpicture}[baseline={([yshift=-0.8ex]current bounding box.center)},every node/.style={font=\tiny}]
	\draw [decoration={markings, mark=at position 0.15 with \arrow{latex},mark=at position 0.35 with \arrow{latex},mark=at position 0.75 with \arrow{latex},mark=at position 0.94 with \arrow{latex}},postaction=decorate] (2.0,0) node[above=0pt] {$n$} -- (-0.5,0) node[above=0pt] {$\text{soft}\, q$};
	\draw [decorate,decoration={coil,segment length=4pt}] (1.5,0) -- ++(0,0.6) node[above=0pt] {$n{-}1$};
	\draw  [decorate,decoration={coil,segment length=4pt}] (1.0,0) -- ++(0,0.6) node[above=0pt] {$\cdots$};
	\draw [decorate,decoration={coil,segment length=4pt}] (0.5,0) -- ++(0,0.6) node[above=0pt] {$2$};
	\draw [decorate,decoration={coil,segment length=4pt}] (0,0) -- ++(0,0.6) node[above=0pt] {$1$};
	\end{tikzpicture}} & \text{kinematic factor} \\
(T^{a_1} T^{a_2}\cdots T^{a_{n-1}})_{i_q}^{\bar{\imath}_n} &  & \langle q|J_{\varepsilon_1}\star J_{\varepsilon_2}\star \cdots \star J_{\varepsilon_{n-1}}|n\rangle                
\end{array}\,.
\end{align}
This is analogous to how a color factor of $n$ adjoint particles can be constructed using the Lie algebra generators $T^{a_i}$, and where the color of leg $n$ is only implicitly represented through the bi-fundamental indices $(\bar{\imath}_n,i_q)$.\footnote{To obtain a physical state one needs to project the bi-fundamental $ (\bar{\imath}_n,i_q)$ onto the adjoint representation, and likewise project the kinematical bi-spinor $\langle q| \cdots |n\rangle$ onto a gluon state.} Note that the kinematic factor in \eqn{eq:OrderProduct} cannot be directly calculated, as we will not give an explicit representation for the generators. Instead, we need to use the algebraic properties of the fusion product to reduce it down to simpler objects involving fewer generators.

By convention, we always evaluate a fusion product from left to right,
\begin{align}\label{eq:orderproduct}
J_{\varepsilon_1}\star J_{\varepsilon_2}\star \cdots \star J_{\varepsilon_{n-1}} \equiv \left(\cdots \left((J_{\varepsilon_1}\star J_{\varepsilon_2})\star J_{\varepsilon_3}\right)\cdots \right)\star J_{\varepsilon_{n-1}}\,,
\end{align}
as this will reduce the number of fusion product coefficients that we need to specify. 
At every step we encounter a two-to-one map $X \star J_{\varepsilon_i} \rightarrow Y $, where one generator is a physical gluon current. Evaluating such fusion products recursively will simplify the expression down to terms with a single current sandwiched between the spinors, which we can identify with an external state corresponding to leg $n$.

The kinematic factor in eq.~(\ref{eq:OrderProduct}) is not yet the BCJ numerator of a pure YM diagram, since pictorially it resembles more a diagram with a massless fermion. Indeed, we will call the result of the above ordered fusion product the \emph{pre-numerator}, denoted by $\mathcal{N}$,
\begin{align}\label{eq:npre}
\npre(1,2,\cdots, n-1,n)\equiv\langle q|J_{\varepsilon_1}\star J_{\varepsilon_2}\star \cdots \star J_{\varepsilon_{n-1}}|n\rangle\Big|_{q ~ {\rm soft}}\,.
\end{align}
Kinematically one should consider the momentum $q$ to be soft compared to $p_n$, thus we can identify $p_n+ q\approx p_n$, but $q$ will provide a reference direction for the polarization vector $\varepsilon_n=\varepsilon_n(p_n,q)$. Note that we do not aim at identifying the pre-numerator with a fermion diagram in full detail, the figure in eq.~(\ref{eq:OrderProduct}) only serves as a motivation for understanding the general properties of the pre-numerator. 

Following the computational steps outlined, the pre-numerator becomes a linear superposition of individual vector and tensor currents sandwiched between the two spinors. To identify such objects with external states in the final step, we make the choice to not distinguish between currents of different $w$ when they are sandwiched between two spinors, hence we drop this index. Then, for a vector current $J_a$, it is natural to evaluate it as 
\begin{align}\label{eq:vector}
\langle q|J_{a}|n\rangle\big|_{q\text{ soft}}=\bar{v}_q\gamma_\mu u_n a^{\mu}\big|_{q\text{ soft}}\equiv a\Cdot\varepsilon_n\,,
\end{align}
and for a tensor current $J_{a_1\otimes a_2\otimes\cdots\otimes a_r}$, we identify it as
\begin{align}\label{eq:tensor}
\langle q|J_{a_1\otimes a_2\otimes\cdots\otimes a_r}|n\rangle\big|_{q~\rm soft}=(\bar{v}_q\gamma_{\mu_1}\ldots\gamma_{\mu_r}u_n)a_1^{\mu_1}\ldots a_r^{\mu_r}\big|_{q~\rm soft}\,.
\end{align}
This object is natural. If we antisymmetrize the Lorentz indices, then the object $(\bar{v}_q\gamma_{[\mu_1}\ldots\gamma_{\mu_r]}u_n)\big|_{q~\rm soft}$ is just the on-shell polarization tensor of an $r$-form field. It is automatically transverse to both $p_n$ and $q$ hence it encodes the tensor structure of the $\text{SO}(D-2)$ little group.  

For completeness, we may consider an analogous decomposition of the color factor in eq.~(\ref{eq:OrderProduct}) into representations of the Lie algbra, but we are primarily interested in the adjoint representation. This is obtained through the projection
\begin{align}\label{eq:Cf}
\mathcal{C}(1,2,\ldots,n)\equiv (T^{a_1} T^{a_2}\cdots T^{a_{n-1}})_{i_q}^{\bar{\imath}_n}(T^{a_n})_{i_n}^{\bar{\imath}_q}=\text{tr}(T^{a_1}T^{a_2}\cdots T^{a_n})\,,
\end{align}
which gives a standard trace factor. Thus, by analogy, one may be tempted to think of the pre-numerator as a kinematic analog of the trace of $n$ generators. However, a slight obstruction is that the cyclic symmetry of the pre-numerator is not guaranteed, and we will not assume it in general, thus we keep leg $n$ in a fixed position in $\npre(1,2,\ldots,n)$. See ref.~\cite{Bern:2011ia}, where an alternative interpretation of kinematic traces were given.

Next, we use the analogy between $\mathcal{C}(1,2,\ldots,n)$ and $\npre(1,2,\ldots,n)$ to realize the color-kinematics duality of pure YM theory. As is by now well known~\cite{Bern:2008qj}, one can write color-dressed pure YM tree amplitudes as a sum over $(2n-5)!!$ cubic graphs,
\begin{align}
\mathcal{A}_n=\sum_{\Gamma\in\text{cubic}}\frac{C_\Gamma\,N_\Gamma}{D_\Gamma}\,,
\end{align}
where $C_\Gamma$, $N_\Gamma$ and $1/D_\Gamma$ are respectively the color factor, kinematic numerator factor and propagator factor of the graph $\Gamma$. Throughout this paper we set gauge coupling constants, and other overall numerical factors in the amplitude, to unity. 

It is convenient to represent an $n$-point cubic graph $\Gamma$ as a nested commutator of the elements $1,\ldots, n{-}1$, where the number of commutators used is $n{-}2$. For example, it can be recursively constructed starting from the seed $[1,2]$ and then, for $i=3,\ldots,n-1$, substitute any Lie-valued object (particle label or commutator) using the rule $* \rightarrow [*,i]$. Consider the next step, there are three Lie-valued objects in $[1,2]$, namely $\{1,2, [1,2]\}$. Applying rule for $i=3$ then gives: $\{[[1,3],2],[1,[2,3]], [[1,2],3]\}$. At multiplicity $n$, one can count that there are $1\times 3 \times 5 \times \ldots \times (2n-5) = (2n-5)!!$ nested commutator expressions generated by this rule, precisely matching the number of cubic graphs. Note that the nested commutator representation has the feature that leg $1$ always appears in the first position, legs $2\,\ldots,n{-}1$ can appear in any order, whereas leg $n$ does not appear at all. For a graph $\Gamma$, we identify its nested commutator representation with the graph itself; e.g. $\Gamma= [\ldots[[[1,2],3],4],\ldots,n{-}1]$. 

We can use this notation to define the inverse propagator factor more precisely. We have $D_\Gamma = \prod_{\gamma \in \Gamma} s_\gamma$, where $\gamma$ is a nested commutator subgraph of $\Gamma$ containing at least two particle labels, and $s_\gamma= \sum_{i,j \in \gamma} p_i {\cdot} p_j$ gives the invariant momentum square of this subgraph. Similarly, using the color traces $\mathcal{C}(1,2,\ldots,n)$ defined in eq.~\eqref{eq:Cf} one can define the color factor of a cubic graph as $C_{\Gamma}$,
\begin{align}
C_{\Gamma}=\mathcal{C}(\commut,n)\,.
\end{align} 
where the commutators are expanded according to the natural rule $\mathcal{C}(\ldots [A,B]\ldots ,n) = \mathcal{C}(\ldots A,B\ldots ,n) -\mathcal{C}(\ldots B, A \ldots ,n)$. As a consequence of this commutator rule, the color factors will satisfy Jacobi identities. For example, at $n{=}4$ the color factor for the $s$-channel diagram can be written as
\begin{align}
C_s=C\left(\!\begin{tikzpicture}[baseline={([yshift=-0.8ex]current bounding box.center)},every node/.style={font=\tiny}]
\draw [decorate,decoration={coil,segment length=4pt}] (0,0) -- ++(60:0.5) node [right=-2pt]{$3$};
\draw [decorate,decoration={coil,segment length=4pt}] (0,0) -- ++(-60:0.5) node [right=-2pt]{$4$};
\draw [decorate,decoration={coil,segment length=4pt}] (0,0) -- (-0.6,0);
\draw [decorate,decoration={coil,segment length=4pt}] (-0.6,0) -- ++(120:0.5) node[left=-2pt]{$2$};
\draw [decorate,decoration={coil,segment length=4pt}] (-0.6,0) -- ++(-120:0.5) node[left=-2pt]{$1$};
\end{tikzpicture}\!\right)&=\mathcal{C}\big(\overbrace{[[1,2],3]}^{\commut},4\big)\nonumber\\
&=\mathcal{C}(1,2,3,4)-\mathcal{C}(2,1,3,4)-\mathcal{C}(3,1,2,4)+\mathcal{C}(3,2,1,4)\nonumber\\
&=f^{a_1a_2b}f^{ba_3a_4}\,.
\end{align}
Similarly for the other two channels, we have
\begin{align}
C_t=\mathcal{C}\big([1,[2,3]],4\big)=\mathcal{C}(1,2,3,4)-\mathcal{C}(1,3,2,4)-\mathcal{C}(2,3,1,4)+\mathcal{C}(3,2,1,4)\,,\nonumber\\
C_u=\mathcal{C}\big([[1,3],2],4\big)=\mathcal{C}(1,3,2,4)-\mathcal{C}(3,1,2,4)-\mathcal{C}(2,1,3,4)+\mathcal{C}(2,3,1,4)\,,
\end{align}
and the Jacobi identity $C_s-C_u=C_t$ is automatic. 

Analogously, the color-kinematics duality will be manifest by construction if the numerators are obtained from the pre-numerators through the same nested commutator expression,
\begin{align}\label{eq:AfromFusionRules}
N_{\Gamma}=\npre(\commut,n)\,.
\end{align}

Of particular importance are the numerators of half-ladder diagrams with leg $1$ and $n$ fixed, which are obtained through the left-nested commutators
\begin{align}
N(1,2,3,\ldots,n-1,n)&\equiv N\left(\!\begin{tikzpicture}[baseline={([yshift=-0.8ex]current bounding box.center)},every node/.style={font=\tiny}]
\draw [decorate,decoration={coil,segment length=4pt}] (0,0) node [left=-2pt]{$1$} -- (0.5,0);
\draw [decorate,decoration={coil,segment length=4pt}] (0.5,0) -- (1,0);
\draw [decorate,decoration={coil,segment length=4pt}] (1,0) -- (2,0);
\draw [decorate,decoration={coil,segment length=4pt}] (2,0) -- (2.5,0) node[right=-2pt]{$n$};
\draw [decorate,decoration={coil,segment length=4pt}](0.5,0) -- ++(0,0.6) node[above=-2pt]{$2$};
\draw [decorate,decoration={coil,segment length=4pt}] (1,0) -- ++(0,0.6) node[above=-2pt]{$3$};
\node at (1.5,0) [above=0.25cm]{$\cdots$};
\draw [decorate,decoration={coil,segment length=4pt}] (2,0) -- ++(0,0.6) node[above=-2pt]{$n{-}1$};
\end{tikzpicture}\!\right)\nonumber\\
&\equiv\mathcal{N}\big([\ldots[[1,2],3],\ldots,n-1],n\big)\,.
\end{align}
The relabelings of this half-ladder (or master) numerator, $N(1,\sigma_2, \sigma_3,\cdots, \sigma_{n-1},n)$, form a $(n{-}2)!$ basis under Jacobi identities, also known as the Del-Duca-Dixon-Maltoni (DDM) basis~\cite{DelDuca:1999rs}. Therefore, the half-ladder numerator $N(1,2,\ldots,n-1,n)$, as well as the pre-numerator $\mathcal{N}(1,2,\ldots,n-1,n)$, are the central objects that we study in this paper. Other BCJ numerators can be inferred by Jacobi relations and permutations. 

Consider the same examples as for the color factors at multiplicity $n{=}4$. The DDM basis is given by the $s$- and $u$-channel numerators,
\begin{align}
N_s=N(1,2,3,4)&=\npre\big([[1,2],3],4\big)\nonumber\\
&=\npre(1,2,3,4)-\npre(2,1,3,4)-\npre(3,1,2,4)+\npre(3,2,1,4)\,,\nonumber\\
N_u=N(1,3,2,4)&=\npre\big([[1,3],2],4\big)\nonumber\\
&=\npre(1,3,2,4)-\npre(3,1,2,4)-\npre(2,1,3,4)+\npre(2,3,1,4)\,.
\end{align}
The $t$-channel numerator is not a basis element, it is written as $N_t=\mathcal{N}\big([1,[2,3]],4\big)$, which automatically implies the kinematic Jacobi identity $N_s-N_u=N_t$. 

Using the BCJ numerators and color factors in the DDM basis, we can re-write the color-dressed YM amplitude as \cite{Bern:2008qj,Vaman:2010ez}
\begin{align} \label{amplitude_in_basis}
\mathcal{A}_n=\sum_{\Gamma\in\text{cubic}}\frac{C_\Gamma\,N_\Gamma}{D_\Gamma} = \sum_{\sigma,\rho \in S_{n-2}}  C(1,\sigma_2,\cdots,\sigma_{n-1},n) m(\sigma | \rho) N(1,\rho_2\cdots,\rho_{n-1},n)\,,
\end{align}
where the $(n{-}2)!$-by-$(n{-}2)!$ matrix $m(\sigma | \rho)$ is built out of linear combinations of the scalar-type  propagators $1/D_\Gamma$, as given by the decomposition of the BCJ numerators and color factors into the DDM basis. It goes by many names in the literature, it is called the ``propagator matrix''~\cite{Vaman:2010ez}, the ``inverse of the KLT kernel''~\cite{Kawai:1985xq,Cachazo:2013iea}, or the ``bi-adjoint scalar amplitude''~\cite{Cachazo:2013iea}. It may also be identified as the double-partial amplitudes of ``dual-scalar theory''~\cite{Bern:2010yg,BjerrumBohr:2012mg}, ``color-scalar theory''~\cite{Du:2011js} or ``scalar $\phi^3$ theory'' \cite{Bern:1999bx,Chiodaroli:2014xia}. 

The color-ordered partial amplitudes follow from \eqn{amplitude_in_basis} as the kinematic factor multiplying each independent color factor, which gives a (non-invertible) map between BCJ numerators and partial amplitudes
\begin{align}
A(1,\sigma,n) =& \sum_{\rho \in S_{n-2}} m(\sigma | \rho) N(1,\rho,n)\,.
\end{align}
The propagator matrix is not invertible for on-shell momenta and hence it has a kernel (or null space). This implies that one can find contributions to BCJ numerators that live in this kernel, and do not feed into the partial amplitudes. Hence, BCJ numerators are in general not unique.  The ambiguity is called \emph{generalized gauge freedom}~\cite{Bern:2008qj,Bern:2010ue} and it corresponds to shifting the existing numerators by what we call \emph{pure gauge} numerators
\begin{align}\label{eq:DefinePureGauge}
\num(1,2,\ldots, n-1,n)\sim \num(1,2,\ldots,n-1,n)+\num^{\text{gauge}}(1,2,\ldots,n-1,n)\,,
\end{align} 
where the pure gauge numerators are annihilated by the propagator matrix,
\begin{align}
\sum_{\rho\in S_{n-2}}m(\sigma|\rho)N^{\text{gauge}}(1,\rho,n)=0\,. 
\end{align} 
The generalized gauge freedom subsume the standard gauge freedom for vector fields, and it generalizes it to also include any other operations, such as field redefinitions, that changes the cubic diagram numerators but leaves the amplitude invariant. 

Finally, we note that the BCJ numerators constructed from pre-numerators will automatically satisfy crossing symmetry relations in the legs $1,\ldots,n{-}1$, because these legs are on equal footing in the pre-numerator, meaning that any permutations of them are allowed to appear. We will refer to this property as exhibiting (manifest) $S_{n-1}$ crossing symmetry, and we note that this symmetry is larger than what is naively obtained from a standard DDM basis of numerators, which naturally admits a manifest $S_{n-2}$ crossing symmetry. See, e.g., ref.~\cite{Edison:2020ehu} for an all-multiplicity form of BCJ numerators for YM that exhibit $S_{n-2}$ crossing symmetry.

\section{Fusion product of currents in the NMHV sector}\label{sec:FusionProd}
In this section, we  explicitly construct the fusion product rules which are used to compute the duality-satisfying kinematic numerators up to the quadratic order in $\varepsilon_i\Cdot\varepsilon_j$ factors. Before we get into the details, let us briefly clarify the decomposition of the kinematic numerator into different sectors based on the structure of the variables that it contains.

As previously mentioned, we consider only local kinematic numerators which are polynomials of Lorentz products of momenta and polarization vectors. We classify the terms in the numerators according to the number of $\varepsilon_i\Cdot\varepsilon_j$ factors, which we call \emph{polarization power}. Terms of different polarization power splits into independent sectors in the BCJ numerator; that is, the sectors do not mix under: permutations of labels, momentum identities, or Jacobi identities.  Since these sectors are in a one-to-one correspondence with the 4D helicity sectors, denoted by N${}^k$MHV, via the gauge choices for the polarization vectors explained in ref.~\cite{Chen:2019ywi}, we will often use the 4D language when referring to these sectors. 

In this paper, we will mostly focus our attention to terms of polarization power two (NMHV), as terms of polarization power one (MHV) are already uniquely known~\cite{Chen:2019ywi}, and terms of polarization power $k{+}1$ (N${}^k$MHV) are beyond our scope. Schematically, the two simplest sectors look as
\begin{equation}\setlength\arraycolsep{5pt} \label{polarization_power}
	\begin{array}{ c c c c c }
		\!\!\!\!  \text{polarization power one:} & (\varepsilon_{i} \Cdot \varepsilon_j) \prod (\varepsilon_k \Cdot p_l) \phantom{\Big|}
		& \!  \!  \rightarrow  & \text{MHV sector} \\ 
		\!\!\!\! \text{polarization power two:} & \! \!  (\varepsilon_{i_1} \Cdot \varepsilon_{j_1}) (\varepsilon_{i_2} \Cdot \varepsilon_{j_2}) (p_{i_3} \Cdot p_{j_3}) \prod (\varepsilon_k \Cdot p_l) \phantom{\Big|}
		& \! \!  \rightarrow & \text{NMHV sector}\,. 
	\end{array} 
\end{equation}
Recall that the mass dimension of kinematic numerators in YM must be $n-2$. Thus the requirement of locality induces a correspondence between the number of $\varepsilon_i\Cdot\varepsilon_j$ factors and $p_i\Cdot p_j$ factors. The polarization power one terms contain no $p_i\Cdot p_j$ factors whereas those of polarization power two are linear in the Mandelstam variables. This explains why BCJ numerators of  polarization power two are not unique, since the $p_i\Cdot p_j$ factors may conspire with the propagator denominators and produce a contact interaction, which cannot be uniquely attributed to a cubic diagram.

\subsection{Currents in NMHV kinematic algebra}\label{sec:currents}
Ref.~\cite{Chen:2019ywi} introduced the fusion products necessary to compute the kinematic numerators in the so-called bi-scalar sector of YM, in which the terms have polarization power up to two, and a common fixed factor $\varepsilon_1\Cdot\varepsilon_n$. In order to generalize to the full NMHV sector, we need additional fusion rules that compute the $\varepsilon_n\Cdot p_i$ and $\varepsilon_i\Cdot\varepsilon_n$ terms. As we restrict our discussion to terms of polarization power one and two, we only need to consider currents whose tensor ranks are at most three. A consideration based on mass dimension of the tensor currents indicates that we can have at most two momenta in the tensor labels~\cite{Chen:2019ywi}. In hindsight, we claim that the relevant tensor currents in the NMHV sector are as follows,
\begin{gather}\label{eq:alltensors}
J_{\varepsilon_i}(p)\,, \qquad J_{p_i}(p)\,, \qquad J_{\varepsilon_i\otimes\varepsilon_j\otimes\varepsilon_k}(p)\,, \qquad J_{p_i\otimes\varepsilon_j\otimes\varepsilon_k}(p)\,,\nonumber\\
J_{\varepsilon_i\otimes\varepsilon_j\otimes p}(p)\,, \qquad J_{p_i\otimes \varepsilon_j \otimes p}(p)\,.
\end{gather}
where we use $p$ without subscript to denote the momentum carried by the tensor.  When computing the ordered fusion products in the pre-numerator~\eqref{eq:npre}, we only need those tensors with $i{<}j{<}k$. The tensor currents also depend on the momentum of the polarization vectors in the tensor label. The superscripts that label the vector and tensor types are suppressed in eq.~\eqref{eq:alltensors}, which will be later worked out based on their behavior in fusion products. The current $J_{\varepsilon_i\otimes\varepsilon_j\otimes p}(p)$ and $J_{p_i\otimes\varepsilon_j\otimes p}(p)$, where the last tensor label agrees with its momentum, will play a special role and provide a simple realization of the fusion products.

In the on-shell limit, we impose a Clifford-algebra like relation among tensors,
\begin{align}\label{eq:Clifford}
J_{a_i\otimes a_j\otimes a_k}+J_{a_i\otimes a_k\otimes a_j}=2a_j\Cdot a_k J_{a_i}\,.
\end{align}
Note that the on-shell representations of vectors and tensors are agnostic to types.
One can use it to put tensor currents into a minimal basis. Different basis choices may lead to different vector currents. As we will see later, on-shell vector currents correspond to the kinematic numerators, while the difference caused by the tensor basis choice is just part of the generalized gauge freedom. 
This is our motivation of introducing the relation~\eqref{eq:Clifford} to the tensors.

\subsection{Method to determine the fusion product}\label{sec:methods}
Here we try to construct the fusion products through an ansatz approach. More precisely, we make an ansatz for every fusion product that contributes to eq.~\eqref{eq:npre}. 
We then use eq.~\eqref{eq:AfromFusionRules} to convert the resultant pre-numerators into the kinematic numerator associated with a graph $\Gamma$. The ansatz can be solved by matching with the amplitude on the support of maximal factorization, namely, all the propagators of $\Gamma$ are on-shell,
\begin{align}\label{eq:MaximalCutCond}
\mathcal{C}(\commut,n)\mathcal{N}(\commut,n)\Big|_{\substack{\text{maximal} \\  \text{factorization}}}=\text{Res}_{\Gamma}\mathcal{A}_n \,,
\end{align}
where $\mathcal{A}_n$ is the full color-dressed amplitude and $\text{Res}_{\Gamma}\mathcal{A}_n$ is the residue when all the propagators of the graph $\Gamma$ are taken on-shell.
We note that eq.~\eqref{eq:MaximalCutCond} is a necessary condition and sufficient only up to polarization power two. For higher polarization powers, we need to consider such conditions for non-maximal factorization channels due to higher powers of Mandelstam variables.

In practice, we impose additional constraints on the fusion products to reduce the number of free parameters and simplify the solution. First, we require that the fusion products relevant to eq.~\eqref{eq:npre} do not reduce the tensor rank. Since only vectors and rank-three tensors are present in the NMHV sector, this means that the fusion between two vectors can lead to both vectors and rank-three tensors, while the fusion between a rank-three tensor and a vector can only give rank-three tensors.
Then following a similar consideration in~\cite{Chen:2019ywi}, we fix one particular fusion product as
\begin{align}\label{eq:fusionSpecial}
J_{\varepsilon_i\otimes\varepsilon_j\otimes p}(p)\star J_{\varepsilon_k}(p_k)&=\frac{1}{2}p^2 J^{(1)}_{\varepsilon_i\otimes\varepsilon_j\otimes \varepsilon_k}(p+p_k)\,,\nonumber\\
J_{p_i\otimes\varepsilon_j\otimes p}(p)\star J_{\varepsilon_k}(p_k)&=\frac{1}{2}p^2 J^{(1)}_{p_i\otimes\varepsilon_j\otimes \varepsilon_k}(p+p_k)\,,
\end{align}
We also assume that there is only one type of the tensor $J_{\varepsilon_i\otimes\varepsilon_j\otimes p}$ and $J_{p_i\otimes\varepsilon_j\otimes p}$ so that we drop the superscript. Since this is the first time that the tensor $J_{\varepsilon_i\otimes\varepsilon_j\otimes \varepsilon_k}$ and $J_{p_i\otimes\varepsilon_j\otimes \varepsilon_k}$ appear, we can freely assign a type label to them, as have been done above. For other currents, the tensor type is determined in an iterative way.
At a given multiplicity $n$, suppose we have two currents $J^{(w_1)}_U$ and $J^{(w_2)}_U$ in the output of $J_{\varepsilon_1}\star J_{\varepsilon_2}\star\cdots\star J_{\varepsilon_{n-2}}$ that have identical tensor labels, we say they are of the same type, namely, $w_1{=}w_2$, if their fusion products with $J_{\varepsilon_{n-1}}(p_{n-1})$ are identical when evaluated on-shell, 
\begin{align}\label{eq:tensorid}
\langle q|J_U^{(w_1)}\star J_{\varepsilon_{n-1}}(p_{n-1})|n\rangle\Big|_{q\, {\rm soft}}=\langle q|J_U^{(w_2)}\star J_{\varepsilon_{n-1}}(p_{n-1})|n\rangle\Big|_{q\, {\rm soft}}\,.
\end{align}
In other words, we ignore the possibility that the fusion products above may differ off-shell. Such difference includes, for example, some tensors in the results are of different types, and thus have different fusion products at higher multiplicity.
Under this simplification, new types of currents only appear in the last stage of the consecutive fusion products of eq.~\eqref{eq:npre} at each multiplicity. This choice keeps the number of current types minimal and reduces the complexity of our ansatz to a manageable level.

\subsection{Constructing NMHV kinematic algebra}\label{sec:construction}
Now we start to construct the fusion products relevant for the pre-numerator~\eqref{eq:npre}. The first type of currents we encounter is $J_{\varepsilon_i}(p_i)$, which corresponds to the on-shell particle $i$. 
As already discussed in section~\ref{sec:kinAlg}, we omit the type label for these special currents. The fusion products that will appear in the process~\eqref{eq:npre} are of the form
\begin{align}
J^{(w)}_{a_i}(p)\star J_{\varepsilon_i}(p_i)\,, \quad J^{(w)}_{a_i\otimes a_k\otimes a_l}(p)\star J_{\varepsilon_i}(p_i)\,,
\end{align} 
where the subscripts $a_i$ follow those in eq.~\eqref{eq:alltensors}, and we will further determine the tensor type $w$. At three points, we encounter only one fusion product, and we make the following ansatz for it, \begin{align}\label{eq:Nod3}
J_{\varepsilon_1}(p_1)\star J_{\varepsilon_2}(p_2)&=\varepsilon_2\Cdot p_1 J_{\varepsilon_1}(p_{12})+x_1\varepsilon_1\Cdot p_2 J_{\varepsilon_2}^{(w_0)}(p_{12})\nonumber\\
&\quad+x_2\varepsilon_1\Cdot\varepsilon_2 J_{p_2}^{(1)}(p_{12})-\frac{x_0}{2}J _{\varepsilon_1\otimes\varepsilon_2\otimes p_{12}}(p_{12})\,,
\end{align}
where $p_{12}{=}p_1{+}p_2$. It is an extension of the one in the bi-scalar sector~\cite{Chen:2019ywi}. As we have preluded in eq.~\eqref{eq:fusionSpecial}, we only use one type of the tensor $J_{\varepsilon_1\otimes\varepsilon_2\otimes p_{12}}$ such that we drop its type label.  We expect that the $J_{\varepsilon_1}(p_{12})$ on the right hand side of eq.~\eqref{eq:Nod3} is of the same type as the $J_{\varepsilon_i}$'s on the left hand side, as inferred from the bi-scalar sector algebra~\cite{Chen:2019ywi}. On the other hand, $J_{p_2}^{(1)}(p_{12})$ is the first current of this type that appears in the algebra so that we can freely assign its type. Thus $J_{\varepsilon_2}^{(w_0)}(p_{12})$ is the only current with undetermined type. The three-point numerator is then computed by 
\begin{align}\label{eq:N3}
\npre([1,2],3)=\npre(1,2,3)-\npre(2,1,3)=\langle q|J_{\varepsilon_1}(p_1)\star J_{\varepsilon_2}(p_2)|3\rangle\Big|_{q\,{\rm soft}}-(1\leftrightarrow 2)\,,
\end{align}
where the vector and tensor currents are evaluated according to eq.~\eqref{eq:vector} and~\eqref{eq:tensor}. In particular, the tensor vanishes due to momentum conservation and Dirac equation,
\begin{align}
\langle q|J_{\varepsilon_1\otimes\varepsilon_2\otimes p_{12}}|3\rangle\Big|_{q\, {\rm soft}}=\bar{v}_{q}\slashed{\varepsilon}_1\slashed{\varepsilon}_2\slashed{p}_{12}u_3=-\bar{v}_q\slashed{\varepsilon}_1\slashed{\varepsilon}_2\slashed{p}_3u_3=0\,.
\end{align}
At three points, the numerator~\eqref{eq:N3} should match directly with the amplitude
\begin{align}
\mathcal{A}_3=f^{a_1a_2a_3}(\varepsilon_2\Cdot p_1 \varepsilon_1\Cdot\varepsilon_3 - \varepsilon_1\Cdot p_2 \varepsilon_2\Cdot\varepsilon_3+\varepsilon_1\Cdot \varepsilon_2 p_2\Cdot\varepsilon_3)\,,
\end{align} 
which gives $x_1=0$ and $x_2=\frac{1}{2}$ while $x_0$ remains free. Thus $J_{\varepsilon_2}^{(w_0)}$ drops out so that we do not need to fix the type.

At four points, the $s$ channel numerator is given by 
\begin{align}\label{eq:N4p}
\npre([[1,2],3],4) = \npre(1,2,3,4) - \npre(2,1,3,4)- \npre(3,1,2,4)+ \npre(3,2,1,4)\,.
\end{align}
We compute $ \npre(1,2,3,4)= \langle q|J_{\varepsilon_1} \star J_{\varepsilon_2} \star J_{\varepsilon_3}|4\rangle\big|_{q\, {\rm soft}}$ from the algebra and obtain the other three by permuting indices. 
While $J_{\varepsilon_1}\star J_{\varepsilon_2}$ is known from eq.~\eqref{eq:Nod3}, we just need to compute the fusion with $J_{\varepsilon_3}$ from the right. This leaves us to determine\footnote{We use the somewhat unusual convention: $s_{ij}=p_i\Cdot p_j$ and $s_{ij\ldots k}=\frac{1}{2}(p_i{+}p_j{+}\ldots{+}p_k)^2$.}
\begin{align}
&\text{vector-vector fusion:} & &J_{\varepsilon_1}(p_{12})\star J_{\varepsilon_3}(p_3)\,,\quad 
J^{(1)}_{p_2}(p_{12})\star J_{\varepsilon_3}(p_3)\,,\nonumber\\
&\text{tensor-vector fusion:} & &J_{\varepsilon_1\otimes\varepsilon_2\otimes p_{12}}(p_{12})\star J_{\varepsilon_3}(p_3)=s_{12} J_{\varepsilon_1\otimes\varepsilon_2\otimes\varepsilon_3}^{(1)}(p_{123})\,,
\end{align}
where $p_{123}{=}p_1{+}p_2{+}p_3$. The tensor-vector fusion here follows immediately our assumption~\eqref{eq:fusionSpecial}. 
We can write down the result of $J_{\varepsilon_{1}}(p_{12})\star J_{\varepsilon_3}(p_3)$ by simply assuming that eq.~\eqref{eq:Nod3} still holds when the momentum of $J_{\varepsilon_{1}}$ is off-shell. Thus we only need to make an ansatz for the last vector-vector fusion,
\begin{align}\label{eq:N4}
	J^{(1)}_{p_{2}}(p_{12})\star J_{\varepsilon_3}(p_3)&=x_3\varepsilon_3\Cdot p_{12} J^{(w_1)}_{p_2}(p_{123})+x_4 s_{23}J^{(w_2)}_{\varepsilon_3}(p_{123})+x_5 \varepsilon_3\Cdot p_{2} J^{(w_3)}_{p_3}(p_{123})\nonumber\\
	&\quad +x_6 J_{p_2\otimes\varepsilon_3\otimes p_{123}}(p_{123})\,.
\end{align}
To solve the unknowns, we evaluate eq.~\eqref{eq:N4p} in the limit $s_{12}\rightarrow 0$ and then equate it to the residue of the four-point amplitude on the $s_{12}$ pole, following eq.~\eqref{eq:MaximalCutCond}.
After a proper rearrangement using the relation~\eqref{eq:Clifford}, the only tensor that appears in eq.~\eqref{eq:N4p} in the on-shell limit is $J_{\varepsilon_1\otimes\varepsilon_2\otimes\varepsilon_3}$, which then drops out when $s_{12}\rightarrow 0$ since it is multiplied by $s_{12}$. It can be verified that this tensor term is in the kernel of the propagator matrix. This will be the case for all tensor terms at higher multiplicities as well, although the maximal factorization condition does not seem to directly imply this property. More details will be discussed later in the paper.
The $s_{12}$ factorization channel constraint gives
\begin{align}
	x_3=1, && x_4=x_0-1, &&x_5=1\,,
\end{align}
while $x_6$ remains undetermined. We cannot determine the superscript $w_{1,2,3}$ at four points, since in the on-shell limit these vector currents are indistinguishable. Instead, they will be retrospectively determined at five points according to their fusion products with $J_{\varepsilon_4}(p_4)$. Here $x_0$ is still a free parameter, and it will remain free even at higher multiplicities. In the following, we choose $x_0=\frac{1}{4}$. This choice keeps the fusion products and the form of pre-numerator the simplest. We will leave the more general expressions with a free $x_0$ to appendix~\ref{sec:AlgebraDiscussion}.

At five points, we need to set up ansatze for the fusion of $J_{p_2}^{(w_1)}$, $J_{\varepsilon_3}^{(w_2)}$, $J_{p_3}^{(w_3)}$ and $J_{\varepsilon_1\otimes\varepsilon_2\otimes\varepsilon_3}$ with $J_{\varepsilon_4}$, and solve them by imposing the ``maximal factorization'' condition~\eqref{eq:MaximalCutCond}. This gives $x_6=-\frac{1}{4}$ and fixes the tensor type
\begin{align}
J_{p_2}^{(w_1)}=J_{p_2}^{(1)}\,,\quad J_{p_3}^{(w_3)}=J_{p_3}^{(1)}\,,\quad J_{\varepsilon_3}^{(w_2)}=J_{\varepsilon_3}^{(2)}\,,
\end{align}
along with solutions to new parameters introduced at five points. Here the identification is made under the assumption of eq.~\eqref{eq:tensorid}.
We iterate this process at each multiplicity and observe the closure of the algebra at seven points. That is, assuming no new currents are generated, we verify that the set of fusion products determined up to seven points computes the correct numerator at higher multiplicities. We have checked explicitly that the algebra gives the correct amplitude at eight points up to polarization power two terms. At nine points, we have checked that the BCJ numerators generate the correct maximal factorization for each cubic graph.

To summarize, there are in all 13 relevant vector and tensor currents,
\begin{align}
&\text{vector:} & & J_{\varepsilon_i}\,,\quad J_{\varepsilon_i}^{(2)}\,,\quad J_{p_i}^{(1)}\,,\quad J_{p_i}^{(2)}\nonumber\\
&\text{tensor:} & & J_{\varepsilon_i\otimes\varepsilon_j\otimes p}\,,\quad J_{p_i\otimes\varepsilon_j\otimes p}\,,\quad J_{\varepsilon_i\otimes\varepsilon_j\otimes\varepsilon_k}^{(1)}\,,\quad J_{\varepsilon_i\otimes\varepsilon_j\otimes\varepsilon_k}^{(2)}\,,\nonumber\\
& & & J_{p_i\otimes\varepsilon_j\otimes\varepsilon_k}^{(1)}\,, \ldots\,,J_{\varepsilon_i\otimes\varepsilon_j\otimes\varepsilon_k}^{(5)}\,.
\end{align}
The fusion products involving vector currents are
 \begin{align}\label{eq:FRV}
	J_{\varepsilon_i}(p)\star J_{\varepsilon_j}(p_j)&=\varepsilon_j\Cdot p J_{\varepsilon_i}(p{+}p_j)+\frac{1}{2}\varepsilon_i\Cdot\varepsilon_j J^{(1)}_{p_j}(p{+}p_j)-\frac{1}{8}J _{\varepsilon_i\otimes\varepsilon_j\otimes (p{+}p_{j})}(p{+}p_j), \nn\\
	J^{(1)}_{p_i}(p)\star J_{\varepsilon_j}(p_j)&=\varepsilon_j\Cdot p J^{(1)}_{p_i}(p{+}p_j)+\varepsilon_j\Cdot p_i J^{(1)}_{p_j}(p{+}p_j)-\frac{3}{4}s_{ij}J_{\varepsilon_j}^{(2)}(p{+}p_j)\nonumber\\
	&\quad-\frac{1}{4}J _{p_i\otimes\varepsilon_j\otimes (p+p_{j})}(p{+}p_j), \nn\\
	J_{\varepsilon_i}^{(2)}(p)\star J_{\varepsilon_j}(p_j)&=\varepsilon_j\Cdot p J_{\varepsilon_i}^{(2)}(p{+}p_j)-\varepsilon_i\Cdot p_{j} J_{\varepsilon_j}^{(2)}(p{+}p_j)+\frac{2}{3}\varepsilon_i\Cdot\varepsilon_j J_{p_j}^{(2)}(p{+}p_j), \nn\\
	J_{p_i}^{(2)}(p)\star J_{\varepsilon_j}(p_j)&=\varepsilon_j\Cdot p J_{p_i}^{(2)}(p{+}p_j)+\varepsilon_j\Cdot p_i J_{p_j}^{(2)}(p{+}p_j).
\end{align}
The fusion rules involving $J _{\varepsilon_i\otimes\varepsilon_j\otimes p}$ and $J _{p_i\otimes\varepsilon_j\otimes p}$ are assumed as in eq.~\eqref{eq:fusionSpecial},
\begin{align}
	J_{\varepsilon_i\otimes\varepsilon_j\otimes p}(p)\star J_{\varepsilon_k}(p_k)&=\frac{1}{2}p^2 J^{(1)}_{\varepsilon_i\otimes\varepsilon_j\otimes \varepsilon_k}(p{+}p_k)\,,\nonumber\\
	J_{p_i\otimes\varepsilon_j\otimes p}(p)\star J_{\varepsilon_k}(p_k)&=\frac{1}{2}p^2 J^{(1)}_{p_i\otimes\varepsilon_j\otimes \varepsilon_k}(p{+}p_k)\,,
\end{align}
The fusion rules involving $J^{(1)}_{\varepsilon_i\otimes\varepsilon_{i_1}\otimes \varepsilon_{i_2}}$ and $J^{(1)}_{p_i\otimes\varepsilon_{i_1}\otimes \varepsilon_{i_2}}$ are similar such that they can be combined into
\begin{align}
  J^{(1)}_{a\otimes\varepsilon_{i_1}\otimes \varepsilon_{i_2}}(p)\star  J_{\varepsilon_j}(p_j)&=\varepsilon_j\Cdot p  J _{a\otimes\varepsilon_{i_1}\otimes \varepsilon_{i_2}}^{(2)}(p{+}p_j)+\varepsilon_j\Cdot a J _{p_j\otimes\varepsilon_{i_1}\otimes \varepsilon_{i_2}}^{(4)}(p{+}p_j)\nonumber\\
                                                                                              &\quad+\varepsilon_{i_2}\Cdot a J _{p_{i_2}\otimes\varepsilon_{i_1}\otimes \varepsilon_j}^{(3)}(p{+}p_j)-\varepsilon_{i_1}\Cdot a J _{p_{i_1}\otimes\varepsilon_{i_2}\otimes \varepsilon_j}^{(3)}(p{+}p_j)\nonumber\\
                                                                                              &\quad+\varepsilon_{i_2}\Cdot p  J _{a\otimes\varepsilon_{i_1}\otimes \varepsilon_j}^{(1)}(p{+}p_j)-\varepsilon_{i_1}\Cdot p  J _{a\otimes\varepsilon_{i_2}\otimes \varepsilon_j}^{(1)}(p{+}p_j)\,,
\end{align}
where the $a=\varepsilon_i$ or $p_i$. There are new types of tensors generated on the right hand side. The tensor $J^{(2)}_{\varepsilon_i\otimes\varepsilon_j\otimes\varepsilon_k}$ and $J^{(2)}_{p_i\otimes\varepsilon_j\otimes\varepsilon_k}$ again have similar fusion rules,
\begin{align}
  J^{(2)}_{a\otimes\varepsilon_{i_1}\otimes \varepsilon_{i_2}}(p)\star  J_{\varepsilon_j}(p_j)&=\varepsilon_j\Cdot p  J _{a\otimes\varepsilon_{i_1}\otimes \varepsilon_{i_2}}^{(2)}(p{+}p_j)+\varepsilon_j\Cdot a J _{p_j\otimes\varepsilon_{i_1}\otimes \varepsilon_{i_2}}^{(4)}(p{+}p_j)\nonumber\\
	&\quad+\varepsilon_{i_2}\Cdot a J _{p_{i_2}\otimes\varepsilon_{i_1}\otimes \varepsilon_j}^{(3)}(p{+}p_j)-\varepsilon_{i_1}\Cdot a J _{p_{i_1}\otimes\varepsilon_{i_2}\otimes \varepsilon_j}^{(3)}(p{+}p_j)\,,
\end{align}
where the $a=\varepsilon_i$ or $p_i$. 
The fusion rule for $J^{(3)}_{p_r\otimes\varepsilon_{i_1}\otimes \varepsilon_{i_2}}$ gives one new type of tensor, 
\begin{align}
	J^{(3)}_{p_r\otimes\varepsilon_{i_1}\otimes \varepsilon_{i_2}}(p)\star  J_{\varepsilon_j}(p_j)&=\varepsilon_j\Cdot p  J _{p_r\otimes\varepsilon_{i_1}\otimes \varepsilon_{i_2}}^{(5)}(p{+}p_j)+\varepsilon_j\Cdot p_r J _{p_j\otimes\varepsilon_{i_1}\otimes \varepsilon_{i_2}}^{(4)}(p{+}p_j)\nonumber\\
	&\quad+\varepsilon_{i_2}\Cdot p_r J _{p_{i_2}\otimes\varepsilon_{i_1}\otimes \varepsilon_j}^{(3)}(p{+}p_j)-\varepsilon_{i_1}\Cdot p  J _{p_r\otimes\varepsilon_{i_2}\otimes \varepsilon_j}^{(1)}(p{+}p_j)\,,
\end{align}
while $J^{(4)}_{p_r\otimes\varepsilon_{i_1}\otimes \varepsilon_{i_2}}$ is self-closed under the fusion with $J_{\varepsilon_j}$,
\begin{align}
	J^{(4)}_{p_r\otimes\varepsilon_{i_1}\otimes \varepsilon_{i_2}}(p)\star  J_{\varepsilon_j}(p_j)&=\varepsilon_j\Cdot p  J _{p_r\otimes\varepsilon_{i_1}\otimes \varepsilon_{i_2}}^{(4)}(p{+}p_j) +\varepsilon_j\Cdot p_r J _{p_j\otimes\varepsilon_{i_1}\otimes \varepsilon_{i_2}}^{(4)}(p{+}p_j)\,.
\end{align}
Finally, the fusion rule of $J^{(5)}_{p_r\otimes\varepsilon_{i_1}\otimes\varepsilon_{i_2}}$ does not have new type of tensors,
\begin{align}
  J^{(5)}_{p_r\otimes\varepsilon_{i_1}\otimes \varepsilon_{i_2}}(p)\star  J_{\varepsilon_j}(p_j)&=\varepsilon_j\Cdot p  J _{p_r\otimes\varepsilon_{i_1}\otimes \varepsilon_{i_2}}^{(5)}(p{+}p_j)+\varepsilon_j\Cdot p_r J _{p_j\otimes\varepsilon_{i_1}\otimes \varepsilon_{i_2}}^{(4)}(p{+}p_j)\nonumber\\
	&\quad+\varepsilon_{i_2}\Cdot p_r J _{p_{i_2}\otimes\varepsilon_{i_1}\otimes \varepsilon_j}^{(3)}(p{+}p_j)\,,
\end{align}
which closes our algebra for the NMHV sector pre-numerators.

\section{BCJ numerators from the fusion product}\label{sec:numerator}
The final result for the NMHV sector algebra contains vector currents $J_{a_i}$ and rank-three tensor currents $\Je{a_{i_1}}.{a_{i_2}}.{a_{i_2}} $, where $a_i$'s are momenta or polarization vectors. Although we have different types of vector and tensor currents participating in the fusion products, we consider their on-shell representations to only depend on their Lorentz structure. Thus we evaluate them in the on-shell limit as eq.~\eqref{eq:vector} for vectors and eq.~\eqref{eq:tensor} for tensors. In particular, for rank-three tensors, we have
\begin{align}\label{eq:tensoronshell}
\langle q|\Je{a_{i_1}}.{a_{i_2}}.{a_{i_2}}|n\rangle\Big|_{q\, {\rm soft}}&=\bar{v}_q\gamma_{\mu_1}\gamma_{\mu_2}\gamma_{\mu_3}u_n a^{\mu_1}_1 a^{\mu_2}_2 a^{\mu_3}_3\Big|_{q\, {\rm soft}}\equiv\varepsilon_n(a_1,a_2,a_3)\,.
\end{align}
The tensor $\varepsilon_n(a_1,a_2,a_3)$ satisfies
\begin{align}\label{eq:rank3rel}
&\varepsilon_{n}(a_1,a_2,a_3)+\varepsilon_n(a_2,a_1,a_3)=2a_1\Cdot a_2\,a_3\Cdot\varepsilon_n\,,\nonumber\\
&\varepsilon_{n}(a_1,a_2,a_3)+\varepsilon_n(a_1,a_3,a_2)=2a_2\Cdot a_3\,a_1\Cdot\varepsilon_n\,,
\end{align}
which is the on-shell version of the Clifford algebra relation~\eqref{eq:Clifford}. We can use these relations to bring tensors into an irreducible basis. Once in this irreducible basis, they must reside in the null space of the propagator matrix since they are not physical states of YM, and thus they can be removed by hand. However, even if the tensors are not in an irreducible basis, their contribution must cancel out of the amplitude through the propagator matrix. This implies that there are certain vector contributions that live in the null space of the propagator matrix, and they can be exposed by changing the tensor basis. Therefore, as shown in ref.~\cite{Chen:2019ywi}, the basis choice encodes a certain subset of the generalized gauge freedom for numerators of physical gluon amplitudes.

We will now study the output of the kinematic algebra constructed in the previous section. Remarkably, the fusion products that we found lead to a closed all-multiplicity formula for the NMHV pre-numerators, 
\begin{align}\label{eq:BCJAll}
\npre(1,2,\ldots,n)=\Big(1+\frac{1}{2}Q_n\Big)\npre^{(1)}_V+\frac{1}{4}(1+Q_n)\npre_T^{(2)}+\npre^{(2)}_V\,,
\end{align}
where $\npre^{(1)}_V$ and $\npre_T^{(2)}$ are given by
\begin{subequations}\label{eq:NvNt}
\begin{align}\label{eq:NBCJSP}
  \npre^{(1)}_V&=(\varepsilon_1\Cdot\varepsilon_n)\prod _{j=2}^{n-1} \varepsilon_j\Cdot \rmom_j \,,\\
  \label{eq:NT}
  \npre_T^{(2)}&=
  -\frac{1}{4} \sum_{i=2}^{n-2}\sum_{\substack{\ell,m=i \\ m>\ell}}^{n-1}(-1)^{\ell-i}\rmom^2_{i+1}\Bigg[\prod_{j\notin\{1,i,i+1,\ldots,\ell,m,n\}}\varepsilon_j\Cdot \rmom_{j}\Bigg]\\
  &\quad\times(\varepsilon\Cdot X)^{i,i+1,i+2\ldots,\ell-1}_{[\ldots[[i+2,i+3],i+4]\ldots,\ell+1]}\varepsilon_n(\varepsilon_1,\varepsilon_\ell,\varepsilon_m)\,.\nonumber
\end{align}
\end{subequations}
Here $\rmom_j=\sum_{i=1}^{j-1}p_i$ denotes the region momentum~\cite{Stieberger:2016lng,Nandan:2016pya, Chiodaroli:2017ngp}, and we define
\begin{align}\label{eq:eX}
(\varepsilon\Cdot\rmom)^{i,i+1,i+2\ldots,\ell-1}_{[\ldots[[i+2,i+3],i+4]\ldots,\ell+1]}=\varepsilon_{i,\mu}\varepsilon_{i+1,\nu}\varepsilon_{i+2,\rho}\ldots\varepsilon_{\ell-1,\lambda}\rmom^{\mu}_{[\ldots[[i+2}\rmom^{\nu\vphantom{\mu}}_{\ldots i+3]}\rmom^{\rho\vphantom{\mu}}_{i+4]}\ldots\rmom^{\lambda\vphantom{\mu}}_{\ldots\ell+1]},
\end{align}
where the indices of $\rmom$ are in the left-nested commutator. For example, the following term appears at $n{=}7$ and beyond, 
\begin{align}
(\varepsilon\Cdot X)^{2,3,4}_{[[4,5],6]}
&=\varepsilon_2\Cdot\rmom_{4}\varepsilon_{3}\Cdot\rmom_{5}\varepsilon_{4}\Cdot\rmom_{6}-\varepsilon_2\Cdot\rmom_{5}\varepsilon_{3}\Cdot\rmom_{4}\varepsilon_{4}\Cdot\rmom_{6}\nonumber\\
&\quad-\varepsilon_2\Cdot\rmom_{6}\varepsilon_{3}\Cdot\rmom_{4}\varepsilon_{4}\Cdot\rmom_{5}+\varepsilon_2\Cdot\rmom_{6}\varepsilon_{3}\Cdot\rmom_{5}\varepsilon_{4}\Cdot\rmom_{4}\,.
\end{align}
We note that when $\ell{=}i$ the commutator does not exist and we set it to unity. When $\ell{=}i{+}1$, it simply becomes $\varepsilon_i\Cdot X_{i+2}$. The superscripts in $\npre_V$ and $\npre_T$ denote the polarization power. We note that $\npre^{(1)}_V$ and $\npre^{(2)}_T$ are exactly the vector and tensor part of the bi-scalar-YM sector numerators obtained in a previous work~\cite{Chen:2019ywi}, whereas the notations therein are slightly different. The connection between the pre-numerator here and the bi-scalar sector numerator in~\cite{Chen:2019ywi} will be discussed in appendix~\ref{sec:AlgebraDiscussion}. Next, $Q_n$ is a differential operator, 
\begin{align}
Q_n&=\varepsilon_{1}\Cdot \mathcal D_{2}\Cdot\mathcal D_{3}\cdots \mathcal D_{n-1}\Cdot \frac{\partial}{\partial \varepsilon_{1}}-1\nn\\
&=\sum_{r=1}^{n-2}\sum_{2\leqslant i_1<i_2<\ldots<i_r\leqslant n-1}(\varepsilon_1\Cdot\varepsilon_{i_1})\Bigg[\prod_{\ell=1}^{r-1}(p_{i_{\ell}}\Cdot\varepsilon_{i_{\ell+1}})\frac{\partial}{\partial\varepsilon_{i_{\ell+1}}\Cdot p_1}\Bigg]\frac{\partial}{\partial \varepsilon_{i_1}\Cdot p_1}\left(p_{i_r}\Cdot\frac{\partial}{\partial \epsilon_1}\right)\,,
\end{align}
where $(\mathcal D_{i})_{\mu\nu}=\varepsilon_{i,\mu} p_{i,\nu}\frac{\partial}{ \partial \varepsilon_{i}\Cdot p_1}+\eta_{\mu\nu}$. 
The action of $Q_n$ on $\npre^{(1)}_V$ and $\npre_T^{(2)}$ does not change the polarization power, and it can be evaluated directly as,
\begin{align}
\Big(1+\frac{1}{2}Q_n\Big)\npre^{(1)}_V&=\frac{1}{2}\Bigg[(\varepsilon_1\Cdot\varepsilon_n)\prod _{j=2}^{n-1} \varepsilon_j\Cdot \rmom_j+\varepsilon_1\Cdot G_{2,3,\ldots,n-1}\Cdot \varepsilon_n\Bigg], \\
\label{eq:QNT}
(1+Q_n)\npre_T^{(2)}&=
  -\frac{1}{4} \sum_{i=2}^{n-2}\sum_{\substack{\ell,m=i \\ m>\ell}}^{n-1}(-1)^{\ell-i}\rmom^2_{i+1}\\
&\quad\times\varepsilon_n\big(\varepsilon_1\Cdot G_{2,\ldots,i-1}\Cdot G^{i,i+1,\ldots,\ell-1}_{[\ldots[i+2,i+3],\ldots,\ell+1]}\Cdot G_{\ell+1,\ldots,\hat{m},\ldots,n-1},\varepsilon_\ell,\varepsilon_m\big)\,.\nonumber
\end{align}
We first introduce $(G^{i}_{j})_{\mu\nu}=\varepsilon_{i,\mu}p_{i,\nu}+\varepsilon_i\Cdot \rmom_{j}\,\eta_{\mu\nu}$ and $G_{i}\equiv G^{i}_{i}$. Then in the second line of eq.~\eqref{eq:QNT}, we have the shorthand notation
\begin{align}
G_{i,\ldots,k}=G_{i}\Cdot G_{i+1}\cdots G_{k-1}\Cdot G_{k}\,,
\end{align}
and a hatted index should be skipped in this product. We also have
\begin{align}
G^{i,\ldots,\ell-1}_{[\ldots[i+2,i+3],\ldots,\ell+1]}=G^{i}_{[\ldots[{i+2}}\Cdot G^{i+1}_{{i+3}]}\cdots G^{\ell-1}_{{\ell+1}]}\,.
\end{align}
The subscripts of $G$ are in the same left-nested commutator as in eq.~\eqref{eq:eX}. Finally, $\npre^{(2)}_V$ contains additional polarization power two terms that are generated from the fusion products of the vector currents,
\begin{align}
\npre^{(2)}_V&=-\frac{1}{4}\Bigg[\sum_{j=2}^{n-3}(\varepsilon_1\Cdot G_{2,\ldots,j}\Cdot p_{j+1})\sum_{\ell=j+1}^{n-2}(\varepsilon_{j+1}\Cdot \tilde{G}_{j+2,\ldots,\ell}\Cdot \varepsilon_{\ell+1})(p_{\ell+1}\Cdot G_{\ell+2,\ldots,n-1}\Cdot\varepsilon_n)\Bigg]_{(\varepsilon\Cdot\varepsilon)^2} \nn\\
&\quad -\frac{3}{8} \Bigg[\sum_{j=2}^{n-2}(\varepsilon_1\Cdot G_{2,\ldots,j}\Cdot  p_{j+1})(\varepsilon_{j+1}\Cdot \tilde{G}_{j+2,\ldots,n-1}\Cdot\varepsilon_n)\Bigg]_{(\varepsilon\Cdot\varepsilon)^2}\,,
\end{align} 
where $\tilde{G}_{i,\ldots,k}=\tilde{G}_i\Cdot \tilde{G}_{i+1}\cdots \tilde{G}_{k-1}\Cdot \tilde{G}_k$ and $(\tilde{G}_{i})_{\mu\nu}=-p_{i,\mu}\varepsilon_{i,\nu}+\varepsilon_i\Cdot \rmom_{i}\,\eta_{\mu\nu}$. Note that we only need to keep the polarization power two terms in the above equation.

As discussed in section~\ref{sec:kinAlg}, the BCJ numerators obtained from the pre-numerators satisfy the $S_{n-1}$ crossing symmetry. At this point, the numerators still contain on-shell tensors, and the crossing symmetry is realized even with these extra contributions. To obtain the conventional on-shell numerators, we need to expand the tensors in a minimal basis and then remove them using BCJ amplitude relations. This process will contribute additional terms due to the relation~\eqref{eq:rank3rel}. Such contributions resulted from the tensor basis choice are part of the generalized gauge freedom~\cite{Chen:2019ywi}. More specifically, the tensors will be removed by the (generalized) binary BCJ relations~\cite{Chen:2019ywi}. Certain basis choices can retain the crossing symmetry in the conventional on-shell numerators. One particularly convenient choice is to antisymmetrize the labels of polarization vectors in the on-shell tensor~\eqref{eq:tensoronshell}. We will discuss pure gauge terms in more details in section~\ref{sec:PureGaugeNMHV}.

We have checked up to eight points that the tensors are indeed in the null space of the propagator matrix when put into a basis. We have also checked that the resulting amplitude is gauge invariant through the NMHV sector. Namely, after the replacement $\varepsilon_i\rightarrow p_i$, the polarization power zero and one terms are in the null space of the propagator matrix.
Finally, we present the explicit pre-numerators up to five points, 
\begin{align}
\npre(1,2,3,4)&=\frac{1}{2}\Big(\varepsilon_1\Cdot\varepsilon_4\, \varepsilon_2\Cdot \rmom_2 \,\varepsilon_3\Cdot \rmom_3+\varepsilon_1\Cdot G_{2}\Cdot G_{3}\Cdot \varepsilon_4\Big)-\frac{1}{2} s_{12} \varepsilon_4(\varepsilon_1,\varepsilon_2,\varepsilon_3) \\
\npre(1,2,3,4,5)&=\frac{1}{2}\Big[(\varepsilon_1\Cdot\varepsilon_5) \varepsilon_2\Cdot \rmom_2 \varepsilon_3\Cdot \rmom_3\varepsilon_4\Cdot \rmom_4+\varepsilon_1\Cdot G_{2}\Cdot G_{3}\Cdot G_{4}\Cdot \varepsilon_5\Big]\nonumber\\
&\quad-\frac{1}{8}s_{12}\Big[\varepsilon_5\big(\varepsilon_1\Cdot G_{4},\varepsilon_2,\varepsilon_3\big)+\varepsilon_5\big(\varepsilon_1\Cdot G_{3},\varepsilon_2,\varepsilon_4\big) -\varepsilon_5\big(\varepsilon_1\Cdot G^{2}_{4},\varepsilon_3,\varepsilon_4\big)\Big]\nonumber\\
&\quad-\frac{1}{8}s_{123}\varepsilon_5\big(\varepsilon_1\Cdot G_{2} ,\varepsilon_3,\varepsilon_4\big)-\left[\frac{1}{4}(\varepsilon_1\Cdot G_{2}\Cdot p_3)(\varepsilon_3\Cdot \varepsilon_4)(p_4\Cdot \varepsilon_5)\right]_{(\varepsilon\Cdot\varepsilon)^2} \nn\\
&\quad-\frac{3}{8}\left[\vphantom{\frac{1}{4}}(\varepsilon_1\Cdot G_{2}\Cdot p_3)(\varepsilon_3\Cdot \tilde{G}_{4}\Cdot \varepsilon_5)+(\varepsilon_1\Cdot G_{2}\Cdot G_{3}\Cdot p_4)(\varepsilon_4\Cdot \varepsilon_5)\right]_{(\varepsilon\Cdot\varepsilon)^2}\,.
\end{align}
In the five-point pre-numerator, the tensors can be expanded by, for example,
\begin{align}
&\varepsilon_5\big(\varepsilon_1\Cdot G_4,\varepsilon_2,\varepsilon_3\big)=\varepsilon_1\Cdot\varepsilon_4\varepsilon_5(p_4,\varepsilon_2,\varepsilon_3)+\varepsilon_4\Cdot\rmom_4\varepsilon_5(\varepsilon_1,\varepsilon_2,\varepsilon_3)\,,\nonumber\\
&\varepsilon_5\big(\varepsilon_1\Cdot G^{2}_4,\varepsilon_3,\varepsilon_4\big)=\varepsilon_1\Cdot\varepsilon_2\varepsilon_5(p_2,\varepsilon_3,\varepsilon_4)+\varepsilon_2\Cdot\rmom_4\varepsilon_5(\varepsilon_1,\varepsilon_3,\varepsilon_4)\,.
\end{align}
The $(\varepsilon\Cdot\varepsilon)^2$ contribution from the last two brackets are given by
\begin{align}
&-\left[\frac{1}{4}(\varepsilon_1\Cdot G_{2}\Cdot p_3)(\varepsilon_3\Cdot \varepsilon_4) (p_4\Cdot \varepsilon_5)+\frac{3}{8}(\varepsilon_1\Cdot G_{2}\Cdot p_3)(\varepsilon_3\Cdot \tilde{G}_{4}\Cdot \varepsilon_5)+\frac{3}{8}(\varepsilon_1\Cdot G_{2}\Cdot G_{3}\Cdot p_4)(\varepsilon_4\Cdot \varepsilon_5)\right]_{(\varepsilon\Cdot\varepsilon)^2}\nonumber\\
&=-\frac{1}{4}s_{23}\varepsilon_1\Cdot\varepsilon_2\Big[\varepsilon_3\Cdot\varepsilon_4\varepsilon_5\Cdot p_4+\frac{3}{2}(\varepsilon_3\Cdot\varepsilon_5\varepsilon_4\Cdot\rmom_4-\varepsilon_4\Cdot\varepsilon_5\varepsilon_3\Cdot p_4)\Big]-\frac{3}{8}s_{24}\varepsilon_1\Cdot\varepsilon_2\varepsilon_4\Cdot\varepsilon_5\varepsilon_3\Cdot\rmom_3\nonumber\\
&\quad-\frac{3}{8}s_{34}\varepsilon_4\Cdot\varepsilon_5\Big[\varepsilon_1\Cdot\varepsilon_2\varepsilon_3\Cdot p_2+\varepsilon_1\Cdot\varepsilon_3\varepsilon_2\Cdot\rmom_2\Big]\,.
\end{align}

\section{Group algebra actions for numerators}\label{eq:gaaction}

In this section, we briefly review the mathematical topic called group algebra (or group ring), which provides a convenient formal language for operating on numerators and polynomial building blocks for kinematics.
As we will see in the next section, this language provides an essential tool for studying generalized gauge freedom in local BCJ numerators.

A \emph{group algebra} $Z[S]$ (see e.g. refs.~\cite{algebra}) is defined over a commutative ring $Z$ and a group $S$ with two binary operations (addition and multiplication). In a representation space, the action of an group algebra element $\mathbb{V}\in Z[S]$ is given by
\begin{align}\label{eq:gadef}
  \mathbb{V}=\sum_{\sigma\in S} a^{(\sigma)}\mathbb{P}_{\sigma}\,,
\end{align}
where $a^{(\sigma)}\in Z$ and $\mathbb{P}_{\sigma}$ is the action of the group element $\sigma$. The identity $\mathbb{I}$ of the group $S$ is also the identity of the group algebra $Z[S]$. The sum and product are inherited from the ring $Z$ and group $S$,
\begin{subequations}
\begin{align}
  \label{eq:gasum}
  &\mathbb{V}+\mathbb{U}=\sum_{\sigma\in S} (a^{(\sigma)}+b^{(\sigma)})\mathbb{P}_{\sigma}\,,\\
  \label{eq:gamulti}
  &\mathbb{V} \mathbb{U}=\sum_{\sigma,\gamma\in S}a^{(\sigma)}b^{(\gamma)}\mathbb{P}_{\sigma}\mathbb{P}_{\gamma}=\sum_{\sigma,\gamma\in S}a^{(\sigma\gamma^{-1})}b^{(\gamma)}\mathbb{P}_{\sigma}\,.
\end{align}
\end{subequations}
The multiplication is associative and distributive over the sum. The multiplication by a scalar $c\in Z$ can be defined as $c\mathbb{V}\equiv \sum_{\sigma\in S}(ca^{(\sigma)})\mathbb{P}_{\sigma}$ and thus is compatible with the multiplication~\eqref{eq:gamulti}. 
If there exist two nonzero elements $\mathbb{V},\mathbb{U}\in Z[S]$ such that $\mathbb{VU}=0$, then $\mathbb{V}$ is a \emph{left zero divisor} of $\mathbb{U}$ (and $\mathbb{U}$ is a \emph{right zero divisor} of $\mathbb{V}$).  
We will focus on the cases that $S$ is a permutation group and $Z$ the integer ring. The action of a permutation group element $\sigma\in S_n$ on a function $f=f(1,2,3,\ldots,n)$ is defined as 
\begin{align}
\mathbb{P}_{\sigma}\circ f(1,2,3,\cdots, n)=f(\sigma_1,\sigma_2,\sigma_3,\cdots,\sigma_n)\,,
\end{align}
and products of functions obey $\mathbb{P}_{\sigma}\circ(fg)=(\mathbb{P}_{\sigma}\circ f)(\mathbb{P}_{\sigma}\circ g)$.

We now give some group algebra elements that will be useful for later discussions. First, due to the crossing symmetry, we can obtain the generic DDM basis numerator $\num(1,\beta_2,\ldots,\beta_{n-1},n)$ by the action of $\mathbb{P}_{\beta}$,
\begin{align}
\num(1,\beta_2,\ldots,\beta_{n-1},n)=\mathbb{P}_{\beta}\circ \num (1,2,\ldots,n-1,n)\,,\quad~~\beta\in S_{n-2}\,.
\end{align}
Next, nested commutators can be generated by the action of\footnote{We use the cycle decomposition for the permutations appearing in the group action. For example, the cycle $(a_1a_2\ldots a_m)$ corresponds to the permutation $a_1\rightarrow a_2,a_2\rightarrow a_3,\ldots,a_m\rightarrow a_1$.}  
\begin{align}
\neco(i_1, i_2,\ldots, i_m)\equiv \Big[\mathbb{I}-\mathbb{P}_{(i_2i_1)}\Big]\Big[\mathbb{I}-\mathbb{P}_{(i_3i_2i_1)}\Big]\cdots \Big[ \mathbb{I}-\mathbb{P}_{(i_m\cdots i_2i_1)}\Big]\,.
\end{align}
Indeed, one can check with explicit computations that
\begin{align}\label{eq:Roperator}
\neco(i_1,i_2,\ldots, i_m)\circ f(i_1,i_2,\ldots, i_m)&= f([\ldots[[i_1,i_2],i_3]\ldots, i_m])\\
&=f(i_1,i_2,i_3\cdots, i_m)-f(i_2,i_1,i_3\cdots, i_m)\nn\\
&\quad-f(i_3,i_1,i_2\cdots, i_m)+f(i_3,i_2,i_1\cdots, i_m) + \cdots\,.\nonumber
\end{align}
Therefore, we can use it to construct BCJ numerators from pre-numerators,
\begin{align}\label{eq:preNtoN}
\num(1,2,\ldots,n-1,n)=\neco(1,2\ldots,n-1)\circ \npre(1,2,\ldots,n-1,n)\,.
\end{align}
We note that the operator $\frac{1}{m}\neco(i_1, i_2,\cdots, i_m)$ is satisfies
\begin{align}
\left[\frac{1}{m}\neco(i_1, i_2,\cdots, i_m)\right]^2=\frac{1}{m}\neco(i_1, i_2,\cdots, i_m)\,,
\end{align}
which implies that it is a projector. It is also known as a Lie idempotent in the topic of free Lie algebras~\cite{reutenauer2003free}. Using the unit operator $\mathbb{I}$, we can construct a mutually orthogonal operator
\begin{align}\label{eq:I-R}
\left[\mathbb{I}-\frac{1}{m}\neco(i_1,\ldots,i_m)\right]\neco(i_1,\ldots,i_m)=\neco(i_1,\ldots,i_m)\left[\mathbb{I}-\frac{1}{m}\neco(i_1,\ldots,i_m)\right]=0\,.
\end{align}
From this orthogonality, it is clear that the operator $\mathbb{I}-\frac{1}{m}\neco(i_1,\ldots,i_m)$ generates the principal ideals of both the left and right zero divisors of $\neco(i_1,\ldots, i_m)$. And, conversely, $\neco(i_1,\ldots,i_m)$ generates both the left and right zero divisors of $\mathbb{I}-\frac{1}{m}\neco(i_1,\ldots,i_m)$.

Another interesting group algebra element is
\begin{align}
\mathbb{B}(i_1,i_2,\ldots,i_m)\equiv\Big[\mathbb{I}-\mathbb{P}_{(i_1i_2\ldots i_m)}\Big]\Big[\mathbb{I}-\mathbb{P}_{(i_1i_2\ldots i_{m-1})}\Big]\cdots \Big[ \mathbb{I}-\mathbb{P}_{(i_1i_2)}\Big]\,,
\end{align}
which can be used to generate the so-called binary BCJ relations~\cite{Chen:2019ywi}. These BCJ amplitude relations are equivalent to other versions of the BCJ relations, but they are nicely organized into $2^m$ terms, as is clear from the iterated product of two ``binary'' terms.  In ref.~\cite{Chen:2019ywi}, the binary BCJ relation was expressed as the following sum
\begin{align}
\mathcal{B}_{[i,\ldots,n-1]}(1,2,\ldots,n)\equiv\sum_{j=i}^{n-1}\sum_{\substack{\gamma\in\{j-1,\ldots,i\} \\ \shuffle\{j+1,\ldots,n-1\}}}\sum_{\substack{\sigma\in\{2,\ldots,i-1\} \\ \shuffle\{j,\gamma\}}}\!\!\!(-1)^{j-i}p_j\Cdot X_j(\sigma)A(1,\sigma,n)\cong 0\,, 
\end{align}
where $X_{i}(\sigma)=\sum_{\sigma_k<i}p_{\sigma_k}$ is the region momenta. The ``$\cong 0$'' means that the expression should be equal to zero for physical amplitudes evaluated for on-shell kinematics. Here we only care about the abstract functional form and hence ``BCJ relation''  is used to denote a polynomial in kinematic variables and $A(1,\sigma,n)$. We mostly drop the ``$\cong 0$'' in the remainder of the paper. Using the group algebra the above binary BCJ relation can be simplified to
\begin{align}
\mathcal{B}_{[i,\ldots,n-1]}(1,2,\ldots,n)=\mathbb{B}(i,\ldots,n-1)\circ\sum_{\substack{\sigma\in\{2,\ldots,i-1\} \\ \shuffle\{i,\ldots,n-1\}}}p_i\Cdot X_i(\sigma)A(1,\sigma,n)\,.
\end{align}
Relations with more general arguments can be obtained by acting by an arbitrary permutation $\beta\in S_{n-2}$ onto the above relation,
\begin{align}
\mathcal{B}_{[\beta_i,\ldots,\beta_{n-1}]}(1,\beta_2,\ldots,\beta_{n-1},n)=\mathbb{P}_{\beta}\circ \mathcal{B}_{[i,\ldots,n-1]}(1,2,\ldots,n)\,.
\end{align}
In appendix~\ref{sec:binaryBCJ}, we will give an independent basis for the binary BCJ relations over local coefficients that are linear in $s_{ij}$. Binary BCJ relations are important for parametrizing all the local pure gauge freedom in the BCJ numerators of the NMHV sector, as we will get to shortly. 

We can expand the product in $\mathbb{B}(i_1,i_2,\ldots,i_m)$ and arrive at a form that follows the definition~\eqref{eq:gadef},
\begin{align}
\mathbb{B}(i_1,i_2,\ldots,i_m)=\sum_{a=1}^{m}\sum_{\rho\in\{i_a,\{i_{a-1},\ldots,i_1\}\shuffle\{i_{a+1}\ldots i_m\}\}}(-1)^{a-1}\mathbb{P}_{\rho}\,.
\end{align}
Interestingly, the operator $\neco(i_1,i_2,\ldots,i_m)$ has an identical expansion after making the replacement $\mathbb{P}_{\rho}\rightarrow \mathbb{P}_{\rho^{-1}}$ in the above equation,
\begin{align}\label{eq:Lexpansion}
\neco(i_1,i_2,\ldots,i_m)=\sum_{a=1}^{m}\sum_{\rho\in\{i_a,\{i_{a-1},\ldots,i_1\}\shuffle\{i_{a+1}\ldots i_m\}\}}(-1)^{a-1}\mathbb{P}_{\rho^{-1}}\,.
\end{align}
Here, $\rho^{-1}$ is the inverse of $\rho$ such that $\rho^{-1}_i=a$ if $\rho_a=i$. For convenience, we abbreviate the two equations as
\begin{align}\label{eq:BR}
\mathbb{B}(i_1,i_2,\ldots,i_m)=\hskip-0.5cm \sum_{\sigma\in K[i_1,\ldots,i_m]} \hskip-0.3cm \mu^{(\sigma)}\mathbb{P}_{\sigma}\,,&&\hskip0.5cm
\neco(i_1,i_2,\ldots,i_m)=\hskip-0.5cm \sum_{\sigma\in K[i_1,\ldots,i_m]}\hskip-0.3cm \mu^{(\sigma)}\mathbb{P}_{\sigma^{-1}}\,,
\end{align}
where $K[i_1,\ldots,i_m]$ is the set of relevant permutations and $\mu^{(\sigma)}=\pm 1$ is the coefficient associated to each permutation. As we will see later, the connection between $\mathbb{B}$ and $\neco$ is the key to understand the pure gauge terms in a BCJ numerator.

Generic numerators consist of a linear combination of \emph{kinematic monomials} $\mathsf{m}$, defined as a product of $\varepsilon\Cdot\varepsilon$, $\varepsilon\Cdot p$ and $p\Cdot p$. In addition, our algebraic construction detailed in section~\ref{sec:FusionProd} also gives tensors in the numerators. Without losing generality, we use tensors in the basis that the $p$ labels and $\varepsilon$ labels are antisymmetrized respectively,
\begin{align}\label{eq:Bbeta}
\hat\varepsilon_n(p_{i_1},{\ldots}, p_{i_k},\varepsilon_{j_{k+1}},{\ldots},\varepsilon_{j_{2m+1}})\equiv\varepsilon_n(p_{[i_1},{\ldots}, p_{i_k]},\varepsilon_{[j_{k+1}},{\ldots},\varepsilon_{j_{2m+1}]})
\end{align}
as additional building blocks of the kinematic monomials.
For example, in the NMHV sector we have $\hat\varepsilon_n(p_{i_1},\varepsilon_{i_2},\varepsilon_{i_3})$ and $\hat\varepsilon_n(\varepsilon_{i_1},\varepsilon_{i_2},\varepsilon_{i_3})$.

We classify monomials by the orbit of the $S_{n-2}$ group action. Two monomials belong to the same \emph{monomial equivalence class} (MEC) if they are related by an $S_{n-2}$ permutation.\footnote{MECs can be classified by the topologies of certain diagrams, a terminology used in ref.~\cite{Chen:2019ywi}.} Therefore, each MEC forms a representation space of the permutation groups.
For example, the following two monomials belong to the same MEC, 
\begin{align}
	p_{3}\Cdot p_{7} \varepsilon_3\Cdot\varepsilon_4 \varepsilon_5\Cdot p_4 \hat\varepsilon_8(\varepsilon_1,\varepsilon_6,\varepsilon_7)\quad\text{and}\quad  p_{6}\Cdot p_{7} \varepsilon_6\Cdot\varepsilon_4 \varepsilon_5\Cdot p_4 \hat\varepsilon_8(\varepsilon_1,\varepsilon_3,\varepsilon_7)\,,
\end{align}
since they are related by the permutation $3\leftrightarrow 6$. 

More specifically, we now consider the independent building blocks for the kinematic monomials in $S_{n-2}$ crossing symmetric numerators. It is obvious that all the $\varepsilon_i\Cdot\varepsilon_j$ are independent as there are no relations among them. As to $\varepsilon_i\Cdot p_j$, momentum conservation and transversality together allow us to take $\varepsilon_i\Cdot p_j$ with $j\neq i$ and $n$ as independent variables. For $\varepsilon_n\Cdot p_j$, we instead require $2{\leqslant} j{\leqslant} n-1$ to accommodate the $S_{n-2}$ crossing symmetry. We choose those Mandelstam variables $p_i\Cdot p_j$ that do not contain leg $p_n$ as independent variables. This leaves us with $\frac{(n-1)(n-2)}{2}$ Mandelstam variables. We do not use the on-shell condition $p_n^2=0$ to reduce the number to $\frac{n(n-3)}{2}$ since this minimal basis does not respect the $S_{n-2}$ crossing symmetry. Finally, we simply require that $p_n$ does not appear in the arguments of the tensor $\hat{\varepsilon}_n$.

Numerators are just polynomials composed of a linear combination of the kinematic monomials. Then, in general, we can expand the numerators as
\begin{align}
	\num(1,2,\ldots,n-1,n)=\sum_{\mec\in \text{MECs}}\sum_{\sigma\in S_{n-2}}a^{(\sigma)}_{\mec}\mathbb{P}_{\sigma}\circ \mathsf{m}_{\mec}\,.
\end{align}
Here $\mathsf{m}_{\mec}$ is a representative monomial of the MEC $\mec$, and the group algebra element $\sum_{\sigma\in S_{n-2}}a^{(\sigma)}_{\mec}\mathbb{P}_{\sigma}$ generates a polynomial that is solely composed of the monomials in $\mec$.

\section{Pure gauge degrees of freedom for the NMHV sector}\label{sec:PureGaugeNMHV}
Recall that a pure gauge numerator, called $\num^{\text{gauge}}$ in eq.~(\ref{eq:DefinePureGauge}), resides in the null space of the propagator matrix and thus does not contribute to the amplitude. 
Equivalently, the pure gauge terms should give zero when fed into some version of a double copy formula. They must thus satisfy the local constraint
\begin{align}\label{eq:pgcondition}
\sum_{\beta\in S_{n-2}}A(1,\beta_2,\ldots,\beta_{n-1},n)\num^{\text{gauge}}(1,\beta_2,\ldots,\beta_{n-1},n)=\sum_{i}c_i\mathcal{B}_i\cong 0\,,
\end{align} 
where $\mathcal{B}_i$ are a set of independent BCJ relations and $c_i$ are some local kinematic coefficients. If we treat the partial amplitudes on the left hand side as formal objects, then the above double copy formula indirectly suggests a local parametrization of the pure gauge terms using the BCJ relations $\mathcal{B}_i$.
 
This brings us to the central question that we will answer in this and the next section: How do we  classify and construct the pure gauge terms that can reside in local numerators that respect color-kinematics duality?  Specifically, can we count the number of free parameters corresponding to pure gauge degrees of freedom, and how does the count change when we impose stronger manifest crossing symmetries on the numerators? 
A keyword of this section is ``local'', as we are only interested in studying numerators that are polynomials of the kinematic Lorentz invariants, and thus contain no physical or spurious poles. 

Due to our objective of finding local numerators, we need to distinguish between independent BCJ relations belonging to the vector space over: polynomials in $s_{ij}$ versus rational functions in $s_{ij}$. Recall that in ref.~\cite{Bern:2008qj}, a complete solution to all BCJ relations was constructed, which gave a basis of $(n-3)!$ partial amplitudes. Since the Kleiss-Kuijf basis of partial amplitudes is of size $(n-2)!$~\cite{Kleiss:1988ne}, one is justified to say that there exist $(n-3)(n-3)! =(n-2)!-(n-3)!$ independent BCJ relations. More precisely the BCJ relations live in a $(n-3)(n-3)!$-dimensional vector space over the rational functions in $s_{ij}$. This vector space is spanned by the explicit solution in ref.~\cite{Bern:2008qj}, or alternatively, by the permutations of the simplest such relation at each multiplicity. The latter were dubbed the ``fundamental BCJ relations''~\cite{Feng:2010my}, because of their ability to span the full vector space. 

With \eqn{eq:pgcondition} in mind, it is important that we consider {\it all} BCJ relations that have local coefficients, that is
\be
\sum_{\sigma \in S_2}   a_{\sigma}\, A(1,\sigma,n) =0\,,~~~~~a_{\sigma} \in {\rm Poly}[s_{ij}]
\ee
While it is still true that this set of local BCJ relations live in the $(n-3)(n-3)!$-dimensional vector space over the rational functions in $s_{ij}$, it is not a useful characterization for identifying them. Instead we need to consider a basis of local BCJ relations that spans a vector space over the polynomials in $s_{ij}$. This vector space has a larger dimension than the corresponding vector space over the rational functions in $s_{ij}$. In appendix~\ref{sec:binaryBCJ}, we argue that the binary BCJ relations give such a basis for all linear-in-$s_{ij}$ BCJ relations, and its dimension is $(n-2)!\sum_{i=2}^{n-2} (n-i)^{-1}$ (the generalized Stirling numbers). For BCJ relations with polynomial coefficients of higher degrees, there is as of yet no established basis for a vector space over the polynomials. 

The binary BCJ relations are important because they provide exactly the basis needed for the local expansion~\eqref{eq:pgcondition} in the NMHV sector.  With these at our disposal, we can proceed as follows to simplify the construction of pure gauge numerators:  pure gauge terms are split according to the independent MEC sectors, and then kinematic polynomials are constructed such that they are
\begin{enumerate*}[label=(\arabic*)]
	\item composed only of monomials from the MEC; 
	\item are invariant under the $\neco$ operator action defined in eq.~\eqref{eq:Roperator}.
\end{enumerate*}
We will show that when those $\neco$-invariant polynomials are appropriately assembled and feed into \eqn{eq:pgcondition} they will reassemble into a superposition of the binary BCJ relations, and hence they correspond to pure gauge terms in the numerator, and will not contribute to the amplitude.

This section will focus on numerators exhibiting $S_{n-2}$ crossing symmetry, and subsequently we discuss imposing manifest $S_{n-1}$ and  $S_{n}$ crossing symmetry in the next section. The case study is the NMHV sector of YM, where we have good control over the building blocks, and we will only briefly touch on the N${}^2$MHV in the next section. We remind the reader of the important result from ref.~\cite{Chen:2019ywi}: the MHV sector does not contain any local generalized gauge freedom, and the local numerators of this sector are thus unique.

\subsection{Pure gauge terms and group algebra invariants}\label{sec:pgandBCJ}
We start with proving that the following sum contains only pure gauge degrees of freedom in the sense of eq.~\eqref{eq:pgcondition},
\begin{align}\label{eq:PureGauge}
\num_{[i,\ldots,n-1]}^{\text{gauge}}(1,2,\ldots,n-1,n)&=\sum_{\substack{\sigma^{-1}\in\{2,\ldots,i-1\} \\ \shuffle\{i,\ldots,n-1\}}}p_{\sigma_i}\Cdot X_{\sigma_i}(\mathbb{I})\,\mathcal{I}_{\neco(i,\ldots,n-1)}(1,\sigma,n)\,,
\end{align} 
where $2\leqslant i\leqslant n-2$ and $X_{\sigma_i}(\mathbb{I})=\sum_{j<\sigma_i}p_j$. The summation is over all the permutations whose inverse are in the shuffle $\{2,\ldots,i-1\}\shuffle\{i,\ldots,n-1\}$. For example, at $n=5$ eq.~\eqref{eq:PureGauge} gives
\begin{subequations}\label{eq:pg5}
	\begin{align}
	\num^{\text{gauge}}_{[2,3,4]}(1,2,3,4,5)&=p_2\Cdot p_1\,\mathcal{I}_{\neco(2,3,4)}(1,2,3,4,5)\,,\\
	\num^{\text{gauge}}_{[3,4]}(1,2,3,4,5)&=p_3\Cdot (p_{1}+p_2)\,\mathcal{I}_{\neco(3,4)}(1,2,3,4,5)+p_2\Cdot p_1\,\mathcal{I}_{\neco(3,4)}(1,3,2,4,5)\nonumber\\
	&\quad+p_2\Cdot p_1\,\mathcal{I}_{\neco(3,4)}(1,4,2,3,5)\,.
	\end{align}
\end{subequations}
In particular, the second equation comes from summing over 
\begin{align}
\sigma^{-1}\in\{2\}\shuffle\{3,4\}=\big\{\{2,3,4\},\{3,2,4\},\{3,4,2\}\big\}\,,
\end{align}
which implies that $\sigma\in\big\{\{2,3,4\},\{3,2,4\},\{4,2,3\}\big\}$.
The function $\mathcal I_{\neco(i,\ldots,n-1)}(1,\sigma,n)$ is obtained from a permutation 
\begin{align}\label{eq:IRsigma}
\mathcal I_{\neco(i,\ldots,n-1)}(1,\sigma,n)=\mathbb{P}_{\sigma}\circ \mathcal I_{\neco(i,\ldots, n-1)}\,,
\end{align}
where the canonically labeled object $\mathcal I_{\neco(i,\ldots, n-1)}$ is an eigenfunction, or {\it group algebra invariant}, of the $\neco$ operator. Specifically, we take it to be a kinematic polynomial multilinear in the polarizations of the external states and of mass dimension $(n{-}4)$, and it solves the eigenfunction equation 
\begin{align}\label{eq:GAInv}
  \frac{1}{n-i}\neco(i,\ldots, n-1)\circ\mathcal I_{\neco(i,\ldots,n-1)}=\mathcal I_{\neco(i,\ldots,n-1)}\,.
\end{align}
Using the orthogonality relation~\eqref{eq:I-R}, it is straightforward to see that any solution to eq.~\eqref{eq:GAInv} can be written as
\begin{align}\label{eq:solutionofAI}
  \mathcal I_{\neco(i,\ldots,n-1)}=\neco(i,\ldots,n-1) \circ \mathsf{m}\,,
\end{align} 
where $\mathsf{m}$ is a generic kinematic monomial, which we take to be of mass dimension $(n{-}4)$ and multilinear in the polarizations. 

Let us show that the numerator given by eq.~\eqref{eq:PureGauge}, with the general solution \eqref{eq:solutionofAI}, indeed satisfies the pure gauge condition~\eqref{eq:pgcondition}. We start with the definition~\eqref{eq:PureGauge} and obtain the full DDM basis of numerators through permutations, 
\begin{align}\label{eq:pgproof1}
&\sum_{\beta\in S_{n-2}} A(1,\beta,n)\,\num^{\text{gauge}}_{[\beta_i,\ldots,\beta_{n-1}]}(1,\beta,n)\nonumber\\
&=\sum_{\beta\in S_{n-2}}A(1,\beta,n)\,\mathbb{P}_{\beta}\circ \num^{\text{gauge}}_{[i,\ldots,n-1]}(1,2,\ldots,n-1,n)\nonumber\\
&=\sum_{\beta\in S_{n-2}}\sum_{\substack{\sigma^{-1}\in\{2,\ldots,i-1\} \\ \shuffle\{i,\ldots,n-1\} }}A(1,\beta,n)\,p_{\beta\sigma_i}\Cdot X_{\beta\sigma_i}(\beta)\,\mathcal{I}_{\neco(i,\ldots,n-1)}(1,\beta\sigma,n)\,.
\end{align}
Using the form of $\neco$ given in eq.~\eqref{eq:BR}, we have
\begin{align}
\mathcal{I}_{\neco(i,\ldots,n-1)}(1,\beta\sigma,n)&=\mathbb{P}_{\beta\sigma}\circ \neco(i,\ldots,n-1)\circ \mathsf{m}=\sum_{\gamma\in K[i,\ldots,n-1]}\mu^{(\gamma)}\, \mathbb{P}_{\beta\sigma\gamma^{-1}}\circ \mathsf{m}\,.
\end{align}
We plug it back into eq.~\eqref{eq:pgproof1}, and redefine dummy variables as $\beta\sigma\gamma^{-1} \rightarrow \beta$, then we obtain
\begin{align}\label{eq:pgproof2}
&\sum_{\beta\in S_{n-2}}A(1,\beta,n) \, \num^{\text{gauge}}_{[\beta_i,\ldots,\beta_{n-1}]}(1,\beta,n)\nonumber\\
&=\sum_{\gamma\in K[i,\ldots,n-1]}\sum_{\beta\in S_{n-2}}\sum_{\substack{\sigma^{-1}\in\{2,\ldots,i-1\} \\ \shuffle\{i,\ldots,n-1\} }}A(1,\beta\gamma\sigma^{-1},n)\,p_{\beta\gamma_i}\Cdot X_{\beta\gamma_i}(\beta\gamma\sigma^{-1})\,\mu^{(\gamma)}\, \mathbb{P}_{\beta}\circ  \mathsf{m}\nonumber\\
&=\sum_{\beta\in S_{n-2}}\Big[\mathbb{P}_{\beta}\circ\mathbb{B}(i,\ldots,n-1)\,\circ \!\!\!\!\!\!\sum_{\substack{\sigma\in\{2,\ldots,i-1\} \\ \shuffle\{i,\ldots,n-1\} }}p_i\Cdot X_i(\sigma)A(1,\sigma,n)\Big]\mathbb{P}_{\beta}\circ  \mathsf{m}\nonumber\\
&=\sum_{\beta\in S_{n-2}}\big[\mathbb{P}_{\beta}\circ\mathcal{B}_{[i,\ldots,n-1]}(1,2,\ldots,n-1,n)\big] \,( \mathbb{P}_\beta\circ \mathsf{m})
\end{align}
To obtain the third line, we have again used the abbreviated form of $\mathbb{B}$ given in eq.~\eqref{eq:BR}, and renamed the dummy variable $\sigma^{-1}  \rightarrow \sigma$. We see that eq.~\eqref{eq:pgproof2} indeed vanishes on-shell due to the binary BCJ relations $\mathcal{B}_{[i,\ldots,n-1]}(1,2,\ldots,n{-}1,n)\cong 0$.

In principle, the above construction is not restricted to the NMHV sector, it should apply more generally to pure gauge terms in any sector of YM. However, there is a slight complication when going beyond NMHV since we do not yet have a complete basis of local BCJ relations over non-linear polynomials. Nevertheless, if we restrict our attention to the NMHV sector, one can argue that eq.~\eqref{eq:PureGauge} forms a complete basis for the pure gauge terms in this sector. We have explicitly and independently checked this statement up to seven points using a direct Ansatz construction of the pure gauge numerator.  Such a brute force calculation of pure gauge terms becomes computationally intractable at eight points and beyond.

The argument that \eqn{eq:PureGauge} captures all NMHV pure gauge freedom goes as follows. We defined the NMHV sector numerators to have at most one power of the $p_i\Cdot p_j$ factors. All such factors are explicit in \eqn{eq:PureGauge}, and hence the invariant function $\mathcal{I}_{\neco}$ cannot contain such factors. This implies that when plugging in the numerators into the pure-gauge constraint (\ref{eq:pgcondition}) the only BCJ relations that are expected to show up on the right-hand side are those that have coefficients linear in $p_i\Cdot p_j$. Since we know the complete basis of local BCJ relations linear in in $p_i\Cdot p_j$, namely the binary BCJ relations, one can conclude that we have fully taken into account the local freedom associated with the constraint (\ref{eq:pgcondition}).

 A caveat to this argument is that we do not yet have a mathematical proof that the binary BCJ relations span the vector space of local BCJ relations with linear coefficients. What we have checked up to multiplicity nine is that the binary BCJ relations span the space of all currently known such BCJ relations. This includes all permutations of the fundamental BCJ relations, and generalized BCJ relations~\cite{Bern:2008qj,BjerrumBohr:2009rd,Chen:2011jxa}. The fact that the dimension of this basis happens to non-trivially coincide with the generalized Stirling numbers, gives support to the presumption that the basis is complete.

As our next step, we explore the basis structure of pure gauge terms in the  NMHV sector. One can consider a fixed monomial $\mathsf{m}$ which belongs to a certain MEC, and then we count the number of independent $\neco$-invariant polynomials for this MEC.

\subsection{Independent group algebra invariants}\label{sec:countO}
Given a MEC denoted by $\mec$, the group algebra invariants $\mathcal{I}_{\neco}=\neco\circ \mathsf{m}$, for different $\neco$'s, will constitute a subset of the polynomial functions that can be built out of the monomials $\mathsf{m}\in\mec$. As discussed, $\neco$ has right zero divisors, or, equivalently, $\neco$ is a projector. This implies that the number of independent $\mathcal{I}_{\neco}$ should be equal to the \emph{rank of $\neco$ in the representation space $\mec$.}

We start with a few examples. At $n=4$, we consider $\neco(2,3)=\mathbb{I}-\mathbb{P}_{(23)}$. We can find three MECs that are of mass dimension zero,
\begin{align}
&\mec_{4,1}=\Big[\varepsilon_2\Cdot\varepsilon_3\, \varepsilon_1\Cdot\varepsilon_4\Big]\,,& &\mec_{4,1'}=\Big[\hat{\varepsilon}_{4}(\varepsilon_1,\varepsilon_2,\varepsilon_3)\Big],& &\mec_{4,2}=\left[\begin{array}{c}
\varepsilon_1\Cdot\varepsilon_2\,\varepsilon_3\Cdot\varepsilon_4 \\
\varepsilon_1\Cdot\varepsilon_3\,\varepsilon_2\Cdot\varepsilon_4
\end{array}\right].
\end{align}
The subscripts of $\mec$ indicate the number of particles and the dimension of the representation, respectively. Extra labels may be used to distinguish representations of the same dimension, and a prime is used to signal the presence of tensors. The permutations have the following representation matrices: 
\begin{align}
\begin{array}{>{\centering $}p{2cm}<{$}|>{\centering $}p{2cm}<{$}|>{\centering $}p{2cm}<{$}|>{\centering\arraybackslash$}p{2cm}<{$}}
 & \mec_{4,1} & \mec_{4,1'} & \mec_{4,2} \\ \hline
\mathbb{I} & 1 & 1 & \begin{pmatrix} 1 & 0 \\ 0 & 1 \end{pmatrix} \\ \hline
\mathbb{P}_{(23)} & 1 & -1 & \begin{pmatrix} 0 & 1 \\ 1 & 0 \end{pmatrix}  
\end{array}
\end{align}
The ranks of $\neco(2,3)$ in these representations are
\begin{align}
\text{Rank}_{\mec_{4,1}}\big[\neco(2,3)\big]=0\,,& &\text{Rank}_{\mec_{4,1'}}\big[\neco(2,3)\big]=1\,,& &\text{Rank}_{\mec_{4,2}}\big[\neco(2,3)\big]=1\,.
\end{align}
It is clear that no invariant arises from $\mec_{4,1}$, since $\neco(2,3)\circ (\varepsilon_2\Cdot\varepsilon_3\, \varepsilon_1\Cdot\varepsilon_4)=0$. And there is one invariant each in $\mec_{4,1'}$ and $\mec_{4,2}$,
\begin{align}\label{eq:fourPG}
&\mathcal{I}_{\neco(2,3)}^{(4,1')}=\neco(2,3)\circ \hat{\varepsilon}_4(\varepsilon_1,\varepsilon_2,\varepsilon_3)=2\hat{\varepsilon}_4(\varepsilon_1,\varepsilon_2,\varepsilon_3)\,,\nonumber\\
&\mathcal{I}_{\neco(2,3)}^{(4,2)}=\neco(2,3)\circ (\varepsilon_1\Cdot\varepsilon_2\,\varepsilon_3\Cdot\varepsilon_4)=\varepsilon_1\Cdot\varepsilon_2\,\varepsilon_3\Cdot\varepsilon_4-\varepsilon_1\Cdot\varepsilon_3\,\varepsilon_2\Cdot\varepsilon_4\,.
\end{align}
Plugging them into eq.~\eqref{eq:PureGauge}, we get two sets of independent pure gauge numerators,
\begin{align}
&\begin{bmatrix}
N^{\text{gauge}}(1,2,3,4) \\ N^{\text{gauge}}(1,3,2,4)
\end{bmatrix} = \begin{bmatrix}
p_1\Cdot p_2\, \hat{\varepsilon}_4(\varepsilon_1,\varepsilon_2,\varepsilon_3) \\
-p_1\Cdot p_3\, \hat{\varepsilon}_4(\varepsilon_1,\varepsilon_2,\varepsilon_3)
\end{bmatrix}\,,\nonumber\\
&\begin{bmatrix}
N^{\text{gauge}}(1,2,3,4) \\ N^{\text{gauge}}(1,3,2,4)
\end{bmatrix} = \begin{bmatrix}
p_1\Cdot p_2\, (\varepsilon_1\Cdot\varepsilon_2\,\varepsilon_3\Cdot\varepsilon_4-\varepsilon_1\Cdot\varepsilon_3\,\varepsilon_2\Cdot\varepsilon_4) \\
-p_1\Cdot p_3\, (\varepsilon_1\Cdot\varepsilon_2\,\varepsilon_3\Cdot\varepsilon_4-\varepsilon_1\Cdot\varepsilon_3\,\varepsilon_2\Cdot\varepsilon_4)
\end{bmatrix}\,.
\end{align}
There are no other $S_2$ crossing symmetric pure gauge degrees of freedom in the local four-point numerators, with particles $1,2,3$ being vectors. If we drop the unphysical tensor, then there is only one degree of freedom, as is familiar from the literature~\cite{Bern:2008qj}.

For the simplicity of presentation, at $n=5$, we restrict the explicit examples to dimension-one monomials proportional to Lorentz contractions between $\varepsilon_1$ and $\varepsilon_5$ (including tensors). They fall into the following four MECs, which can be viewed as representation spaces for $S_3$,
\begin{align}\label{eq:5pMEC}
&\mec_{5,3a}=\left[\begin{array}{c}
\varepsilon_2\Cdot p_1\,\varepsilon_3\Cdot\varepsilon_4\,\varepsilon_1\Cdot\varepsilon_5 \\
\varepsilon_3\Cdot p_1\,\varepsilon_2\Cdot\varepsilon_4\,\varepsilon_1\Cdot\varepsilon_5 \\
\varepsilon_4\Cdot p_1\,\varepsilon_2\Cdot\varepsilon_3\,\varepsilon_1\Cdot\varepsilon_5
\end{array}\right], & &\mec_{5,3a'}=\left[\begin{array}{c}
\varepsilon_2\Cdot p_1\,\hat{\varepsilon}_5(\varepsilon_1,\varepsilon_3,\varepsilon_4) \\
\varepsilon_3\Cdot p_1\,\hat{\varepsilon}_5(\varepsilon_1,\varepsilon_2,\varepsilon_4) \\
\varepsilon_4\Cdot p_1\,\hat{\varepsilon}_5(\varepsilon_1,\varepsilon_2,\varepsilon_3)
\end{array}\right],\nonumber\\
&\mec_{5,6a}=\left[\begin{array}{c}
\varepsilon_2\Cdot p_3\,\varepsilon_3\Cdot\varepsilon_4\,\varepsilon_1\Cdot\varepsilon_5 \\
\varepsilon_2\Cdot p_4\,\varepsilon_3\Cdot\varepsilon_4\,\varepsilon_1\Cdot\varepsilon_5 \\
\varepsilon_3\Cdot p_2\,\varepsilon_2\Cdot \varepsilon_4\,\varepsilon_1\Cdot\varepsilon_5 \\
\varepsilon_3\Cdot p_4\,\varepsilon_2\Cdot\varepsilon_4\,\varepsilon_1\Cdot\varepsilon_5 \\
\varepsilon_4\Cdot p_2\,\varepsilon_2\Cdot\varepsilon_3\,\varepsilon_1\Cdot\varepsilon_5 \\
\varepsilon_4\Cdot p_3\,\varepsilon_2\Cdot\varepsilon_3\,\varepsilon_1\Cdot\varepsilon_5 
\end{array}\right],& &\mec_{5,6a'}=\left[\begin{array}{c}
\varepsilon_2\Cdot p_3\,\hat{\varepsilon}_5(\varepsilon_1,\varepsilon_3,\varepsilon_4) \\
\varepsilon_2\Cdot p_4\,\hat{\varepsilon}_5(\varepsilon_1,\varepsilon_3,\varepsilon_4) \\
\varepsilon_3\Cdot p_2\,\hat{\varepsilon}_5(\varepsilon_1,\varepsilon_2,\varepsilon_4) \\
\varepsilon_3\Cdot p_4\,\hat{\varepsilon}_5(\varepsilon_1,\varepsilon_2,\varepsilon_4) \\
\varepsilon_4\Cdot p_2\,\hat{\varepsilon}_5(\varepsilon_1,\varepsilon_2,\varepsilon_3) \\
\varepsilon_4\Cdot p_3\,\hat{\varepsilon}_5(\varepsilon_1,\varepsilon_2,\varepsilon_3) 
\end{array}\right].
\end{align}
We consider the invariants $\mathcal{I}_{\neco}$ given by $\neco(2,3,4)=\mathbb{I}-\mathbb{P}_{(23)}-\mathbb{P}_{(243)}+\mathbb{P}_{(24)}$. The rank of $\neco(2,3,4)$ thus depends on the representation matrices of these permutations. Explicit computations show that
\begin{align}\label{eq:rankR5}
&\text{Rank}_{\mec_{5,3a}}\big[\neco(2,3,4)\big]=1\,, & &\text{Rank}_{\mec_{5,3a'}}\big[\neco(2,3,4)\big]=1\,, \nonumber\\
&\text{Rank}_{\mec_{5,6a}}\big[\neco(2,3,4)\big]=2\,,& &\text{Rank}_{\mec_{5,6a'}}\big[\neco(2,3,4)\big]=2\,.
\end{align}
It turns out that all the other MECs also form three- or six-dimensional representations for $S_3$, and one can check that eq.~\eqref{eq:rankR5} holds generically at $n=5$: $\text{Rank}\big[\neco(2,3,4)\big]=1$ for all the three-dimensional MECs and $\text{Rank}\big[\neco(2,3,4)\big]=2$ for all the six-dimensional ones. Additionally, we have to consider the invariants $\mathcal{I}_{\neco}$ generated by $\neco(3,4)$. In this case, the MECs will decompose into representations of $S_2=\{\mathbb{I},\mathbb{P}_{(34)}\}$, and the calculation is identical to the $n=4$ example.

To compute the rank of $\neco$ operators systematically, we may start with the rank-trace relation of an idempotent matrix~\cite{matrix}. Namely, the rank of an idempotent matrix equals its trace,
\begin{align}\label{eq:rankR}
	\text{Rank}_{\mec}\big[\neco(i_1,i_2,\ldots,i_r)\big]=\frac{1}{r}\text{Tr}_{\mec}\big[\neco(i_1,i_2,\ldots,i_r)\big]\,.
\end{align}
Recall that $\neco/r$ is the idempotent operator, not $\neco$ itself. 
Since the trace is a linear operation, we can now more easily utilize the expansion of $\neco$ in terms of permutations, following eq.~\eqref{eq:Lexpansion}. 
The trace of a permutation $\sigma$ in $\mec$ equals the character $\chi_{\mec}$ of the conjugacy class $\mathsf{K}(\lambda)$ it belongs to, $\text{Tr}_{\mec}\big[\mathbb{P}_{\sigma}\big]=\chi_{\mec}\big[\mathsf{K}(\lambda)\big]$ for all $\sigma\in\mathsf{K}(\lambda)$. Here $\lambda$ is a label for conjugacy classes of permutations groups, $\lambda=(k_1,k_2,\ldots,k_m)$, where $m$ is the number of cycles and $k_i$ is the length of the $i$-th cycle.
For example, both $\sigma{=}(12)(34)$ and $(13)(24)$ belongs to the conjugacy class with two length-two cycles, $\mathsf{K}(2,2)$, such that $\text{Tr}_{\mec}\big[\mathbb{P}_{(13)(24)}\big]=\text{Tr}_{\mec}\big[\mathbb{P}_{(13)(24)}\big]=\chi_{\mec}\big[\mathsf{K}(2,2)\big]$. For low multiplicities, we can simply expand $\neco$ and compute the trace explicitly, for example,
\begin{align}\label{eq:rankLexample}
\text{Rank}_{\mec}\big[\neco(2,3,4)\big]&=\frac{1}{3}\Big(\Dc-\chi_{\mec}\big[\mathsf{K}(3)\big]\Big)\,,\nonumber\\
\text{Rank}_{\mec}\big[\neco(2,3,4,5)\big]&=\frac{1}{4}\Big(\Dc-\chi_{\mec}\big[\mathsf{K}(2,2)\big]\Big)\,,\\
\text{Rank}_{\mec}\big[\neco(2,3,4,5,6)\big]&=\frac{1}{5}\Big(\Dc-\chi_{\mec}\big[\mathsf{K}(5)\big]\Big)\,,\nonumber\\
\text{Rank}_{\mec}\big[\neco(2,3,4,5,6,7)\big]&=\frac{1}{6}\Big(\Dc-\chi_{\mec}\big[\mathsf{K}(2,2,2)\big]-\chi_{\mec}\big[\mathsf{K}(3,3)\big]+\chi_{\mec}\big[\mathsf{K}(6)\big]\Big)\,.\nonumber
\end{align}
Note that the identity $\mathbb{I}$ forms its own conjugacy class, and its trace gives the dimension of the representation, $\Dc=\text{Tr}_{\mec}[\mathbb{I}]$. In table~\ref{tab:mec6}, we give the rank of $\neco(2,3,4,5)$ in some typical MECs at six points. The general formula for the rank of $\neco$ will be derived and presented in eq.~\eqref{eq:rankRgeneral}.

\begin{table}
\centering
\renewcommand{\arraystretch}{1.1}
\begin{tabular}{>{\centering $}p{1.5cm}<{$}|>{\centering $}p{5cm}<{$}|>{\centering $}p{1.5cm}<{$}|>{\centering $}p{2.5cm}<{$}|>{\centering\arraybackslash$}p{1.75cm}<{$}}
  \text{MEC} & \text{Representative} & \Dc &\text{Tr}_{\mec}[\mathbb{P}_{(25)(34)}]& \text{Rank}_{\mec}[\neco] \\ \hline
  	\mec_{6,24} & \varepsilon_2\Cdot p_4\,\varepsilon_3\Cdot p_1\,\varepsilon_4\Cdot\varepsilon_5\,\varepsilon_1\Cdot\varepsilon_6 & 24& 0 & 6\\
   	\mec_{6,12a} &\varepsilon _2\Cdot p_1\, \varepsilon _3\Cdot p_2\,\varepsilon_4\Cdot\varepsilon _5\,\varepsilon_1\Cdot\varepsilon_6 & 12& 0 & 3\\
   	\mec_{6,12a'} & \varepsilon _2\Cdot p_1\, \varepsilon _3\Cdot p_2\,\hat\varepsilon_6(\varepsilon _1,\varepsilon _4,\varepsilon _5)  & 12& 0 & 3\\
  	\mec_{6,12b} &  \varepsilon _2\Cdot p_5\,\varepsilon _3\Cdot p_4 \varepsilon _4\Cdot \varepsilon _5\,\varepsilon_1\Cdot\varepsilon_6 & 12& 4 & 2\\
  	\mec_{6,12b'} &	\varepsilon _2\Cdot p_5\, \varepsilon _3\Cdot p_4\,\hat\varepsilon_6(\varepsilon _1,\varepsilon _4,\varepsilon _5)  & 12& -4& 4  \\
  	\mec_{6,6a} &\varepsilon _2\Cdot p_1 \,\varepsilon _3\Cdot p_1\,\varepsilon _4\Cdot\varepsilon _5\,\varepsilon_1\Cdot\varepsilon _6& 6&  2 & 1\\
  	\mec_{6,6a'} &\varepsilon _2\Cdot p_1 \,\varepsilon _3\Cdot p_1\,\hat\varepsilon_6(\varepsilon _1,\varepsilon _4,\varepsilon _5)  & 6& -2 & 2\\
  	\mec_{6,4'} & \varepsilon_1\Cdot p_2\,\varepsilon_2\Cdot p_1\,\hat{\varepsilon}_6(\varepsilon_3,\varepsilon_4,\varepsilon_5) & 4& 0  & 1\\
\end{tabular}
\caption{Examples of MECs at $n=6$ and the rank of $\neco(2,3,4,5)$.}
\label{tab:mec6}
\end{table}

\subsection{Counting MECs}\label{sec:countMEC}
A remaining task is to find all the MECs to which the kinematic monomials belong. For simplicity, in this section \emph{we will focus on the monomials without tensors},\footnote{This will complete the counting for pure gauge freedom in the NMHV sector of YM, for standard vector states. Tensors are natural in our algebraic construction, and can be argued to be  mathematically interesting for understanding the wider structure of the kinematic algebra. But a complete counting for tensor pure gauge freedom is less important.} which is a reasonable restriction since vectors are the physical states. Given a monomial $\mathsf{m}$ that consists of $\varepsilon\Cdot\varepsilon$ and $\varepsilon\Cdot p$ factors, we can construct polynomials invariant under any action of a finite group $G$,
\begin{align}\label{eq:RGm}
\mathsf{m}\longrightarrow \reyop_{G}\circ\mathsf{m}\equiv\frac{1}{|G|}  \sum_{\sigma\in G}\mathbb{P}_{\sigma}\circ\mathsf{m}\,.
\end{align}
Considering the group $G=S_{n-2}$,  the two monomials $\mathsf{m}$ and $\mathsf{m}'$ belong to the same MEC if and only if the quotient of the two polynomials $ \reyop_{S_{n-2}}\circ\mathsf{m}$ and $ \reyop_{S_{n-2}}\circ\mathsf{m}'$ is a constant. Thus the distinct invariants $ \reyop_{S_{n-2}}\circ\mathsf{m}$ are in one-to-one correspondence to the MECs.

Let us refine the analysis by factorizing the monomial $\mathsf{m}$ as
\begin{align}
\mathsf{m}=\mathsf{m}^{(\varepsilon\varepsilon)}\mathsf{m}^{(\varepsilon p)}\,,
\end{align}
where, in the NMHV sector, $\mathsf{m}^{(\varepsilon\varepsilon)}$ contains two powers of $\varepsilon\Cdot\varepsilon$ and $\mathsf{m}^{(\varepsilon p)}$ has $(n{-}4)$ powers of $\varepsilon\Cdot p$. Based on the group action of $S_{n-2}$, we can recognize five classes that $\mathsf{m}^{(\varepsilon\varepsilon)}$ could belong to,
\begin{align}\label{eq:m0class}
\renewcommand{\arraystretch}{1.1}
\begin{array}{>{\centering $}p{1.5cm}<{$}|>{\centering $}p{2cm}<{$}|>{\centering $}p{2cm}<{$}|>{\centering $}p{2cm}<{$}|>{\centering $}p{2cm}<{$}|>{\centering\arraybackslash$}p{3.25cm}<{$}}
 & \mathfrak{C}_1 & \mathfrak{C}_2 & \mathfrak{C}_3 & \mathfrak{C}_4 & \mathfrak{C}_5 \\ \hline
\begin{array}{c}
\mathsf{m}^{(\varepsilon\varepsilon)}
\end{array} & \varepsilon_a\Cdot\varepsilon_b\,\varepsilon_1\Cdot\varepsilon_n & \varepsilon_1\Cdot\varepsilon_a\,\varepsilon_b\Cdot\varepsilon_n & \varepsilon_a\Cdot\varepsilon_b\,\varepsilon_c\Cdot\varepsilon_n & \varepsilon_1\Cdot\varepsilon_a\,\varepsilon_b\Cdot\varepsilon_c & \varepsilon_a\Cdot\varepsilon_b\,\varepsilon_c\Cdot\varepsilon_d \\ \hline
G^{\text{inv}} & S_2{\times} S_{n-4} & S_{n-4} & S_2{\times} S_{n-5} & S_2{\times} S_{n-5} & S_2{\times} S_2{\times} S_2{\times} S_{n-6} \\ 
\end{array} \,,
\end{align}
where $a,b,c,d\in\{2,\ldots,n{-}1\}$. An individual monomial $\mathsf{m}^{(\varepsilon\varepsilon)}\in\mathfrak{C}_{i}$ is associated with a subgroup $G^{\text{inv}}_i\subset S_{n-2}$ that leaves it invariant.\footnote{More precisely, the subgroup that leaves $\mathsf{m}^{(\varepsilon\varepsilon)}\in \mathfrak{C}_i$ invariant is isomorphic to $G^{\text{inv}}_i$.} For example, $\mathsf{m}^{(\varepsilon\varepsilon)}=\varepsilon_a\Cdot\varepsilon_b\,\varepsilon_1\Cdot\varepsilon_n\in\mathfrak{C}_1$ is invariant under the exchange $\mathbb{P}_{(ab)}$ and the permutations of the remaining $(n{-}4)$ labels, thus $G^{\text{inv}}_{1}=S_2\times S_{n-4}$. 

The classes $\mathfrak{C}_i$ are coarse-grained versions of the MECs, namely the classes will split into MECs with no overlap. If we denote the number of MECs contained in $\mathfrak{C}_i$ as $d_{\mathfrak{C}_i}$, the total number of tensor-free MECs is 
\begin{align}
\text{\# MECs}=\degree_{\mathfrak{C}_1}+\degree_{\mathfrak{C}_2}+\degree_{\mathfrak{C}_3}+\degree_{\mathfrak{C}_4}+\degree_{\mathfrak{C}_5}\,.
\end{align}
Let us now fine grain $\mathfrak{C}_i$ into MECs by studying the action of $G^{\text{inv}}_{i}$ on $\mathsf{m}^{(\varepsilon p)}$. For a given $\mathsf{m}^{(\varepsilon\varepsilon)}\in \mathfrak{C}_i$, we have the allowed monomials
\begin{align}\label{eq:mep}
\mathsf{m}^{(\varepsilon p)}\in\left\{\prod_{l=1}^{n-4}\varepsilon_{i_l}\Cdot p_{j_l}\middle|\begin{array}{c}
\varepsilon_{i_l}\notin \mathsf{m}^{(\varepsilon\varepsilon)}\text{ and distinct,} \\
j_l\notin\{i_l,n\}\text{ if }i_l\neq n\,,j_l\notin\{1,n\}\text{ if }i_n=n
\end{array}\right\}.
\end{align}
Since there are $(n{-}2)$ choices for each $p_{j_l}$, $\mathsf{m}^{(\varepsilon p)}$ forms a $(n{-}2)^{n{-}4}$-dimensional representation space for $G^{\text{inv}}_i$. The number of MECs contained in $\mathfrak{C}_i$ equals the number of distinct group algebra invariants under the action $ \reyop_{G^{\text{inv}}_i}\circ\mathsf{m}^{(\varepsilon p)}$, which is given by~\cite{Sturmfels} 
\begin{align}\label{eq:dCi}
\degree_{\mathfrak{C}_i}=\frac{1}{|G_i^{\text{inv}}|}\sum_{\sigma\in G_i^{\text{inv}}}\text{Tr}[\mathbb{P}_{\sigma}]\,.
\end{align}
Here the trace is taken in the representation space~\eqref{eq:mep}, and $\text{Tr}[\mathbb{P}_{\sigma}]$ equals the number of $\mathsf{m}^{(\varepsilon p)}$ monomials that are left invariant under the permutation $\mathbb{P}_{\sigma}$. Next, we illustrate the above general considerations with some examples.

\paragraph{Five points:} 
From eq.~\eqref{eq:m0class}, the nonempty classes are $\mathfrak{C}_{1,2,3,4}$. We first choose the typical $\mathsf{m}^{(\varepsilon\varepsilon)}$ from each class, which determines $G^{\text{inv}}$ and the $\mathsf{m}^{(\varepsilon p)}$ space, 
\begin{align}\label{eq:eeclass}
\renewcommand{\arraystretch}{1.1}
\begin{array}{>{\centering $}p{2cm}<{$}|>{\centering $}p{2cm}<{$}|>{\centering $}p{2cm}<{$}|>{\centering $}p{2cm}<{$}|>{\centering\arraybackslash$}p{2.5cm}<{$}}
& \mathfrak{C}_1 & \mathfrak{C}_2 & \mathfrak{C}_3 & \mathfrak{C}_4  \\ \hline
\mathsf{m}^{(\varepsilon\varepsilon)} & \varepsilon_2\Cdot\varepsilon_3\,\varepsilon_1\Cdot\varepsilon_5 & \varepsilon_1\Cdot\varepsilon_2\,\varepsilon_3\Cdot\varepsilon_5 & \varepsilon_2\Cdot\varepsilon_3\,\varepsilon_4\Cdot\varepsilon_5 & \varepsilon_1\Cdot\varepsilon_2\,\varepsilon_3\Cdot\varepsilon_4 \\ \hline
G^{\text{inv}} & \{\mathbb{I},\mathbb{P}_{(23)}\} & \{\mathbb{I}\} & \{\mathbb{I},\mathbb{P}_{(23)}\} & \{\mathbb{I},\mathbb{P}_{(34)}\}  \\ \hline
\mathsf{m}^{(\varepsilon p)}\text{ space} & \begin{bmatrix}
\varepsilon_4\Cdot p_1 \\
\varepsilon_4\Cdot p_2 \\
\varepsilon_4\Cdot p_3
\end{bmatrix} & \begin{bmatrix}
\varepsilon_4\Cdot p_1 \\
\varepsilon_4\Cdot p_2 \\
\varepsilon_4\Cdot p_3
\end{bmatrix} & \begin{bmatrix}
\varepsilon_1\Cdot p_2 \\
\varepsilon_1\Cdot p_3 \\
\varepsilon_1\Cdot p_4
\end{bmatrix} & \begin{bmatrix}
\varepsilon_5\Cdot p_2 \\
\varepsilon_5\Cdot p_3 \\
\varepsilon_5\Cdot p_4
\end{bmatrix}
\end{array} \,.
\end{align}
The representation matrices of $G^{\text{inv}}$ are
\begin{align}
\begin{array}{>{\centering $}p{2cm}<{$}|>{\centering $}p{2cm}<{$}|>{\centering $}p{2cm}<{$}|>{\centering $}p{2cm}<{$}|>{\centering $}p{2cm}<{$}|>{\centering\arraybackslash$}p{2.5cm}<{$}}
\text{space} & \multicolumn{2}{c|}{\text{group elements}} & \text{space} & \multicolumn{2}{c}{\text{group elements}} \\ \hline
\multirow{2}{*}[-1em]{$\begin{bmatrix}
	\varepsilon_4\Cdot p_1 \\
	\varepsilon_4\Cdot p_2 \\
	\varepsilon_4\Cdot p_3
	\end{bmatrix}$} & \mathbb{I}               & \mathbb{P}_{(23)}              & \multirow{2}{*}[-1em]{$\begin{bmatrix}
	\varepsilon_1\Cdot p_2 \\
	\varepsilon_1\Cdot p_3 \\
	\varepsilon_1\Cdot p_4
	\end{bmatrix}$} & \mathbb{I} & \mathbb{P}_{(23)}              \\ \cline{2-3} \cline{5-6} 
& \begin{pmatrix}
1 & 0 & 0 \\
0 & 1 & 0 \\
0 & 0 & 1
\end{pmatrix} & \begin{pmatrix}
1 & 0 & 0 \\
0 & 0 & 1 \\
0 & 1 & 0
\end{pmatrix} &  & \begin{pmatrix}
1 & 0 & 0 \\
0 & 1 & 0 \\
0 & 0 & 1
\end{pmatrix}               & \begin{pmatrix}
0 & 1 & 0 \\
1 & 0 & 0 \\
0 & 0 & 1
\end{pmatrix} \\ \hline
\multirow{2}{*}[-1em]{$\begin{bmatrix}
	\varepsilon_4\Cdot p_1 \\
	\varepsilon_4\Cdot p_2 \\
	\varepsilon_4\Cdot p_3
	\end{bmatrix}$} & \multicolumn{2}{c|}{\mathbb{I}} & \multirow{2}{*}[-1em]{$\begin{bmatrix}
	\varepsilon_5\Cdot p_2 \\
	\varepsilon_5\Cdot p_3 \\
	\varepsilon_5\Cdot p_4
	\end{bmatrix}$} & \mathbb{I} & \mathbb{P}_{(34)} \\ \cline{2-3} \cline{5-6} 
& \multicolumn{2}{c|}{\begin{pmatrix}
	1 & 0 & 0 \\
	0 & 1 & 0 \\
	0 & 0 & 1
	\end{pmatrix}}  & & \begin{pmatrix}
1 & 0 & 0 \\
0 & 1 & 0 \\
0 & 0 & 1
\end{pmatrix} & \begin{pmatrix}
1 & 0 & 0 \\
0 & 0 & 1 \\
0 & 1 & 0
\end{pmatrix}                
\end{array}
\end{align}
Following eq.~\eqref{eq:dCi}, we get
\begin{align}
& \degree_{\mathfrak{C}_1}=2\,,& & \degree_{\mathfrak{C}_2}=3\,,& \degree_{\mathfrak{C}_3}=2\,,& & \degree_{\mathfrak{C}_4}=2\,.
\end{align}
Therefore, there are in all nine MECs without tensors. We list them explicitly using the notations established in section~\ref{sec:countO},
\begin{align}\label{eq:5pMECs}
\begin{array}{>{\centering $}p{1cm}<{$}|>{\centering $}p{2cm}<{$}|>{\centering $}p{3cm}<{$}|>{\centering $}p{1cm}<{$}>{\centering $}p{2cm}<{$}>{\centering\arraybackslash $}p{3cm}<{$}}
& \text{MEC} & \text{Representitive} & \multicolumn{1}{l|}{}                    & \multicolumn{1}{c|}{\text{MEC}} & \text{Representative} \\ \hline
\multirow{2}{*}{$\mathfrak{C}_1$} & \mec_{5,3a} & \varepsilon_4\Cdot p_1\,\varepsilon_2\Cdot\varepsilon_3\,\varepsilon_1\Cdot\varepsilon_5 & \multicolumn{1}{c|}{\multirow{2}{*}{$\mathfrak{C}_3$}} & \multicolumn{1}{c|}{\mec_{5,3b}} & \varepsilon_1\Cdot p_4\,\varepsilon_2\Cdot\varepsilon_3\,\varepsilon_4\Cdot\varepsilon_5             \\ \cline{2-3} \cline{5-6} 
& \mec_{5,6a} & \varepsilon_4\Cdot p_2\,\varepsilon_2\Cdot\varepsilon_3\,\varepsilon_1\Cdot\varepsilon_5 & \multicolumn{1}{c|}{} & \multicolumn{1}{c|}{\mec_{5,6b}} &\varepsilon_1\Cdot p_2\,\varepsilon_2\Cdot\varepsilon_3\,\varepsilon_4\Cdot\varepsilon_5 \\ \hline
\multirow{3}{*}{$\mathfrak{C}_2$} & \mec_{5,6c} & \varepsilon_4\Cdot p_1\,\varepsilon_1\Cdot\varepsilon_2\, \varepsilon_3\Cdot\varepsilon_5  & \multicolumn{1}{c|}{\multirow{2}{*}{$\mathfrak{C}_4$}} & \multicolumn{1}{c|}{\mec_{5,3c}} & \varepsilon_5\Cdot p_2\,\varepsilon_1\Cdot\varepsilon_2\,\varepsilon_3\Cdot\varepsilon_4  \\ \cline{2-3} \cline{5-6} 
& \mec_{5,6d} & \varepsilon_4\Cdot p_2\,\varepsilon_1\Cdot\varepsilon_2\, \varepsilon_3\Cdot\varepsilon_5 & \multicolumn{1}{l|}{}                    & \multicolumn{1}{c|}{\mec_{5,6f}} & \varepsilon_5\Cdot p_3\,\varepsilon_1\Cdot\varepsilon_2\,\varepsilon_3\Cdot\varepsilon_4  \\ \cline{2-6} 
& \mec_{5,6e} & \varepsilon_4\Cdot p_3\,\varepsilon_1\Cdot\varepsilon_2\, \varepsilon_3\Cdot\varepsilon_5 &   &   &                \\ \cline{1-3}
\end{array}
\end{align}
Note that both $\mec_{5,3a}$ and $\mec_{5,6a}$ have already appeared in eq.~\eqref{eq:5pMEC}.

\paragraph{Six points:} 
The $\mathsf{m}^{(\varepsilon\varepsilon)}$ classes and invariant groups are given by
\begin{align}\label{eq:mee6}
\renewcommand{\arraystretch}{1.1}
\begin{array}{>{\centering $}p{1cm}<{$}|>{\centering $}p{2.6cm}<{$}|>{\centering $}p{1.95cm}<{$}|>{\centering $}p{1.95cm}<{$}|>{\centering $}p{1.95cm}<{$}|>{\centering\arraybackslash$}p{3.65cm}<{$}}
& \mathfrak{C}_1 & \mathfrak{C}_2 & \mathfrak{C}_3 & \mathfrak{C}_4 & \mathfrak{C}_5 \\ \hline
\mathsf{m}^{(\varepsilon\varepsilon)} & \varepsilon_4\Cdot\varepsilon_5\,\varepsilon_1\Cdot\varepsilon_6 & \varepsilon_1\Cdot\varepsilon_4\,\varepsilon_5\Cdot\varepsilon_6 & \varepsilon_3\Cdot\varepsilon_4\,\varepsilon_5\Cdot\varepsilon_6 & \varepsilon_1\Cdot\varepsilon_3\,\varepsilon_4\Cdot\varepsilon_5 & \varepsilon_2\Cdot\varepsilon_3\,\varepsilon_4\Cdot\varepsilon_5 \\ \hline
G^{\text{inv}} & \begin{array}{c}
\{\mathbb{I},\mathbb{P}_{(23)},\mathbb{P}_{(45)},\\
\mathbb{P}_{(23)(45)}\}
\end{array} & \{\mathbb{I},\mathbb{P}_{(23)}\} & \{\mathbb{I},\mathbb{P}_{(34)}\} & \{\mathbb{I},\mathbb{P}_{(45)}\} & \langle\mathbb{P}_{(23)},\mathbb{P}_{(45)},\mathbb{P}_{(24)(35)}\rangle\\ 
\end{array} 
\end{align}
The $G^{\text{inv}}$ for $\mathfrak{C}_5$ is $S_2\times S_2\times S_2$ and here we only give the generators of this group. For each class the space of $\mathsf{m}^{(\varepsilon p)}$ is 16 dimensional. The number of MECs contained in each $\mathsf{m}^{(\varepsilon\varepsilon)}$ class $\mathfrak{C}_i$ is
\begin{align}
& \degree_{\mathfrak{C}_1}=7\,,& &\degree_{\mathfrak{C}_2}=10\,,& & \degree_{\mathfrak{C}_3}=10\,, & & \degree_{\mathfrak{C}_4}=10\,,& & \degree_{\mathfrak{C}_5}=3\,.
\end{align}
There are in all 40 tensor-free MECs. Here we sketch the calculation of $\degree_{\mathfrak{C}_1}$. The group under consideration is $S_2\times S_2$, which contains four permutations. According to eq.~\eqref{eq:dCi}, we need to find the traces of these four elements. An equivalent calculation is to find the number of $\mathsf{m}^{(\varepsilon p)}$ monomials that are invariant under the action. For the identity $\mathbb{I}$ the number is clearly 16, the dimension of the representation. There are four invariant monomials for each of the other three permutations,
\begin{align}
& \{\varepsilon_2\Cdot p_1\,\varepsilon_3\Cdot p_1\,,\varepsilon_2\Cdot p_4\,\varepsilon_3\Cdot p_4\,,\varepsilon_2\Cdot p_5\,\varepsilon_3\Cdot p_5\,,\varepsilon_2\Cdot p_3\,\varepsilon_3\Cdot p_2\}\text{ are invariant under }\mathbb{P}_{(23)}\,, \nonumber\\
& \{\varepsilon_2\Cdot p_1\,\varepsilon_3\Cdot p_1\,,\varepsilon_2\Cdot p_1\,\varepsilon_3\Cdot p_2\,,\varepsilon_2\Cdot p_3\,\varepsilon_3\Cdot p_1\,,\varepsilon_2\Cdot p_3\,\varepsilon_3\Cdot p_2\}\text{ are invariant under }\mathbb{P}_{(45)}\,,\\
& \{\varepsilon_2\Cdot p_1\,\varepsilon_3\Cdot p_1\,,\varepsilon_2\Cdot p_3\,\varepsilon_3\Cdot p_2\,,\varepsilon_2\Cdot p_4\,\varepsilon_3\Cdot p_5\,,\varepsilon_2\Cdot p_5\,\varepsilon_3\Cdot p_4\}\text{ are invariant under }\mathbb{P}_{(23)(45)}\,.\nonumber
\end{align}
Therefore, eq.~\eqref{eq:dCi} gives
\begin{align}
\degree_{\mathfrak{C}_1}=\frac{1}{4}\Big(\text{Tr}[\mathbb{I}]+\text{Tr}[\mathbb{P}_{(23)}]+\text{Tr}[\mathbb{P}_{(45)}]+\text{Tr}[\mathbb{P}_{(23)(45)}]\Big)=\frac{1}{4}(16+4+4+4)=7\,.
\end{align}

\subsection{Example: NMHV pure gauge terms at five points}\label{sec:NMHVex}
Eq.~\eqref{eq:rankR} and~\eqref{eq:dCi} together complete the counting of NMHV (tensor-free) pure gauge freedom. Here we collect the results and present the full NMHV pure gauge freedom at five points.

The relevant MECs are listed in eq.~\eqref{eq:5pMECs}. Then according to eq.~\eqref{eq:pg5} and~\eqref{eq:IRsigma}, the pure gauge numerators can be generated by $\neco(2,3,4)$ and $\neco(3,4)$,
\begin{subequations}
\begin{align}\label{eq:N234}
& \num^{\text{gauge}}_{[2,3,4]}(1,2,3,4,5)=p_2\Cdot p_1\,\neco(2,3,4)\circ\mathsf{m}\,,\\
\label{eq:N34}
& \num^{\text{gauge}}_{[3,4]}(1,2,3,4,5)=\big(p_3\Cdot p_{12}\,\mathbb{I}+p_2\Cdot p_1\,\mathbb{P}_{(23)}+p_2\Cdot p_1\,\mathbb{P}_{(243)}\big)\circ\neco(3,4)\circ\mathsf{m}\,.
\end{align}
\end{subequations}
The ranks of these two operators are computed from eq.~\eqref{eq:rankR}. We find that 
\begin{align}
&\text{Rank}_{\mec_{5,3}}[\neco(2,3,4)]=1\,,& &\text{Rank}_{\mec_{5,3}}[\neco(3,4)]=1\,, \nonumber\\
&\text{Rank}_{\mec_{5,6}}[\neco(2,3,4)]=2\,,& & \text{Rank}_{\mec_{5,6}}[\neco(3,4)]=3\,.
\end{align}
Namely, at five points the rank of $\neco$ only depends on the dimension of the MEC. 
There are 15 degrees of freedom associated with $\neco(2,3,4)$, which can be obtained by acting eq.~\eqref{eq:N234} onto the following monomials,
\begin{align}\label{eq:R234dof}
\begin{array}{|>{\centering $}p{1cm}<{$}|>{\centering $}p{3cm}<{$}|>{\centering $}p{1cm}<{$}|>{\centering $}p{3cm}<{$}|>{\centering $}p{1cm}<{$}|>{\centering\arraybackslash $}p{3cm}<{$}|}
\hline
\mec_{5,3a} & \varepsilon_4\Cdot p_1\,\varepsilon_2\Cdot\varepsilon_3\,\varepsilon_1\Cdot\varepsilon_5  & \mec_{5,3b} & \varepsilon_1\Cdot p_4\,\varepsilon_2\Cdot\varepsilon_3\,\varepsilon_4\Cdot\varepsilon_5  & \mec_{5,3c} & \varepsilon_5\Cdot p_3\,\varepsilon_1\Cdot\varepsilon_3\,\varepsilon_2\Cdot\varepsilon_4  \\ \hline
\multirow{2}{*}{$\mec_{5,6a}$} & \varepsilon_4\Cdot p_2\,\varepsilon_1\Cdot\varepsilon_5\,\varepsilon_2\Cdot\varepsilon_3  & \multirow{2}{*}{$\mec_{5,6b}$} & \varepsilon_1\Cdot p_2\,\varepsilon_2\Cdot\varepsilon_3\,\varepsilon_4\Cdot\varepsilon_5  & \multirow{2}{*}{$\mec_{5,6c}$} & \varepsilon_4\Cdot p_1\,\varepsilon_1\Cdot\varepsilon_2\,\varepsilon_3\Cdot\varepsilon_5  \\ \cline{2-2} \cline{4-4} \cline{6-6} 
& \varepsilon_3\Cdot p_2\,\varepsilon_1\Cdot\varepsilon_5\,\varepsilon_2\Cdot\varepsilon_4  &  & \varepsilon_1\Cdot p_2\,\varepsilon_2\Cdot\varepsilon_4\,\varepsilon_3\Cdot\varepsilon_5  & & \varepsilon_3\Cdot p_1\,\varepsilon_1\Cdot\varepsilon_2\,\varepsilon_4\Cdot\varepsilon_5  \\ \hline
\multirow{2}{*}{$\mec_{5,6d}$} & \varepsilon_4\Cdot p_2\,\varepsilon_1\Cdot\varepsilon_2\,\varepsilon_3\Cdot\varepsilon_5 & \multirow{2}{*}{$\mec_{5,6e}$} & \varepsilon_4\Cdot p_3\,\varepsilon_1\Cdot\varepsilon_2\,\varepsilon_3\Cdot\varepsilon_5 & \multirow{2}{*}{$\mec_{5,6f}$} & \varepsilon_5\Cdot p_3\,\varepsilon_1\Cdot\varepsilon_2\,\varepsilon_3\Cdot\varepsilon_4 \\ \cline{2-2} \cline{4-4} \cline{6-6} 
& \varepsilon_3\Cdot p_2\,\varepsilon_1\Cdot\varepsilon_2\,\varepsilon_4\Cdot\varepsilon_5 & & \varepsilon_3\Cdot p_4\,\varepsilon_1\Cdot\varepsilon_2\,\varepsilon_4\Cdot\varepsilon_5 & & \varepsilon_5\Cdot p_4\,\varepsilon_1\Cdot\varepsilon_2\,\varepsilon_3\Cdot\varepsilon_4 \\ \hline
\end{array}\;.
\end{align}
There are 21 degrees of freedom associated with $\neco(3,4)$, which can be obtained by acting eq.~\eqref{eq:N34} onto the following monomials,
\begin{align}\label{eq:R34dof}
\begin{array}{|>{\centering $}p{1cm}<{$}|>{\centering $}p{3cm}<{$}|>{\centering $}p{1cm}<{$}|>{\centering $}p{3cm}<{$}|>{\centering $}p{1cm}<{$}|>{\centering\arraybackslash $}p{3cm}<{$}|}
\hline
\mec_{5,3a} & \varepsilon_4\Cdot p_1\,\varepsilon_2\Cdot\varepsilon_3\,\varepsilon_1\Cdot\varepsilon_5  & \mec_{5,3b} & \varepsilon_1\Cdot p_4\,\varepsilon_2\Cdot\varepsilon_3\,\varepsilon_4\Cdot\varepsilon_5  & \mec_{5,3c} & \varepsilon_5\Cdot p_3\,\varepsilon_1\Cdot\varepsilon_3\,\varepsilon_2\Cdot\varepsilon_4  \\ \hline
\multirow{3}{*}{$\mec_{5,6a}$} & \varepsilon_2\Cdot p_4\,\varepsilon_1\Cdot\varepsilon_5\,\varepsilon_3\Cdot\varepsilon_4  & \multirow{3}{*}{$\mec_{5,6b}$} & \varepsilon_1\Cdot p_2\,\varepsilon_2\Cdot\varepsilon_3\,\varepsilon_4\Cdot\varepsilon_5  & \multirow{3}{*}{$\mec_{5,6c}$} & \varepsilon_4\Cdot p_1\,\varepsilon_1\Cdot\varepsilon_3\,\varepsilon_2\Cdot\varepsilon_5  \\ \cline{2-2} \cline{4-4} \cline{6-6} 
& \varepsilon_3\Cdot p_2\,\varepsilon_1\Cdot\varepsilon_5\,\varepsilon_2\Cdot\varepsilon_4  &  & \varepsilon_1\Cdot p_3\,\varepsilon_2\Cdot\varepsilon_3\,\varepsilon_4\Cdot\varepsilon_5  & & \varepsilon_3\Cdot p_1\,\varepsilon_1\Cdot\varepsilon_2\,\varepsilon_4\Cdot\varepsilon_5  \\ \cline{2-2} \cline{4-4} \cline{6-6} 
& \varepsilon_3\Cdot p_4\,\varepsilon_1\Cdot\varepsilon_5\,\varepsilon_2\Cdot\varepsilon_4  &  & \varepsilon_1\Cdot p_4\,\varepsilon_3\Cdot\varepsilon_4\,\varepsilon_2\Cdot\varepsilon_5  & & \varepsilon_2\Cdot p_1\,\varepsilon_1\Cdot\varepsilon_3\,\varepsilon_4\Cdot\varepsilon_5  \\ \hline
\multirow{3}{*}{$\mec_{5,6d}$} & \varepsilon_2\Cdot p_3\,\varepsilon_1\Cdot\varepsilon_3\,\varepsilon_4\Cdot\varepsilon_5 & \multirow{3}{*}{$\mec_{5,6e}$} & \varepsilon_2\Cdot p_4\,\varepsilon_1\Cdot\varepsilon_3\,\varepsilon_4\Cdot\varepsilon_5 & \multirow{3}{*}{$\mec_{5,6f}$} & \varepsilon_5\Cdot p_2\,\varepsilon_1\Cdot\varepsilon_3\,\varepsilon_2\Cdot\varepsilon_4 \\ \cline{2-2} \cline{4-4} \cline{6-6} 
& \varepsilon_3\Cdot p_2\,\varepsilon_1\Cdot\varepsilon_2\,\varepsilon_4\Cdot\varepsilon_5 & & \varepsilon_3\Cdot p_4\,\varepsilon_1\Cdot\varepsilon_2\,\varepsilon_4\Cdot\varepsilon_5 & & \varepsilon_5\Cdot p_4\,\varepsilon_1\Cdot\varepsilon_2\,\varepsilon_3\Cdot\varepsilon_4 \\ \cline{2-2} \cline{4-4} \cline{6-6} 
& \varepsilon_3\Cdot p_4\,\varepsilon_1\Cdot\varepsilon_4\,\varepsilon_2\Cdot\varepsilon_5 & & \varepsilon_3\Cdot p_2\,\varepsilon_1\Cdot\varepsilon_4\,\varepsilon_2\Cdot\varepsilon_5 & & \varepsilon_5\Cdot p_4\,\varepsilon_1\Cdot\varepsilon_3\,\varepsilon_2\Cdot\varepsilon_4 \\ \hline
\end{array}\;.
\end{align}
Adding up the numbers, we find that the pure gauge freedom in the NMHV sector is $15+21=36$.

\section{Pure gauge for \texorpdfstring{$S_{n-1}$}{Sn-1} and  \texorpdfstring{$S_{n}$}{Sn} crossing symmetric numerators}\label{sec:Sn-1puregauge}
Assuming that we have obtained pure gauge BCJ numerators with $S_{n-2}$ crossing symmetry, can we enhance their properties? Indeed, we want to further constrain the numerators so that they can be interpreted as originating from a standard quantum-field-theory framework that manifestly preserves crossing symmetry. For example, a Berends-Giele recursion that solves the field equations perturbatively will have manifest $S_{n-1}$ crossing symmetry. And a calculation using Feynman rules from a Lagrangian will have manifest $S_{n}$ crossing symmetry. We will start by imposing $S_{n-1}$ crossing symmetry.

First we discuss what symmetries are missing for numerators $\num(1,\sigma_2,\ldots,\sigma_{n-1},n)$ that only exhibit $S_{n-2}$ crossing symmetry, then we will find a simple way to implement the constraints. Any permutation of legs $2,\ldots,n-1$ is a simple relabeling of the $S_{n-2}$ symmetric numerators. However, if we permute  1 with any of the labels $2,\ldots,n-1$, it is not clear that the resulting expression can be expanded in the $(n-2)!$ DDM basis of numerators. Hence, once should impose this using the following $n-2$ constraint equations:
\begin{align} \label{crossing_n-1}
\num(2,1,\ldots,n-1,n) &= -\num(1,2,\ldots,n-1,n)\,, \nonumber \\
\num(2,3, 1\ldots,n-1,n) &= -\num(1,[2,3],\ldots,n-1,n)\,, \\
\num(2,3, 4, 1\ldots,n-1,n) &= -\num(1,[[2,3],4]\ldots,n-1,n)\,, \nonumber \\
& ~ \vdots  \nonumber \\
\num(2,3, 4, 5,\ldots,n-1,1,n) &= -\num(1,[\ldots[[[2,3],4],5],\ldots],n-1],n)\,. \nonumber
\end{align}
The solution to these equations will give numerators that manifest $S_{n-1}$ crossing symmetry. 

It turns out that there is a simple way of solving the above crossing equations, which recycles the same group algebra ideas used for the $S_{n-2}$ symmetric numerators. Imposing the $S_{n-1}$ crossing symmetry is equivalent to requiring the numerators be invariant under projection by $\neco$,
\begin{align} \label{eigenvalue_eq}
\frac{1}{n-1}\neco(1,2,\ldots, n-1)\circ \num(1,2,\ldots,n-1,n)= \num(1,2,\ldots,n-1,n)\,.
\end{align}
According to eq.~\eqref{eq:I-R}, the solution to  \eqn{eigenvalue_eq} is
\begin{align} \label{num_from_pre-num}
\num(1,2,\ldots,n-1,n)=\neco(1,2,\ldots,n-1)\circ \npre(1,2,\ldots,n-1,n)\,.
\end{align}
Therefore, the pure gauge terms with $S_{n-1}$ crossing symmetry can be constructed from those of the $S_{n-2}$ symmetric pre-numerators by applying the nested commutator $\neco(1,2,3,\ldots, n-1)$. At this point we note that since we are talking about pure gauge terms, which have no physical content, the distinction between an $S_{n-2}$ symmetric numerator or pre-numerator is totally insignificant. Indeed, one can identify the pure-gauge pre-numerator with an $S_{n-2}$ symmetric pure gauge numerator
\be
\npre^{\rm gauge}(1,2,\ldots,n-1,n) =  \frac{1}{n-1} \num^{\rm gauge}_{S_{n-2}}(1,2,\ldots,n-1,n)\,,
\ee
where the normalization factor is unimportant since an overall scale does not matter, but it is there to maintain consistency with previous equations. Nevertheless, let us continue the discussion using the pre-numerator, with the mental note that we can make the identification $\npre^{\rm gauge} \sim \num^{\rm gauge}_{S_{n-2}}$ when discussing pure gauge terms.

Recycling the result in eq.~\eqref{eq:PureGauge} we can write the most generic pure gauge pre-numerator in the NMHV sector,
\begin{align}\label{eq:npregauge}
  \npre^{\text{gauge}}(1,2,\ldots,n-1,n)&=\sum_{i=2}^{n-2}\sum_{\substack{\sigma^{-1}\in\{2,\ldots,i-1\} \\ \shuffle\{i,\ldots,n-1\}}}p_{\sigma_i}\Cdot X_{\sigma_i}(\mathbb{I})\mathcal{I}_{\neco(i,\ldots,n-1)}^{(\text{pre})}(1,\sigma,n)\,.
\end{align}
As before, we will only consider numerators where all external legs are physical vectors. We can use the same method given in section~\ref{sec:PureGaugeNMHV} to construct the invariant function $\mathcal{I}^{\text{pre}}_{\neco}$. Since now the desired crossing symmetry is $S_{n-1}$, we should include $\varepsilon_n\Cdot p_1$ as an additional independent variable for kinematic monomials. 

When projecting pre-numerators into numerators, independent building blocks for pre-numerators may live in the null space of $\neco(1,2,\ldots,n{-}1)$, which will lead to an overcounting of the degrees of freedom. We observe that out of the MECs constructed in section~\ref{sec:countMEC}, only those involving $\varepsilon_1\Cdot\varepsilon_n$ and $\varepsilon_n\Cdot p_1$ are independent. This can be understood by noticing that other $\varepsilon_i\Cdot\varepsilon_n$ and $\varepsilon_n\Cdot p_i$ can be generated by the nested commutator $\neco(1,2,\ldots,n{-}1)$. When on-shell identities are imposed, there are additional linear combinations involving $\mathcal{I}^{\text{pre}}_{\neco(i,\ldots,n-1)}$ with different $i$'s that live in the null space of $\neco(1,2,\ldots,n{-}1)$. Before giving some examples, we note that the above statements are supported by explicit computations up to seven points.

\paragraph{Four points:} The only MEC that satisfy the additional requirement that $\varepsilon_n$ only appears in $\varepsilon_1\Cdot\varepsilon_n$ and $\varepsilon_n\Cdot p_1$ is $\big[\varepsilon_1\Cdot\varepsilon_4\,\varepsilon _2\Cdot \varepsilon _3\big]$. This term lives in the null space of $\neco(2,3)$ as stated in section~\ref{sec:countO}. Thus there are no $S_{n-1}$ crossing symmetric pure gauge degrees of freedom at four points. 

\paragraph{Five points:} The pure gauge pre-numerators are given by 
\begin{align}
\npre^{\text{gauge}}(1,2,3,4,5)=\npre^{\text{gauge}}_{[2,3,4]}(1,2,3,4,5)+\npre^{\text{gauge}}_{[3,4]}(1,2,3,4,5)\,,
\end{align} 
where
\begin{subequations}\label{eq:npre5}
\begin{align}
	& \npre^{\text{gauge}}_{[2,3,4]}(1,2,3,4,5)=p_2\Cdot p_1\,\neco(2,3,4)\circ\mathsf{m}\,,\\
	& \npre^{\text{gauge}}_{[3,4]}(1,2,3,4,5)=\big[p_3\Cdot (p_{1}+p_2)\,\mathbb{I}+p_2\Cdot p_1\,\mathbb{P}_{(23)}+p_2\Cdot p_1\,\mathbb{P}_{(243)}\big]\circ\neco(3,4)\circ\mathsf{m}\,.
\end{align}
\end{subequations}
To find the independent monomials, we start with the $\mathsf{m}^{(\varepsilon\varepsilon)}$ classes as in section~\ref{sec:countMEC}. Out of the four classes shown in eq.~\eqref{eq:eeclass}, the new requirement that $\varepsilon_5$ can only be contracted with $\varepsilon_1$ or $p_1$ leaves us with only $\mathfrak{C}_1$ and $\mathfrak{C}_4$. The $\mathfrak{C}_1$ class contains two MECs $\mec_{5,3a}$ and $\mec_{5,6a}$ as before. They together contribute $7$ degrees of freedom as given in eq.~\eqref{eq:R234dof} and~\eqref{eq:R34dof}. We repeat the independent monomials associated with $\neco(2,3,4)$ and $\neco(3,4)$ below for convenience,
\begin{align}\label{eq:m5}
\begin{array}{>{\centering $}p{1cm}<{$}|>{\centering $}p{3cm}<{$}|>{\centering\arraybackslash $}p{3cm}<{$}}
 & \neco(2,3,4) & \neco(3,4) \\ \hline
\mec_{5,3a} & \varepsilon_4\Cdot p_1\,\varepsilon_2\Cdot\varepsilon_3\,\varepsilon_1\Cdot\varepsilon_5 & \varepsilon_4\Cdot p_1\,\varepsilon_2\Cdot\varepsilon_3\,\varepsilon_1\Cdot\varepsilon_5 \\ \hline
\mec_{5,6a} & \begin{array}{c} \varepsilon_4\Cdot p_2\,\varepsilon_1\Cdot\varepsilon_5\,\varepsilon_2\Cdot\varepsilon_3 \\
\varepsilon_3\Cdot p_2\,\varepsilon_1\Cdot\varepsilon_5\,\varepsilon_2\Cdot\varepsilon_4 \end{array} & \begin{array}{c} \varepsilon_2\Cdot p_4\,\varepsilon_1\Cdot\varepsilon_5\,\varepsilon_3\Cdot\varepsilon_4 \\
\varepsilon_3\Cdot p_2\,\varepsilon_1\Cdot\varepsilon_5\,\varepsilon_2\Cdot\varepsilon_4 \\ \varepsilon_3\Cdot p_4\,\varepsilon_1\Cdot\varepsilon_5\,\varepsilon_2\Cdot\varepsilon_4 \end{array} \\ 
\end{array}\,.
\end{align}
Now the $\mathfrak{C}_4$ class only contains one MEC since $\varepsilon_5$ can only be contracted with $p_1$. The MEC is three dimensional,
\begin{align}
\begin{bmatrix}
\varepsilon_1\Cdot\varepsilon_2\,\varepsilon_3\Cdot\varepsilon_4\,\varepsilon_5\Cdot p_1 \\ \varepsilon_1\Cdot\varepsilon_3\,\varepsilon_2\Cdot\varepsilon_4\,\varepsilon_5\Cdot p_1 \\ \varepsilon_1\Cdot\varepsilon_4\,\varepsilon_2\Cdot\varepsilon_3\,\varepsilon_5\Cdot p_1
\end{bmatrix}\,,
\end{align}
which contributes one independent monomial $\mathsf{m}^{\mathfrak{C}_4}=\varepsilon_1\Cdot\varepsilon_3\,\varepsilon_2\Cdot\varepsilon_4\,\varepsilon_5\Cdot p_1$ for both $\neco(2,3,4)$ and $\neco(3,4)$ if we follow section~\ref{sec:countO}. We plug this $\mathsf{m}^{\mathfrak{C}_4}$ into eq.~\eqref{eq:npre5} and get two pure gauge pre-numerators $\mathcal{N}^{(\mathfrak{C}_4)}_{[2,3,4]}$ and $\mathcal{N}^{(\mathfrak{C}_4)}_{[3,4]}$. When on-shell identities are imposed, we can show that the sum $\mathcal{N}^{(\mathfrak{C}_4)}_{[2,3,4]}+\mathcal{N}^{(\mathfrak{C}_4)}_{[3,4]}$ is in the null space of $\neco(1,2,3,4)$,
\begin{align}
\neco(1,2,3,4)\circ\Big[\mathcal{N}^{(\mathfrak{C}_4)}_{[2,3,4]}(1,2,3,4,5)+\mathcal{N}^{(\mathfrak{C}_4)}_{[3,4]}(1,2,3,4,5)\Big]\cong 0\,.
\end{align}
In particular, the relevant on-shell identities are
\begin{align}
\varepsilon_5\Cdot(p_1+p_2+p_3+p_4)=0\,,& & p_3\Cdot p_4=-p_1\Cdot p_2-p_1\Cdot p_3-p_1\Cdot p_4-p_2\Cdot p_3-p_2\Cdot p_4\,.
\end{align}
Therefore, the independent degrees of freedom is $7{+}1{=}8$, and the independent monomials are given by eq.~\eqref{eq:m5} and $\mathsf{m}^{\mathfrak{C}_4}$.
                                          
\paragraph{Six points:} We again start with the pure gauge degrees of freedom in eq.~\eqref{eq:npregauge}. The allowed $\mathsf{m}^{(\varepsilon\varepsilon)}$ classes at six points are $\mathfrak{C}_1$, $\mathfrak{C}_4$ and $\mathfrak{C}_5$ in eq.~\eqref{eq:mee6}. In all they contain $11$ independent MECs,
\begin{align}
\begin{array}{|>{\centering $}p{1cm}<{$}|>{\centering $}p{1.5cm}<{$}|>{\centering $}p{4cm}<{$}|>{\centering $}p{1cm}<{$}>{\centering $}p{1.5cm}<{$}>{\centering\arraybackslash $}p{4cm}<{$}}
\hline
\text{Class}  & \Dc & \text{Representative} & \multicolumn{1}{c|}{$\text{Class}$}  & \multicolumn{1}{c|}{\Dc} & \multicolumn{1}{c|}{$\text{Representative}$} \\ \hline
\multirow{7}{*}{$\mathfrak{C}_1$} & 6 & \varepsilon_1\Cdot\varepsilon_6\,\varepsilon_4\Cdot\varepsilon_5\,\varepsilon_2\Cdot p_1\,\varepsilon_3\Cdot p_1 & \multicolumn{1}{c|}{\multirow{3}{*}{$\mathfrak{C}_4$}} & \multicolumn{1}{c|}{12}  & \multicolumn{1}{c|}{\varepsilon_1\Cdot\varepsilon_5\,\varepsilon_3\Cdot\varepsilon_4\,\varepsilon_2\Cdot p_1\,\varepsilon_6\Cdot p_1} \\ \cline{2-3} \cline{5-6} 
& 6 & \varepsilon_1\Cdot\varepsilon_6\,\varepsilon_4\Cdot\varepsilon_5\,\varepsilon_2\Cdot p_3\,\varepsilon_3\Cdot p_2 & \multicolumn{1}{l|}{} & \multicolumn{1}{c|}{12} & \multicolumn{1}{c|}{\varepsilon_1\Cdot\varepsilon_5\,\varepsilon_3\Cdot\varepsilon_4\,\varepsilon_2\Cdot p_5\,\varepsilon_6\Cdot p_1} \\ \cline{2-3} \cline{5-6} 
& 12 & \varepsilon_1\Cdot\varepsilon_6\,\varepsilon_4\Cdot\varepsilon_5\,\varepsilon_2\Cdot p_1\,\varepsilon_3\Cdot p_2 & \multicolumn{1}{l|}{} & \multicolumn{1}{c|}{24} & \multicolumn{1}{c|}{\varepsilon_1\Cdot\varepsilon_5\,\varepsilon_3\Cdot\varepsilon_4\,\varepsilon_2\Cdot p_3\,\varepsilon_6\Cdot p_1}  \\ \cline{2-6} 
& 24 & \varepsilon_1\Cdot\varepsilon_6\,\varepsilon_4\Cdot\varepsilon_5\,\varepsilon_2\Cdot p_4\,\varepsilon_3\Cdot p_1 & \multicolumn{1}{c|}{\mathfrak{C}_5} & \multicolumn{1}{c|}{12} & \multicolumn{1}{c|}{\varepsilon_2\Cdot\varepsilon_3\,\varepsilon_4\Cdot\varepsilon_5\,\varepsilon_1\Cdot p_2\,\varepsilon_6\Cdot p_1}  \\ \cline{2-6} 
& 24 & \varepsilon_1\Cdot\varepsilon_6\,\varepsilon_4\Cdot\varepsilon_5\,\varepsilon_2\Cdot p_4\,\varepsilon_3\Cdot p_2 &  &  &  \\ \cline{2-3}
& 12 & \varepsilon_1\Cdot\varepsilon_6\,\varepsilon_4\Cdot\varepsilon_5\,\varepsilon_2\Cdot p_5\,\varepsilon_3\Cdot p_4 &  &  & \\ \cline{2-3}
& 12 & \varepsilon_1\Cdot\varepsilon_6\,\varepsilon_4\Cdot\varepsilon_5\,\varepsilon_2\Cdot p_4\,\varepsilon_3\Cdot p_4 & & &\\ \cline{1-3}
\end{array}
\end{align}
To get the total number of degrees of freedom, we first sum over the ranks of $\neco(2,3,4,5)$, $\neco(3,4,5)$ and $\neco(4,5)$ in these MECs,
\begin{align}
\begin{array}{>{\centering $}p{1cm}<{$}>{\centering $}p{2cm}<{$}|>{\centering $}p{0.75cm}<{$}|>{\centering $}p{0.75cm}<{$}|>{\centering $}p{0.75cm}<{$}|>{\centering $}p{0.75cm}<{$}|>{\centering $}p{0.75cm}<{$}|>{\centering $}p{0.75cm}<{$}|>{\centering $}p{0.75cm}<{$}|>{\centering $}p{0.75cm}<{$}|>{\centering $}p{0.75cm}<{$}|>{\centering $}p{0.75cm}<{$}|>{\centering\arraybackslash $}p{0.75cm}<{$}|}
\cline{3-13}
& & \multicolumn{7}{c|}{\mathfrak{C}_1} & \multicolumn{3}{c|}{\mathfrak{C}_4} & \mathfrak{C}_5 \\ \hline
\multicolumn{2}{|c|}{\Dc} & 6 & 6 & 12 & 24 & 24 & 12 & 12 & 12 & 12 & 24 & 12 \\ \hline
\multicolumn{1}{|c|}{\multirow{3}{*}{Rank}} & \neco(2,3,4,5) & 1 & 1 & 3 & 6 & 6  & 2  & 3  & 3 & 3 & 6 & 3  \\ \cline{2-13} 
\multicolumn{1}{|c|}{}                      & \neco(3,4,5) & 2 & 2 & 4 & 8 & 8  & 4  & 4  & 4 & 4 & 8 & 4 \\ \cline{2-13} 
\multicolumn{1}{|c|}{}                      & \neco(4,5) & 2 & 2 & 5 & 12 & 12 & 6 & 5 & 5 & 5  & 12 & 5  \\ \hline
\multicolumn{2}{|c|}{\text{gauge d.o.f}} & 5 & 5 & 12 & 26 & 26 & 12 & 12 & 12 & 12 & 26 & 12 \\ \hline
\end{array}
\end{align}
Summing over the last row, we get 160 independent degrees of freedom in eq.~\eqref{eq:npregauge}. When converting to numerators using $\neco(1,2,3,4,5)$, we find 12 of those originated from $\mathfrak{C}_4$ and $\mathfrak{C}_5$ live in the null space of $\neco(1,2,3,4,5)$ after imposing on-shell identities. Therefore, the final pure gauge degrees of freedom for $S_{n-1}$ crossing symmetric numerators at $n{=}6$ is 148.

The same calculational steps can be repeated for $n{=}7$, however we refrain from spelling out the details as they are more involved at this mutiplicity. The result of the calculation is that the 16583 gauge degrees of freedom for $S_{n-2}$ will get reduced down to 2734 in the $S_{n-1}$ crossing symmetric numerator. This concludes the discussion on $S_{n-1}$ symmetry. 

\paragraph{Full crossing symmetry:} In order to obtain $S_{n}$ crossing symmetry, we need to impose further equations that involve permutations of leg $n$ with any one of the labels $1,\ldots,n-1$. Naively, it appears that this requires imposing a total of $n-1$ equations. However, one can show that there is one equation that is both necessary and sufficient, it is the reversal symmetry of the numerator
\be \label{crossing_n}
N(1,2,3,\ldots, n-1,n) = (-1)^n N(n,n-1, \ldots, 3, 2, 1)\,.
\ee
It is clearly a necessary equation to impose as it is a standard symmetry of a crossing-symmetric Feynman diagram. But it is also a sufficient constraint since when the $S_{n-1}$ equations~(\ref{crossing_n-1}) are imposed it follows that crossing symmetries related to permuting $1 \leftrightarrow j\in \{2,\ldots,n-1\}$ hold, and the reversal symmetry promotes this to crossing symmetries $n \leftrightarrow j\in \{2,\ldots,n-1\}$. And the $n \leftrightarrow 1$ swap is part of the reversal symmetry itself.  Since imposing $S_{n}$ crossing symmetry only involves one relatively simple constraint, \eqn{crossing_n}, we solve this by direct calculation using the explicitly computed $S_{n-1}$ symmetric numerators. The results are given in the tables below.

Finally, we can collect the data of the $S_{n-2}$, $S_{n-1}$ and $S_{n}$ crossing symmetric pure gauge BCJ numerators in a table. By calculations up to multiplicity $n=7$, we find that the number of free parameters (degrees of freedom) in the NMHV sector of YM are the following:
\begin{align}
\begin{array}{|>{\centering $}p{3.5cm}<{$}|>{\centering $}p{1.65cm}<{$}|>{\centering $}p{1.65cm}<{$}|>{\centering $}p{1.65cm}<{$}|>{\centering\arraybackslash $}p{1.65cm}<{$}|}
\hline
\multirow{2}{*}{crossing symmetry}& \multicolumn{4}{c|}{\text{Total  d.o.f. $N^{ \rm gauge} \sim (\varepsilon {\cdot} \varepsilon)^2 (p {\cdot} p) \prod \varepsilon {\cdot} p$ }} \\ \cline{2-5} 
& n=4    & n=5    & n=6    & n=7      \\ \hline
S_{n-2} & 1      & 36     & 760    & 16583    \\ \hline
S_{n-1} & 0      & 8      & 148    &  2734  \\ \hline   
S_{n} & 0      & 1      & 25   & 381     \\ \hline
\end{array}
\end{align}
We remind the reader that by the NMHV sector of YM, we mean all terms in the $D$-dimensional BCJ numerator that contains exactly two powers of $\varepsilon_i {\cdot} \varepsilon_j$ or equivalently one power of $p_i  {\cdot}  p_j$. Such terms are necessary for computing NMHV amplitudes. We also remind the reader that for terms that have no $p_i  {\cdot}  p_j$ factors, or equivalently have exactly one $\varepsilon_i {\cdot} \varepsilon_j$ factor, the corresponding BCJ numerator is unique to any multiplicity, and exhibit full $S_n$ crossing symmetry. Such terms belong to the MHV sector of YM, see discussion around  \eqn{polarization_power} for further clarification. 

For completeness, we also considered similar calculations of pure gauge BCJ numerators of the N${}^2$MHV sector of YM. The methods used are less powerful because we do not yet have access to a complete basis of local BCJ amplitude relations quadratic in $s_{ij}$, analogous to the binary BCJ relations that form a basis of local relations linear in $s_{ij}$. Nevertheless, one can with some brute force computational effort sidestep this difficultly, and recycle the efficient MEC decomposition for the N${}^2$MHV sector. We will not give the details of this calculation as it is less streamlined, but the conclusion is that we extracted the number of free parameters in the pure gauge BCJ numerator up to seven points. The freedom starts at six points, where the N${}^2$MHV sector first appears, and we get the following table for different manifest crossing symmetries:
\begin{align}
\begin{array}{|>{\centering $}p{3.5cm}<{$}|>{\centering $}p{3.4cm}<{$}|>{\centering\arraybackslash $}p{3.4cm}<{$}|}
\hline
\multirow{2}{*}{crossing symmetry} & \multicolumn{2}{c|}{\text{Total d.o.f.  \!$N^{ \rm gauge}\! \sim\! (\varepsilon{\cdot}\varepsilon)^3 (p{\cdot}p)^2 \prod \varepsilon {\cdot} p$ }} \\ \cline{2-3} 
   & n=6   & n=7      \\ \hline
S_{n-2}   & 137    & 9024    \\ \hline
S_{n-1}      & 26   &  1494  \\ \hline   
S_{n}     & 5   & 314     \\ \hline
\end{array}
\end{align}
Recall, the N${}^2$MHV sector contains terms that are cubic in $\varepsilon_i {\cdot} \varepsilon_j$ or equivalently quadratic in $p_i  {\cdot}  p_j$.  It is interesting that we can now compare to a result in ref.~\cite{Bern:2010yg}, where interaction terms of a  self-BCJ Lagrangian were constructed up to six points, and 30 free parameters were observed. This number exactly matches $25+5$, which are the fully crossing-symmetric degrees of freedom found in the above tables for the $n=6$ NMHV and N${}^2$MHV sector, respectively.

\section{Conclusion and outlook}\label{sec:conclusion}

In this paper, we presented an algebraic framework that generates BCJ numerators of pure YM theory through the NMHV sector; that is, all the terms in the $D$-dimensional numerators that contain at most two powers of $\varepsilon_i\Cdot\varepsilon_j$. A useful steppingstone is the introduction of pre-numerators, from which one obtains the kinematic numerator of any cubic graph by acting with a nested commutator. The kinematic numerators constructed this way enjoy a manifest $S_{n-1}$ crossing symmetry, and automatically satisfy the color-kinematics duality. The non-trivial step in this construction is to work out the details of the the fusion products between vector and tensor currents, which are used for computing the NMHV sector pre-numerators. While there exists some freedom in the definition of the NMHV sector fusion products, our construction guides us towards fusion products that permits an all-multiplicity closed formula for the pre-numerators.

While we have obtained a formulation of the kinematic algebra that describes the NMHV sector of YM numerators, much work remains to be done to fully elucidate the kinematic Lie algebra underlying the color-kinematics duality. Better understanding of the generators of the algebra is warranted, and how they act on each other through fusion products. We use formal objects as generators, the vector and tensor currents, and the fusion products contain the non-trivial information of the kinematic algebra. This is quite analogous to a formal expansion of the operator product in a current algebra~\cite{GellMann:1964tf}, or to the formal expansion of the YM field equations through a Berends-Giele current \cite{Berends:1987me}. Indeed, the origins of our construction are motivated by such analogies, and thus any similarities may not be accidental. For example, in appendix~\ref{sec:feynman}, we use similar vector and tensor currents to formulate a complete set of cubic Feynman rules of pure YM. Such fusion rules are useful for explicit calculations, but Jacobi identities necessary for a kinematic Lie algebra are absent in this example.

The complete kinematic algebra for YM should be an infinite-dimensional version of a Lie algebra. Assuming its existence, it can be realized either by finding a full set of Lie brackets that obey Jacobi identities, or by finding a full set of generators that can be abstractly multiplied, similar to a matrix product, which give an enveloping algebra. The latter approach is what we attempted to emulate through the fusion product of tensor current generators, and the pre-numerator corresponds to an object in the enveloping algebra. Either of theses approaches would give BCJ numerators which enjoy manifest $S_{n-1}$ crossing symmetry, whereas to guarantee manifest $S_{n}$ crossing symmetry a Lagrangian approach using auxiliary fields is perhaps more optimal.

It is an open and difficult problem to realize the complete kinematic Lie algebra using the above mentioned approaches. In our framework, further progress requires working out more general fusion products for vectors and (higher-rank) tensors. For example, the fusion product $J_{\varepsilon_i}\star J_{\varepsilon_j\otimes\varepsilon_k\otimes\varepsilon_r}$ has not yet been defined, since all our fusion rules involved a vector current in the rightmost argument of the product.  Such an additional fusion product should in principle not alter the NMHV sector numerators, but for the consistency and closure of the algebra, we need to know every possible fusion product. For approaching the $\text{N}^{k\geqslant 1}\text{MHV}$ sectors, higher-rank tensor currents and associated fusion products need to be worked out. A problem one encounters in these sectors is that the corresponding gauge-invariant amplitudes are at increasingly high multiplicity, thus it is practically difficult to constrain the algebra using amplitude calculations.

Taking it one step further, it might be wise to realize the kinematic algebra at the Lagrangian level. The vector and tensor currents are then promoted to quantum fields and the fusion products may be mapped to interactions. 
This Lagrangian approach was first attempted in refs.~\cite{Bern:2010yg,Tolotti:2013caa}, but an NMHV sector Lagrangian would likely be similar to an extension of the Cheung-Shen Lagrangian for the MHV sector~\cite{Cheung:2016prv}. 
Another natural direction would be to explore if the NMHV sector tensor currents and fusion products can be represented by derivative operators acting in some function space, which would endow the kinematic algebra with an explicit representation. This is precisely what was achieved by Monteiro and O'Connell for self-dual~YM~\cite{Monteiro:2011pc}, and the simplicity of their generators provides some hope that also the vector and tensor currents of the NMHV sector might admit simple representations. The details of the fusion rules may be different in an explicit generator representation, but the general structure that we have found should likely be preserved. 

Of course, there are many other possible approaches for obtaining mathematical structures that one may rightly call a kinematic algebra for YM. There are promising work involving Berends-Giele currents in so-called ``BCJ gauge''~\cite{Lee:2015upy,Bridges:2019siz}, which potentially may have common ground with the current approach. Similar mathematical studies relying on recursive or infinite-algebraic approaches includes the refs.~\cite{Fu:2016plh,Lopez-Arcos:2019hvg,Reiterer:2019dys,Gomez:2020vat, Fu:2020frx}. Recently, there are several mathematically-advanced attempts to understand the color-kinematics duality and the double copy through BRST symmetry at the Lagrangian level~\cite{Borsten:2019prq, Borsten:2020zgj,Borsten:2020xbt,Borsten:2021hua,Borsten:2021zir}; see also recent work on diffeomorphisms~\cite{Frenkel:2020djn, Campiglia:2021srh}. Indeed, it has been known for  some time that gauge and diffeomorphism symmetries should be key ingredients of a kinematic algebra~\cite{Bern:2008qj,Bern:2010ue,Monteiro:2011pc,Chiodaroli:2016jqw,Chiodaroli:2017ngp}.

In the second part of this paper, we perform a thorough mathematical analysis of the generalized gauge freedom of the BCJ numerators. Interestingly, we find that the commutator structure of the kinematic algebra implies that powerful group-algebraic methods can be used to characterize the freedom. Conversely, we expect that a refined understanding of the pure gauge freedom should give new insights to the kinematic algebra formulation. In section~\ref{sec:PureGaugeNMHV}, we propose that the pure gauge degrees of freedom in the entire NMHV sector, with $S_{n-2}$ crossing symmetric DDM basis numerators, can be characterized by kinematic polynomials that are group-algebra invariants over the permutation group.  Specifically, they are invariants of the $\neco$ operators defined in eq.~\eqref{eq:Roperator}, which generate left nested commutators, closely resembling the half-ladder graph structure of the DDM basis. 

Our method, based on finite-group representation theory, is fully constructive and does not involve building any ansatz for the pure gauge NMHV numerators. We have explicitly checked, by comparing to brute-force calculations up to seven points, that our method give a complete parametrization of the local pure gauge freedom in the $S_{n-2}$ crossing symmetric numerators. As part of this work, we note that the previously introduced binary BCJ relations~\cite{Chen:2019ywi} have a direct connection to the $\neco$ operators. At general multiplicity, the completeness of our construction is equivalent to proving that the binary BCJ relations form a complete basis for local BCJ relations with polynomial coefficients that are linear in the Mandelstam variables.  A completeness proof of the binary BCJ relations is beyond the scope of this paper, but we have explicitly checked up to nine points that all known such BCJ relations are reproduced, and that the basis size correspond to the generalized Stirling numbers.

By direct generalization of the above construction to the NMHV pre-numerator, we obtain pure-gauge BCJ numerators with manifest $S_{n-1}$ crossing symmetry. The result gives a complete but redundant parametrization of the pure-gauge numerators, as the $\neco$ operators give a null space that makes certain parameter combinations redundant. However, the redundancy can be straightforwardly removed. We explicitly count the independent degrees of freedom up to seven points. As a final step, we construct the $S_n$ crossing symmetric pure gauge numerators, which we note are obtained by imposing a reversal symmetry of the half-ladder numerators with $S_{n-1}$ crossing symmetry. The freedom found here will directly correspond to freedom in the Lagrangian formulation of the kinematic algebra.

We also count the pure gauge terms in the N${}^2$MHV sector up to seven points. However,  the calculation is performed partially by brute force, as the more powerful methods introduced are currently incomplete beyond the NMHV sector. As mentioned, the completeness of the $\neco$-invariant polynomials relies on the completeness of the binary BCJ relations. We have checked that the binary BCJ relations alone do not form a complete basis for all the local BCJ relations with non-linear polynomials of Mandelstam variables. Thus, it is an interesting open problem to find such a complete set of local BCJ relations, needed for constructing the pure gauge terms in a general $\text{N}^{k\geqslant 2}\text{MHV}$ sector. We note that the $\neco$ operator, which acts as a nested commutator, is an interesting idempotent operator well known in the literature on free Lie algebras~\cite{reutenauer2003free}. In the literature related to color-kinematics duality, very similar mathematical structures were considered in refs.~\cite{Frost:2019fjn,Frost:2020eoa}. The central role of $\neco$ for constructing the NMHV pure gauge terms is an invitation to further explore these topics.

Finally, we note that it is desirable to generalize the kinematic algebra construction of this paper, by including fundamental matter coupled to YM. This is a natural setting where the pre-numerator can be promoted to a BCJ numerator for a fermion or scalar diagram. Considering the matter amplitudes directly should simplify certain steps, as such numerators are less sensitive to gauge ambiguities~\cite{Johansson:2015oia,Johansson:2014zca, Brandhuber:2021kpo}. This is also an interesting avenue as it directly connects with recent effort to simplify the construction of gravitational amplitudes, via the double copy or by other means, for describing black-hole scattering, and associated effective potential calculations and gravitational wave emission~\cite{Luna:2016due,Goldberger:2016iau,Luna:2017dtq,Shen:2018ebu,Plefka:2018dpa,Bern:2019nnu,Plefka:2019hmz,Bern:2019crd,Bern:2020buy,delaCruz:2020bbn,Bern:2021dqo,Almeida:2020mrg}.  Considering numerators for massive matter is of direct relevance for connections to heavy-mass effective theory~\cite{Bautista:2019evw, Johansson:2019dnu, Edison:2020ehu, Bjerrum-Bohr:2020syg,Plefka:2019wyg, Haddad:2020tvs}. In parallel work to this paper~\cite{Brandhuber:2021kpo}, a novel double-copy prescription for heavy-mass effective theory is obtained, where the BCJ numerators are constructed from generalizations of the tensor currents and fusion products. Interestingly, the BCJ numerators constructed this way are unique and gauge invariant, which suggests that they should be helpful for further detailed studies of the kinematic algebra.
 


\acknowledgments
We would like to thank Maor~Ben-Shahar, Andreas~Brandhuber, Chih-Hao~Fu, Yu-tin~Huang, Carlos~Mafra, Ricardo~Monteiro, Jan~Plefka, Gabriele~Travaglini, Congkao~Wen, Chris~White and Yang~Zhang for useful discussions on related subjects. We also thank Alexander~Edison and Gregor~K\"{a}lin for sharing Mathematica codes and assistance related to computational efficiency. The research is supported by the Knut and Alice Wallenberg Foundation under grants KAW 2013.0235, KAW 2018.0116, KAW 2018.0162, the Swedish Research Council under grant 621-2014-5722, and the Ragnar S\"{o}derberg Foundation (Swedish Foundations' Starting Grant). G.C. is also supported by the Science and Technology Facilities Council (STFC) Consolidated Grants ST/P000754/1 \textit{``String theory, gauge theory \& duality''} and  ST/T000686/1 \textit{``Amplitudes, strings  \& duality''}, and by the European Union's Horizon 2020 research and innovation programme under the Marie Sk\l{}odowska-Curie grant agreement No.~764850 {\it ``\href{https://sagex.org}{SAGEX}''}. T.W. is supported by the German Research Foundation through grant PL457/3-1. F.T. is also supported in part by the U.S. Department of Energy (DOE) under grant no.~DE-SC0013699. Computational resources (project SNIC 2019/3-645) were provided by the Swedish National Infrastructure for Computing (SNIC) at UPPMAX, partially funded by the Swedish Research Council through grant no. 2018-05973.

\appendix

 \section{Independent local BCJ relations }\label{sec:binaryBCJ}
In this appendix, we study linear algebraic relations among binary BCJ relations, and argue that they span the vector space of local BCJ relations that are linear in Mandelstam variables $s_{ij}$. 

A binary BCJ relation, $\mathcal{B}$, is characterized by the length of the nested commutators, which can range from $2$ to $n{-}2$. For a fixed length, we can always write down $(n{-}2)!$ such binary BCJ relations, labeled by $\beta\in S_{n-2}$, see eq.~\eqref{eq:Bbeta}. Thus the binary BCJ relations form the set
\begin{align}
\bigcup_{i=2}^{n-2}\Big\{\mathcal{B}_{[\beta_i,\ldots,\beta_{n-1}]}(1,\beta_2,\ldots,\beta_{n-1},n)\,\Big|\,\beta\in S_{n-2}\Big\}\,.
\end{align}
However, for a fixed $i$, these relations are not independent. There exist the following relations among the $\mathcal{B}$'s, with constant unit coefficients,
\begin{align}\label{eq:rel1}
\sum_{\sigma\in\{\beta_{n-1}\}\shuffle\{\beta_i,\ldots,\beta_{n-2}\}}\mathcal{B}_{[\sigma_{\beta_i},\ldots,\sigma_{\beta_{n-1}}]}(1,\beta_2,\ldots,\beta_{n-1},n)=0\,.
\end{align}
We have explicitly checked up to nine points that there are no other relations among the $\mathcal{B}$'s which involve linear combinations of them with constant coefficients. That is, as a vector space over the rational (constant) numbers, which is a subspace of the corresponding vector space over $s_{ij}$-polynomials, there are no further relations among the $\mathcal{B}$'s. While there should exist further relations involving $s_{ij}$-dependent coefficients, this does not change the counting of independent $\mathcal{B}$'s since in order to make use of them one needs to divide by $s_{ij}$ factors, and this operation is not allowed in the vector space over polynomials that we are considering. 

We can use \eqn{eq:rel1} to fix $\beta_{n-1}$ to be the largest label among $\{\beta_i,\ldots,\beta_{n-1}\}$. Thus the independent binary BCJ relations under eq.~\eqref{eq:rel1} are simply
\begin{align}\label{eq:binaryBasis}
\bigcup_{i=2}^{n-2}\Big\{\mathcal{B}_{[\beta_i,\ldots,\beta_{n-1}]}(1,\beta_2,\ldots,\beta_{n-1},n)\,\Big|\,\beta_{\ell}<\beta_{n-1}\text{ for }i\leqslant \ell\leqslant n-2\Big\}\,.
\end{align}
From this restriction we can directly read off the dimension of the independent basis, which we recognize to be the generalized Stirling numbers,\footnote{See \href{http://oeis.org/A001705}{http://oeis.org/A001705}.}
\begin{align}
\sum_{i=2}^{n-2}\frac{(n-2)!}{(n-i)!}\times(n-i-1)!=(n-2)!\sum_{i=2}^{n-2}\frac{1}{n-i}\,,
\end{align}
To be explicit, we list the dimension of the binary BCJ basis up to nine points:
\begin{align}\label{eq:basiscount}
n&& 4 &&5&&6&&7~&&8~~&&9~~&&\nn\\
\text{dimension} && 1&&5&& 26&& 154&& 1044&& 8028 && 
\end{align}

Of course, we cannot be satisfied by knowing that we have found a basis of binary BCJ relations. We want the basis for any kind of BCJ relation involving coefficients linear in $s_{ij}$ variables. For example, the fundamental BCJ relations are precisely of this type. If we generate the full set of $S_n$ permutations of a given fundamental BCJ relation, we have explicitly checked up to $n=9$ that they together span a vector space over rational numbers which has dimension:
\begin{align}\label{eq:fundamental}
n&& 4 &&5&&6&&7~&&8~&&9~~&&\nn\\
\text{dimension} && 1&&5&& 26&& 154&& 939&& 6453 &&
\end{align}
As is obvious from this count,  this number is smaller than the dimension of the binary BCJ relations starting at multiplicity $n=8$. Thus the fundamental BCJ relations do not span the vector space of all local BCJ relations (however, they still span the BCJ relation vector space over non-local $s_{ij}$-functions). 

We have checked up to $n=9$ that our binary BCJ relations, and the more familiar generalized BCJ relations~\cite{Bern:2008qj,BjerrumBohr:2009rd,Chen:2011jxa}, span the same vector space over polynomials, with the dimensions given in~\eqref{eq:basiscount}.  Indeed, there exists a closed formula that express a binary BCJ relation through the generalized BCJ relations~\cite{Chen:2019ywi}, but the inverse map is not known. A direct  construction of the independent basis for the generalized BCJ relations appears to be quite non-trivial, however under the assumption that they span the same space, the binary BCJ relations provides an easier linear space to work with.

\section{More on fusion products}\label{sec:AlgebraDiscussion}
To obtain the bi-scalar sector numerators from the pre-numerators given in section~\ref{sec:numerator}, one first needs to convert pre-numerators into kinematic numerators following the prescription~\eqref{eq:AfromFusionRules} and then take the terms proportional to $\varepsilon_1\Cdot\varepsilon_n$. The bi-scalar sector is defined~\cite{Chen:2019ywi} to be the terms in the YM numerator that are proportional to $\varepsilon_1\Cdot\varepsilon_n$. One may note that the bi-scalar numerators, as obtained from the closed formula in section~\ref{sec:numerator}, do not precisely match those given in ref.~\cite{Chen:2019ywi}. Instead, they are equivalent after taking into account a generalized gauge transformation. This is somewhat curious, since we have already pointed out in section~\ref{sec:numerator} that the $\mathcal{N}_V^{(1)}$ and $\mathcal{N}_{T}^{(2)}$ given in eq.~\eqref{eq:NvNt} combine exactly into the same form of the bi-scalar numerator as given in~\cite{Chen:2019ywi}. Intuitively, it should be possible to put our pre-numerator in a different gauge such that the terms proportional to $\varepsilon_1\Cdot\varepsilon_n$ agree exactly with the bi-scalar numerator.

In this appendix, we show that the above consideration can indeed be realized by choosing a different value for the free parameter $x_0$, which was fixed to $x_0=\frac{1}{4}$ in section~\ref{sec:construction}. We first give the fusion rules involving vector currents with generic $x_0$,
\begin{align}\label{eq:FRVx}
	J_{\varepsilon_i}(p)\star J_{\varepsilon_j}(p_j)&=\varepsilon_j\Cdot p J_{\varepsilon_i}(p{+}p_j)+\frac{1}{2}\varepsilon_i\Cdot\varepsilon_j J^{(1)}_{p_j}(p{+}p_j)-\frac{x_0}{2}J _{\varepsilon_i\otimes\varepsilon_j\otimes (p+p_{j})}(p{+}p_j), \nn\\
	J^{(1)}_{p_i}(p)\star J_{\varepsilon_j}(p_j)&=\varepsilon_j\Cdot p J^{(1)}_{p_i}(p{+}p_j)+\varepsilon_j\Cdot p_i J^{(1)}_{p_j}(p{+}p_j)+(x_0-1)s_{ij}J_{\varepsilon_j}^{(2)}(p{+}p_j)\nonumber\\
	&\quad-\frac{1}{4}J _{p_i\otimes\varepsilon_j\otimes (p+p_{j})}(p{+}p_j), \nn\\
	J_{\varepsilon_i}^{(2)}(p)\star J_{\varepsilon_j}(p_j)&=\varepsilon_j\Cdot p J_{\varepsilon_i}^{(2)}(p{+}p_j)-\varepsilon_i\Cdot p_{j} J_{\varepsilon_j}^{(2)}(p{+}p_j)-\frac{1}{2(x_0-1)}\varepsilon_i\Cdot\varepsilon_j J_{p_j}^{(2)}(p{+}p_j), \nn\\
	J_{p_i}^{(2)}(p)\star J_{\varepsilon_j}(p_j)&=\varepsilon_j\Cdot p J_{p_i}^{(2)}(p{+}p_j)+\varepsilon_j\Cdot p_i J_{p_j}^{(2)}(p{+}p_j).
\end{align}
Comparing with section~\ref{sec:construction}, the $x_0$ dependence modifies the following fusion rules,
\begin{align}
J^{(1)}_{\varepsilon_i\otimes\varepsilon_{i_1}\otimes \varepsilon_{i_2}}(p)\star  J_{\varepsilon_j}(p_j)&=\varepsilon_j\Cdot p  J _{\varepsilon_i\otimes\varepsilon_{i_1}\otimes \varepsilon_{i_2}}^{(2)}(p{+}p_j)+\varepsilon_{i_2}\Cdot p  J _{\varepsilon_i\otimes\varepsilon_{i_1}\otimes \varepsilon_j}^{(1)}(p{+}p_j)\nonumber\\
&\quad-\varepsilon_{i_1}\Cdot p  J _{\varepsilon_i\otimes\varepsilon_{i_2}\otimes \varepsilon_j}^{(1)}(p{+}p_j)+\frac{1}{4x_0}\Big[\varepsilon_j\Cdot\varepsilon_i J _{p_j\otimes\varepsilon_{i_1}\otimes \varepsilon_{i_2}}^{(4)}(p{+}p_j)\nonumber\\
&\quad+\varepsilon_{i_2}\Cdot\varepsilon_i J _{p_{i_2}\otimes\varepsilon_{i_1}\otimes \varepsilon_j}^{(3)}(p{+}p_j)-\varepsilon_{i_1}\Cdot\varepsilon_i J _{p_{i_1}\otimes\varepsilon_{i_2}\otimes \varepsilon_j}^{(3)}(p{+}p_j)\Big],
\end{align}
\begin{align}
J^{(1)}_{p_i\otimes\varepsilon_{i_1}\otimes \varepsilon_{i_2}}(p)\star  J_{\varepsilon_j}(p_j)&=\varepsilon_j\Cdot p  J _{p_i\otimes\varepsilon_{i_1}\otimes \varepsilon_{i_2}}^{(2)}(p{+}p_j)+\varepsilon_{i_2}\Cdot p  J _{p_i\otimes\varepsilon_{i_1}\otimes \varepsilon_j}^{(1)}(p{+}p_j)\nonumber\\
&\quad-\varepsilon_{i_1}\Cdot p  J _{p_i\otimes\varepsilon_{i_2}\otimes \varepsilon_j}^{(1)}(p{+}p_j)+\varepsilon_j\Cdot p_i J _{p_j\otimes\varepsilon_{i_1}\otimes \varepsilon_{i_2}}^{(4)}(p{+}p_j)\nonumber\\
&\quad+\varepsilon_{i_2}\Cdot p_i J _{p_{i_2}\otimes\varepsilon_{i_1}\otimes \varepsilon_j}^{(3)}(p{+}p_j)-\varepsilon_{i_1}\Cdot p_i J _{p_{i_1}\otimes\varepsilon_{i_2}\otimes \varepsilon_j}^{(3)}(p{+}p_j),
\end{align}
\begin{align}
J^{(2)}_{\varepsilon_i\otimes\varepsilon_{i_1}\otimes \varepsilon_{i_2}}(p)\star  J_{\varepsilon_j}(p_j)&=\varepsilon_j\Cdot p  J _{\varepsilon_i\otimes\varepsilon_{i_1}\otimes \varepsilon_{i_2}}^{(2)}(p{+}p_j)+\frac{1}{4x_0}\Big[\varepsilon_j\Cdot\varepsilon_i J _{p_j\otimes\varepsilon_{i_1}\otimes \varepsilon_{i_2}}^{(4)}(p{+}p_j)\nonumber\\
&\quad+\varepsilon_{i_2}\Cdot\varepsilon_i J _{p_{i_2}\otimes\varepsilon_{i_1}\otimes \varepsilon_j}^{(3)}(p{+}p_j)-\varepsilon_{i_1}\Cdot\varepsilon_i J _{p_{i_1}\otimes\varepsilon_{i_2}\otimes \varepsilon_j}^{(3)}(p{+}p_j)\Big],
\end{align}
\begin{align}
J^{(2)}_{p_i\otimes\varepsilon_{i_1}\otimes \varepsilon_{i_2}}(p)\star  J_{\varepsilon_j}(p_j)&=\varepsilon_j\Cdot p  J _{p_i\otimes\varepsilon_{i_1}\otimes \varepsilon_{i_2}}^{(2)}(p{+}p_j)+\varepsilon_j\Cdot p_i J _{p_j\otimes\varepsilon_{i_1}\otimes \varepsilon_{i_2}}^{(4)}(p{+}p_j)\nonumber\\
&\quad+\varepsilon_{i_2}\Cdot p_i J _{p_{i_2}\otimes\varepsilon_{i_1}\otimes \varepsilon_j}^{(3)}(p{+}p_j)-\varepsilon_{i_1}\Cdot p_i J _{p_{i_1}\otimes\varepsilon_{i_2}\otimes \varepsilon_j}^{(3)}(p{+}p_j),
\end{align}
while the other tensors follow the same fusion rules as in section~\ref{sec:construction}.

The above fusion rules will lead to an $x_0$ dependence in the pre-numerator,
\begin{align}\label{eq:BCJAllx}
\npre(1,2,\ldots,n)=\Big(1+\frac{1}{2}Q_n\Big)\npre^{(1)}_{V}+\Big(x_0+\frac{1}{4}Q_n\Big) \npre^{(2)}_T+\npre^{(2)}_V(x_0)\,,
\end{align}
where $\mathcal{N}^{(1)}_{V}$, $\mathcal{N}^{(2)}_T$ and the operator $Q_n$ are defined as in section~\ref{sec:numerator}, while $\mathcal{N}^{(2)}_{V}$ develops an $x_0$ dependence,
\begin{align}
\npre^{(2)}_V(x_0)&=-\frac{1}{4}\Bigg[\sum_{j=2}^{n-3}(\varepsilon_1\Cdot G_{2,\ldots,j}\Cdot p_{j+1})\sum_{\ell=j+1}^{n-2}(\varepsilon_{j+1}\Cdot \tilde{G}_{j+2,\ldots,\ell}\Cdot \varepsilon_{\ell+1})(p_{\ell+1}\Cdot G_{\ell+2,\ldots,n-1}\Cdot\varepsilon_n)\Bigg]_{(\varepsilon\Cdot\varepsilon)^2} \nn\\
&\quad -\frac{1-x_0}{2} \Bigg[\sum_{j=2}^{n-2}(\varepsilon_1\Cdot G_{2,\ldots,j}\Cdot  p_{j+1})(\varepsilon_{j+1}\Cdot \tilde{G}_{j+2,\ldots,n-1}\Cdot\varepsilon_n)\Bigg]_{(\varepsilon\Cdot\varepsilon)^2}\,.
\end{align} 
The action of $Q_n$ only generates $\varepsilon_n\Cdot p_i$ terms, so that they do not contribute to the bi-scalar sector. The first bracket of $\mathcal{N}^{(2)}_{V}$ contains $\varepsilon_n\Cdot p_i$ terms only, and thus does not contribute to the bi-scalar sector either. On the other hand, the second bracket of $\mathcal{N}^{(2)}_V$ contains $\varepsilon_i\Cdot\varepsilon_n$ terms, which will generate $\varepsilon_1\Cdot\varepsilon_n$ terms under the action of nested commutators. This is the reason that for generic $x_0$, we need to turn the pre-numerators into kinematic numerators before restricting to the bi-scalar sector. However, if we pick $x_0=1$, the second bracket of $\mathcal{N}^{(2)}_V$ vanishes, such that we can obtain the bi-scalar sector numerator by directly taking the terms proportional to $\varepsilon_1\Cdot\varepsilon_n$ in the pre-numerator~\eqref{eq:BCJAllx}. The result is
\begin{align}
N_{\text{bi-scalar}}(1,2,\ldots,n)=\mathcal{N}(1,2,\ldots,n)\Big|^{x_0=1}_{\varepsilon_1\Cdot\varepsilon_n}=\mathcal{N}^{(1)}_V+\mathcal{N}^{(2)}_T\,,
\end{align}
which exactly agrees with the numerator given in~\cite{Chen:2019ywi}. Note that $x_0=1$ is a singular gauge choice. We cannot impose it in the fusion products~\eqref{eq:FRVx}, but only in the pre-numerator~\eqref{eq:BCJAllx}.

\section{The rank of \texorpdfstring{$\neco$}{L}-operators}\label{sec:proof}
In this appendix, we derive the general formula for the rank of the $\neco$-operator. As mentioned in eq.~\eqref{eq:rankR}, due to a rank-trace relation of idempotent operators, the rank of $\neco$ is proportional to the trace of $\neco$. We first formally write $\neco$ as a sum of conjugacy classes $\mathsf{K}(\lambda)$ and the permutations therein, with coefficient $e_{\sigma}$,
\begin{align}\label{eq:rtot2}
\text{Rank}_{\mec}\big[\neco(1,2,\ldots,n)\big]=\frac{1}{n}\text{Tr}_{\mec}\big[\neco(1,2,\ldots,n)\big]&=\sum_{\lambda}\sum_{\sigma\in\mathsf{K}(\lambda)}e_{\sigma}\text{Tr}_{\mec}\big[\mathbb{P}_{\sigma}\big]\nonumber\\
&=\sum_{\lambda}\chi_{\mec}\big[\mathsf{K}(\lambda)\big]\sum_{\sigma\in\mathsf{K}(\lambda)}e_{\sigma}\,.
\end{align}
To obtain the second line, we have used $\text{Tr}_{\mec}\big[\mathbb{P}_{\sigma}\big]=\chi_{\mec}\big[\mathsf{K}(\lambda)\big]$ for all $\sigma\in\mathsf{K}(\lambda)$. Next, for a Lie idempotent, the sum of $e_{\sigma}$ in a conjugacy class is given by Theorem~8.14 of~\cite{reutenauer2003free},
\begin{align}\label{eq:sume}
\sum_{\sigma\in\mathsf{K}(\lambda)}e_{\sigma}=\left\{\begin{array}{ccl}
\frac{\mu(p)}{n} & \quad\quad & \text{if }\lambda= p^{n/p} \\
0 & \quad & \text{otherwise}
\end{array}
\right.\,,
\end{align}
where $p$ is an integer that divides $n$ and $\mu(p)$ is the M\"{o}bius function.\footnote{For an integer $p$, $\mu(p)=0$ if $p$ contains a squared prime factor. Otherwise, $\mu(p)=1$ if $p$ contains an even number of prime factors and $\mu(p)=-1$ if $p$ contains an odd number of prime factors.} Here $p^{n/p}$ denotes the conjugacy class in which the permutations all have $n/p$ cycles with the same length~$p$, 
\begin{align}
p^{n/p}\equiv (\underbrace{p,p,\ldots,p}_{n/p})\,.
\end{align} 
For example, at $n{=}4$, the permutation $\sigma=(12)(34)$ belongs to $\mathsf{K}(2^{4/2})\equiv\mathsf{K}(2,2)$. Similarly, at $n=6$, the permutation $\sigma=(134)(256)$ belongs to $\mathsf{K}(3^{6/3})\equiv\mathsf{K}(3,3)$. In other words, the sum is potentially nonzero only for these conjugacy classes. Plugging eq.~\eqref{eq:sume} back to eq.~\eqref{eq:rtot2}, we get our final result
\begin{align}\label{eq:rankRgeneral}
\text{Rank}_{\mec}\big[\neco(1,2,\ldots,n)\big]=\frac{1}{n}\sum_{p\,|\,n}\mu(p)\,\chi_{\mec}\big[\mathsf{K}(p^{n/p})\big]\,,
\end{align}
where the summation is over those $p$ that divides $n$, with $1$ and $n$ itself included. Now we consider some special cases that further simplify the above formula. For $n$ being a prime number, the only divisors are $1$ and $n$ itself, such that eq.~\eqref{eq:sume} is nonzero only for $\mathsf{K}(\lambda)=\mathsf{K}(n)$ and the identity. Since $\mu(n)=-1$ for all the prime numbers and $\mu(1)=1$, we have
\begin{align}
\text{Rank}_{\mec}\big[\neco(1,2,\ldots,n)\big]=\frac{1}{n}\Big(\Dc-\chi_{\mec}\big[\mathsf{K}(n)\big]\Big)\quad\text{ for prime }n\,.
\end{align}
Next, we consider the case that $n=r^m$ where $r$ is a prime number. All the divisors of $n$ have the form $p=r^q$ with $0\leqslant q\leqslant m$. On the other hand, all the $p=r^q$ with $2\leqslant q\leqslant m$ contains a squared prime factor $r^2$, such that $\mu(p)=0$. Therefore, the only nonzero contribution comes from $\mu(1)=1$ and $\mu(r)=-1$, which leads to
\begin{align}
\text{Rank}_{\mec}\big[\neco(1,2,\ldots,n)\big]&=\frac{1}{n}\Big(\chi_{\mec}\big[\mathbb{I}\big]-\chi_{\mec}\big[\mathsf{K}(r^{n/r})\big]\Big)\nonumber\\
&=\frac{1}{n}\Big(\Dc-\chi_{\mec}\big[\mathsf{K}(\underbrace{r,r,\ldots,r}_{r^{m-1}})\big]\Big)\quad\begin{array}{c}\text{ for }n=r^m \\ \text{ and prime }r\end{array}\,.
\end{align}
The examples in eq.~\eqref{eq:rankLexample} include both these special cases. For $n\leqslant 100$, eq.~\eqref{eq:rankRgeneral} gives at most eight terms.

\section{YM Feynman rules as fusion products}\label{sec:feynman}
To compute a cubic-graph YM numerator, $N_{\Gamma}$ in eq.~\eqref{eq:AfromFusionRules}, a standard approach is to first use Feynman rules and then resolve the quartic contact vertices into cubic ones by spurious propagator insertions. However, the numerator obtained this way does not satisfy the color-kinematics duality beyond four points. The same cubic-graph numerators can be computed from repeated use of fusion products, following much of the framework used in this paper. We represent every edge of the graph $\Gamma$ by some combination of a vector and a tensor current. In particular, every external leg $i$ in $\Gamma$ is assigned a vector current $\mathcal{J}_{\varepsilon_i}(p_i)$, and the internal currents are generated by the fusion rules. The commutator brackets in the graph $\Gamma$ are replaced by fusion products that obey the rules
\begin{align}\label{eq:FeynRule}
&\mathcal{J}_{\varepsilon_i}(p_{\mathsf{L}})\diamond \mathcal{J}_{\varepsilon_j}(p_{\mathsf{R}})= \varepsilon_j\Cdot p_{\mathsf{L}} \mathcal{J}_{\varepsilon_i}(p_{\mathsf{LR}})-\varepsilon_i\Cdot p_{\mathsf{R}} \mathcal{J}_{\varepsilon_j}(p_{\mathsf{LR}})-\frac{1}{2}\varepsilon_i\Cdot \varepsilon_j \big[\mathcal{J}_{p_{\mathsf{L}}}(p_{\mathsf{LR}})-\mathcal{J}_{p_{\mathsf{R}}}(p_\mathsf{LR})\big]\nn\\
&\qquad\qquad\qquad\qquad-\frac{1}{2} \mathcal{J}_{\varepsilon_i\otimes\varepsilon_j\otimes p_{\mathsf{LR}}}(p_{\mathsf{LR}})+\frac{1}{2}\mathcal{J}_{\varepsilon_j\otimes\varepsilon_i\otimes p_{\mathsf{LR}}}(p_{\mathsf{LR}})\,,\nn\\
&\mathcal{J}_{\varepsilon_i\otimes\varepsilon_j\otimes p_{\mathsf{L}}}(p_{\mathsf{L}})	\diamond \mathcal{J}_{\varepsilon_k}(p_{\mathsf{R}}) = -\frac{1}{4}p_{\mathsf{L}}^2 \varepsilon_i\Cdot\varepsilon_k \mathcal{J}_{\varepsilon_j}(p_{\mathsf{LR}})+\frac{1}{4}p_{\mathsf{L}}^2 \varepsilon_j\Cdot\varepsilon_k \mathcal{J}_{\varepsilon_i}(p_{\mathsf{LR}})\,,\nn\\
&\mathcal{J}_{\varepsilon_k}(p_{\mathsf{L}})	\diamond \mathcal{J}_{\varepsilon_i\otimes\varepsilon_j\otimes p_{\mathsf{R}}}(p_{\mathsf{R}}) = \frac{1}{4}p_{\mathsf{R}}^2 \varepsilon_i\Cdot\varepsilon_k \mathcal{J}_{\varepsilon_j}(p_\mathsf{LR})-\frac{1}{4}p_{\mathsf{R}}^2 \varepsilon_j\Cdot\varepsilon_k \mathcal{J}_{\varepsilon_i}(p_{\mathsf{LR}})\,,\nn\\
& \mathcal{J}_{\varepsilon_i\otimes\varepsilon_j\otimes p_{\mathsf{L}}}(p_{\mathsf{L}})	\diamond \mathcal{J}_{\varepsilon_i\otimes\varepsilon_j\otimes p_{\mathsf{R}}}(p_{\mathsf{R}}) =0\,,
\end{align}
where $p_{\mathsf{L}}$ and $p_{\mathsf{R}}$ are the momenta of the two incident legs, and $p_{\mathsf{LR}}=p_{\mathsf{L}}{+}p_{\mathsf{R}}$ is the momentum of the outgoing leg. Only one vector $\mathcal{J}_{\varepsilon_i}$ and one tensor $\mathcal{J}_{\varepsilon_i\otimes\varepsilon_j\otimes p}$ are needed for fully describing (tree-level) pure YM theory. Here the vector $\mathcal{J}_{p}$ is the same object as $\mathcal{J}_{\varepsilon_i}\big|_{\varepsilon_i\rightarrow p}$ and not an independent current. Similar to a commutator $[A,B]$, this fusion product is antisymmetric in its two arguments and not associative, however, unlike a commutator it does not obey a Jacobi identity. 

Finally, the numerator $N_{\Gamma}$ is given by the standard sandwich between spinors defined in \eqn{eq:tensor} and (\ref{eq:vector}),
\begin{align}
N_{\Gamma}=\langle q| \mathcal{J}[\commut]|n\rangle\Big|_{q~\rm soft}\,,
\end{align}
where $\mathcal{J}[\commut]$ is the resulting object after replacing external particles with vector currents, and commutators with fusion products. 
For example, the $s$- and $t$-channel numerator at four points are given by
\begin{align}
N_s=\langle q|\big((\mathcal{J}_{\varepsilon_1}\diamond\mathcal{J}_{\varepsilon_2})\diamond\mathcal{J}_{\varepsilon_3}\big)|n\rangle\Big|_{q~\rm soft}\,,\quad N_t=\langle q|\big(\mathcal{J}_{\varepsilon_1}\diamond(\mathcal{J}_{\varepsilon_2}\diamond\mathcal{J}_{\varepsilon_3})\big)|n\rangle\Big|_{q~\rm soft}\,.
\end{align}
This process is analogous to a local version of Berends-Giele recursion, which involves the vector field $A^\mu$ and an auxiliary tensor field  $B^{\mu\nu\rho}$, where the latter is responsible for removing the quartic vertex of YM. The vector and tensor currents are thus in one-to-one correspondence with these fields.

\bibliographystyle{JHEP}
\bibliography{ScatEq}

\end{document}